\documentclass[12pt]{article}

\usepackage{verbatim}
\usepackage{amsmath}
\usepackage{amssymb}
\usepackage{graphicx}
\usepackage{cite}
\usepackage{subfig}
\usepackage{setspace}
\usepackage{url}
\usepackage[percent]{overpic}
\usepackage{slashed}
\usepackage{authblk}

\usepackage{xspace}

\usepackage{fullpage}
\usepackage{hyperref}
\def\tev{\,{\rm TeV}}
\def\gev{\,{\rm GeV}}
\def\ie{{\it i.e.}}
\def\eg{{\it e.g.}}

\def\to{\rightarrow}

\title{Lessons and Prospects from the pMSSM after LHC \\Run I: Neutralino LSP}
\author{M. Cahill-Rowley$^a$}
\author{J.L. Hewett$^a$}
\author{A. Ismail$^{b,c}$}
\author{T.G. Rizzo$^a$}

\affil{$^a$~SLAC National Accelerator Laboratory, Menlo Park, CA, USA\footnote{mrowley, hewett, rizzo@slac.stanford.edu}}
\affil{$^b$~Argonne National Laboratory, Argonne, IL, USA\footnote{aismail@anl.gov}}
\affil{$^c$~University of Illinois at Chicago, Chicago, IL, USA}

\begin{document}

\rightline{\vbox{\halign{&#\hfil\cr
&SLAC-PUB-15874\cr
}}}


{\let\newpage\relax\maketitle}

\begin{abstract}

We study SUSY signatures at the 7, 8 and 14 TeV LHC employing the 19-parameter, R-Parity conserving p(henomenological)MSSM, in the scenario with
a neutralino LSP. 
Our results were obtained via a fast Monte Carlo simulation of the ATLAS SUSY analysis suite. The flexibility of this framework allows us 
to study a wide variety of SUSY phenomena simultaneously and to probe for weak spots in existing SUSY search analyses. We determine the 
ranges of the sparticle masses that are either disfavored or allowed after the searches with the 7 and 8 TeV data sets are combined.  We find that natural
SUSY models with light squarks and gluinos remain viable. We extrapolate 
to 14 TeV with both 300 fb$^{-1}$ and 3 ab$^{-1}$ of integrated luminosity and determine the expected sensitivity of the jets + MET and stop searches to the pMSSM 
parameter space. We find that the high-luminosity LHC will be powerful in probing SUSY with neutralino LSPs and can provide a more definitive statement
on the existence of natural Supersymmetry.

\end{abstract}

\newpage

\section{Overview of the pMSSM}

Even though the last missing piece of the Standard Model (SM), the Higgs boson, has recently been discovered at the LHC{\cite {ATLASH,CMSH}}, we know that the SM is not a complete description of particle physics. For example, the SM does not provide a candidate particle for Dark Matter. 
In addition, there are good arguments which suggest at least some 
of the deficiencies in the SM may be addressed by new physics at the TeV scale. The LHC has performed extensive searches for
new physics at both 7 and 8 TeV, and these searches will be 
resumed with increased fervor at 14 TeV within approximately one year's time. In most cases, searches for new physics are hampered by large backgrounds from conventional SM processes, and experimenters have developed (and must continue to develop) clever techniques for extracting signals in these increasingly challenging conditions. 
For any new model, it is important to know whether it can be discovered or excluded at the LHC given the
backgrounds. The first crucial step in answering this question is to consider the range of potential signals that the model may exhibit, 
and then determine how well  the experimental analyses can probe its interesting parameter space. Employing this process is particularly important for Supersymmetry (SUSY), which remains the most attractive, well-motivated, and most widely explored new physics framework, despite the persistent absence of any direct experimental evidence for sparticles. Of course, determining which signatures of SUSY may be observed at the LHC is non-trivial, since SUSY can appear in many different guises as it is a theoretical framework rather than a specific model. Even in the simplest manifestation of SUSY, the R-Parity 
conserving Minimal Supersymmetric SM (MSSM), the number of free soft-breaking parameters ($\sim$ 100) is much too large to study in complete generality. Various approaches have been developed to address 
this large obstacle. Historically, the first idea was to reduce the number of independent parameters by postulating high-scale theories with specific mechanisms for SUSY breaking; this predicted specific relationships between the soft parameters and dramatically reduced the dimensionality of the parameter space. Such high-scale theories (an example being mSUGRA{\cite {SUSYrefs}) therefore have only a few parameters, from which all the properties of  the
sparticle spectrum at the TeV scale can be determined and studied in great detail. While such approaches are often extremely valuable~\cite{Cohen:2013kna}, they are somewhat phenomenologically limiting and many of them face 
ever-increasing tension with a wide range of experimental data (including the $\sim 126$ GeV mass of the Higgs boson) as a result of {\it insufficient} parameter freedom.

One possible approach to circumvent such limitations is to employ the more general 19-parameter p(henomenological)MSSM{\cite{Djouadi:1998di}}. The increased dimensionality of the 
parameter space not only allows for a more unprejudiced study of the MSSM, but also yields valuable information on `unusual' scenarios, identifies potential weaknesses (or gaps) 
in the LHC analyses and provides a framework to combine the results obtained from many independent SUSY-related searches. 
With this motivation, we have recently embarked on a detailed study of pMSSM signatures at the 7 and 8 TeV LHC, supplemented by input from Dark Matter 
experiments as well as from precision electroweak and flavor measurements{\cite {us1,us2a,us2b,us4,lhc-snowmass}}, and continue
this effort in this work. The pMSSM is defined as the most general version of the 
R-parity conserving MSSM subjected to a minimal set of experimentally-motivated guiding principles: ($i$) CP conservation, ($ii$) Minimal Flavor Violation 
at the electroweak scale so that flavor physics is essentially controlled by the CKM mixing matrix, ($iii$) degenerate 1\textsuperscript{st} and 2\textsuperscript{nd} 
generation sfermion soft-mass parameters (\eg, right-handed up and charm squarks are degenerate apart from small corrections due to non-zero quark masses), and ($iv$) negligible 
Yukawa couplings and A-terms are assumed for the first two generations. In particular, assumptions about physics at high scales, {\it e.g.}, grand unification or 
the nature of SUSY breaking, are not present in order to capture electroweak scale phenomenology for which
 a UV-complete theory may not yet exist. Imposing these principles 
($i$)-($iv$) results in a decrease in the number of free parameters in the MSSM at the TeV-scale from 105 to 19 for the case of a neutralino LSP (or to 20 when the 
gravitino mass is included as an additional parameter when it plays the role of the LSP.)  In what follows we will only consider the case where the 
neutralino is the LSP, although we will make some comparisons with the case of a gravitino LSP;
corresponding results for models where the LSP is a gravitino will be discussed in detail in a companion paper.

Dark matter constraints play a very important role in restricting the allowed parameter space of any R-Parity conserving SUSY scenario. In performing our analyses, we have not assumed 
that the LSP thermal relic density necessarily saturates the WMAP/Planck value{\cite{Komatsu:2010fb}} in order to allow for the possibility of multi-component DM. For 
example, axions, which can be introduced to solve the strong CP problem, may constitute a substantial amount of dark matter. 
One can imagine considering the possibility of non-thermal mechanisms in the early universe which increased or diluted the LSP density; this would relax constraints arising from overproduction of the LSP (such as a bino LSP without co-annihilation or resonant annihilation). 
However, we have not taken this route here and leave this possibility for future study. The 19/20 pMSSM parameters, and the ranges over which they are scanned, are presented in Table~\ref{ScanRanges}. Like throwing darts, we generate many millions of model points in this space (using SOFTSUSY{\cite{Allanach:2001kg}} and checking for consistency with SuSpect{\cite{Djouadi:2002ze}}). Each point (which we also call a pMSSM model) corresponds to a specific set of values for these parameters. 
These individual models are then subjected to a global set of collider, flavor, precision measurement, dark matter and theoretical\footnote{We note that after our scan was completed, an updated study of vacuum stability in the pMSSM, one of the theoretical constraints included, was performed~\cite{Camargo-Molina:2014pwa}.} constraints~\cite{us1}.  
Roughly $\sim$225k models with either type of LSP (neutralino and gravitino) 
survive this initial selection and can then be used for further physics studies. We calculate decay patterns of the SUSY partners and 
the extended Higgs sector using privately modified versions of SUSY-HIT~\cite{Djouadi:2006bz}, CalcHEP~\cite{calchep}, and MadGraph~\cite{madgraph}. 
Such modifications are necessary to correctly implement the pMSSM in these public codes.
Our scan ranges include sparticle masses up to 4 TeV,  with this upper limit being chosen to enable phenomenological studies at the 14 TeV LHC.
This implies that the neutralinos and charginos in 
our model sets are typically very close to being a pure electroweak eigenstate as the off-diagonal elements of the corresponding mass matrices are at most 
of order $\sim M_W$.

In addition to these two large pMSSM model sets, we have also generated a smaller, specialized set with a neutralino LSP in order to explore the
effectiveness of the LHC in constraining natural Supersymmetry.  This model sample contains $\sim$ 10.2k `natural' models, all of which 
predict $m_h = 126 \pm 3$ GeV, have an LSP that {\it does} saturate the WMAP/Planck relic density, and produce values of fine-tuning (FT) better than $1\%$ using the 
Ellis-Barbieri-Giudice measure~\cite{Ellis:1986yg, Barbieri:1987fn}.  In order to produce this model 
set, we modified the parameter scan ranges as indicated in Table~\ref{ScanRanges} to greatly increase the likelihood that a chosen point will satisfy the combined relic density, 
Higgs mass, and FT constraints.   In addition to these modified scan ranges, we also required $|M_1/\mu|<1.2$ and $|X_t|/m_{\tilde t} >1$, where
$X_t=A_t-\mu\cot\beta$ quantifies the mixing between the stop-squarks and $m_{\tilde t}$ is the geometric mean of the tree-level stop masses.
Amongst other things, satisfying these requirements necessitates a bino LSP, as well as light Higgsinos and 
highly-mixed stops. We generated $\sim 3.3 \times 10^8$ low-FT points in this 19-dimensional
parameter space and subjected them to the global precision, flavor, dark matter and collider 
constraints as with the other model samples. 
Since the requirements are much stricter here than for our two larger model sets, only $\sim$ 10.2k low-FT models survive the constraints and are available
for further study. 

We now subject these sets of pMSSM models to the suite of  SUSY searches performed at the 7 and 8 TeV LHC, as well as planned searches at the 14 TeV LHC, forming the content of the remainder of this paper.

\begin{table}
\centering
\begin{tabular}{|c|c|c|} \hline\hline
Parameter & General Neutralino Set & Low Fine-Tuned Set \\
\hline\hline
$m_{\tilde L(e)_{1/2,3}}$ & $100 \gev - 4 \tev$ & $100 \gev - 4 \tev$ \\ 
$m_{\tilde Q(u,d)_{1/2}}$ & $400 \gev - 4 \tev$ & $100 \gev - 4 \tev$  \\ 
$m_{\tilde Q(u,d)_{3}}$ &  $200 \gev - 4 \tev$ &  $100 \gev - 4 \tev$ \\
$|M_1|$ & $50 \gev - 4 \tev$ & $25 \gev - 552 \gev$ \\
$|M_2|$ & $100 \gev - 4 \tev$ &  $100 \gev - 2.1\tev$ \\
$|\mu|$ & $100 \gev - 4 \tev$ & $100 \gev - 460 \gev$ \\ 
$M_3$ & $400 \gev - 4 \tev$ &  $400 \gev - 4 \tev$ \\ 
$|A_{t,b,\tau}|$ & $0 \gev - 4 \tev$ & $0 \gev - 2.3 \tev$ ($A_t$ only) \\ 
$M_A$ & $100 \gev - 4 \tev$ &  $100 \gev - 4 \tev$ \\ 
$\tan \beta$ & $1 - 60$ & $1 - 60$ \\
\hline\hline
\end{tabular}
\caption{Scan ranges for the 19 parameters of the pMSSM with a neutralino LSP. The parameters are scanned with flat priors; we expect this choice to have little qualitative impact on 
our results for observables~\cite{us}.}
\label{ScanRanges}
\end{table}

\section{7 and 8 TeV LHC SUSY Searches}

We begin this discussion with a short overview of our approach in applying the
searches for Supersymmetry at the 7 and 8 TeV LHC to the pMSSM; the same overall method will also be utilized in our 14 TeV study. 
In general, we follow the suite of ATLAS SUSY analyses as closely as possible employing fast Monte Carlo, and supplement these with several searches 
performed by CMS. The specific analyses applied to the neutralino model set as discussed below are listed in Tables~\ref{SearchList7} and~\ref{SearchList8}. 
We augment the MET-based SUSY searches by including the search for a heavy neutral SUSY Higgs decaying to $\tau^+\tau^-$ performed by CMS~\cite{Chatrchyan:2012vp} as well as 
measurements of the rare decay mode $B_s\to \mu^+\mu^-$ as discovered by CMS and LHCb~\cite{BSMUMU}.  Both of these play distinct and important roles in exploring 
the pMSSM parameter space. We have implemented every relevant ATLAS SUSY search publicly available as of the beginning of March 2013, 
as well as the more 
recently available $\sim 20$ fb$^{-1}$ jets + MET analysis with the latter search being quite powerful at covering SUSY parameter space.

\begin{table}
\centering
\begin{tabular}{|l|l|c|c|c|} \hline\hline
Search & Reference & Neutralino & Gravitino & Low-FT   \\
\hline
2-6 jets & ATLAS-CONF-2012-033  & 21.2\% &  17.4\% & 36.5\% \\
multijets & ATLAS-CONF-2012-037 & 1.6\%  & 2.1\% & 10.6\% \\
1 lepton & ATLAS-CONF-2012-041 & 3.2\%  & 5.3\% & 18.7\%  \\

HSCP      &  1205.0272  & 4.0\% & 17.4\% & $<$0.1\%  \\
Disappearing Track  & ATLAS-CONF-2012-111 & 2.6\%  & 1.2\% & $<$0.1\% \\
Muon + Displaced Vertex  & 1210.7451 & - & 0.5\% & - \\
Displaced Dilepton & 1211.2472 & - & 0.8\% & - \\

Gluino $\to$ Stop/Sbottom   & 1207.4686 & 4.9\% &  3.5\% & 21.2\% \\
Very Light Stop  & ATLAS-CONF-2012-059 & $<$0.1\% & $<$0.1\% & 0.1\%  \\
Medium Stop  & ATLAS-CONF-2012-071 & 0.3\% & 5.1\% & 2.1\% \\
Heavy Stop (0$\ell$)  & 1208.1447 & 3.7\% & 3.0\% & 17.0\% \\
Heavy Stop (1$\ell$)   & 1208.2590 & 2.0\% & 2.2\% & 12.6\% \\
GMSB Direct Stop  & 1204.6736 & $<$0.1\% & $<$0.1\% & 0.7\% \\
Direct Sbottom & ATLAS-CONF-2012-106 & 2.5\% & 2.3\% & 5.1\% \\
3 leptons & ATLAS-CONF-2012-108 & 1.1\% & 6.1\% & 17.6\% \\
1-2 leptons & 1208.4688 & 4.1\% & 8.2\% & 21.0\% \\
Direct slepton/gaugino (2$\ell$)  & 1208.2884 & 0.1\% & 1.2\% & 0.8\% \\
Direct gaugino (3$\ell$) & 1208.3144 & 0.4\% & 5.4\% & 7.5\% \\
4 leptons & 1210.4457 & 0.7\% & 6.3\% & 14.8\% \\
1 lepton + many jets & ATLAS-CONF-2012-140 & 1.3\% & 2.0\% & 11.7\% \\
1 lepton + $\gamma$ & ATLAS-CONF-2012-144 & $<$0.1\% & 1.6\% & $<$0.1\% \\
$\gamma$ + b & 1211.1167 & $<$0.1\% & 2.3\% & $<$0.1\% \\
$\gamma \gamma $ + MET & 1209.0753 & $<$0.1\% & 5.4\% & $<$0.1\% \\

$B_s \to \mu \mu$ & 1211.2674 & 0.8\% & 3.1\% & * \\
$A/H \to \tau \tau$ & CMS-PAS-HIG-12-050 & 1.6\% & $<$0.1\% & * \\

\hline\hline
\end{tabular}
\caption{7 TeV LHC searches included in the present analysis and the corresponding fraction of the neutralino, gravitino and low-FT pMSSM 
model sets excluded by each search channel. Note that in the case of the last two rows, the experimental constraints were included 
in the model generation process for the low-FT model set and therefore are not applicable here.}
\label{SearchList7}
\end{table}

\begin{table}
\centering
\begin{tabular}{|l|l|c|c|c|} \hline\hline
Search & Reference & Neutralino & Gravitino & Low-FT    \\
\hline

2-6 jets   & ATLAS-CONF-2012-109 & 26.7\% & 22.5\% & 44.9\% \\
multijets   & ATLAS-CONF-2012-103 & 3.3\% & 5.6\% & 20.9\% \\
1 lepton     & ATLAS-CONF-2012-104 & 3.3\% & 6.0\% & 20.9\% \\
SS dileptons & ATLAS-CONF-2012-105 & 4.9\% & 12.5\% & 35.5\% \\
2-6 jets   & ATLAS-CONF-2013-047 & 38.0\% & 31.1\% & 56.5\% \\

HSCP      &  1305.0491  & - & 23.0\% & -  \\

Medium Stop (2$\ell$) & ATLAS-CONF-2012-167 & 0.6\% & 8.1\% & 4.9\% \\
Medium/Heavy Stop (1$\ell$) & ATLAS-CONF-2012-166 & 3.8\% & 4.5\% & 21.0\% \\
Direct Sbottom (2b) & ATLAS-CONF-2012-165 & 6.2\% & 5.1\% & 12.1\% \\
3rd Generation Squarks (3b) & ATLAS-CONF-2012-145 & 10.8\% & 9.9\% & 40.8\% \\
3rd Generation Squarks (3$\ell$) & ATLAS-CONF-2012-151 & 1.9\% & 9.2\% & 26.5\% \\
3 leptons & ATLAS-CONF-2012-154 & 1.4\% & 8.8\% &32.3\% \\
4 leptons & ATLAS-CONF-2012-153 & 3.0\% & 13.2\% & 46.9\% \\
Z + jets + MET & ATLAS-CONF-2012-152 & 0.3\% & 1.4\% &6.8\% \\

\hline\hline
\end{tabular}
\caption{Same as in the previous table but now for the 8 TeV ATLAS MET-based SUSY search channels. Note that when all the searches from this table and the previous table are 
combined for the neutralino 
(gravitino, low-FT) model set we find that $\sim 45.5~(61.3,~74.0)\%$ of these models are currently excluded by the LHC.}
\label{SearchList8}
\end{table}

A brief overview of our procedure is as follows:~We generate SUSY events for each model for all relevant (up to 85) production channels in 
PYTHIA 6.4.26~\cite{Sjostrand:2006za}, and then pass the events through fast detector simulation using PGS 4~\cite{PGS}. Both programs have been modified to, \eg, 
correctly deal with gravitinos, multi-body decays, hadronization of stable colored sparticles, and ATLAS b-tagging. We scale our event rates to NLO by calculating 
the relevant K-factors using Prospino 2.1~\cite{Beenakker:1996ch}. The individual searches are then implemented using our customized analysis code{\cite {us}}, which 
follows the published cuts and selection criteria employed by ATLAS as closely as possible. This code is validated for each of the many search regions in every 
analysis employing the benchmark model points provided by ATLAS (and CMS). Models are then excluded using the 95\% CL$_s$ limits as computed by ATLAS (and CMS). 

For the large neutralino model set, these analyses are performed {\it without} implementing the Higgs mass constraint, $m_h = 126 \pm 3$ GeV (combined experimental and theoretical 
errors), so that we can understand its influence on the search results. Note that roughly $20\%$ of models in the large neutralino model set predict a Higgs mass in the above 
range. While there is some variation amongst the individual search channels themselves, we find that, once combined, the total fraction of models surviving the set of all 
LHC searches is essentially {\it independent} of whether or not the Higgs mass constraint has been imposed. Conversely, the fraction of 
neutralino models predicting the correct Higgs mass is also found to remain at $\sim 20\%$, approximately independent of whether the LHC SUSY search constraints have been applied. These results can be seen explicitly in Fig.~\ref{figm1}, which shows the predicted Higgs mass distribution in the large 
neutralino model set both before and after the LHC SUSY search results 
have been imposed. This result is very powerful and demonstrates the approximate decoupling of SUSY search results from the discovery of the Higgs boson;
{\it i.e.}, the LHC SUSY search efficiency is roughly independent of the Higgs mass.  This allows us
to continue examining the effects of the LHC SUSY searches on the 
entire model set for either LSP type (thus increasing our statistical sample) with some reasonable validity. 
We also find that the LHC SUSY searches have a rather small effect on the distributions of values for the 
Higgs branching fractions into various final states within the pMSSM, as demonstrated in {\cite {Higgs}} and {\cite {newHiggs}}.

\begin{figure}[htbp]
\centerline{\includegraphics[width=5.0in]{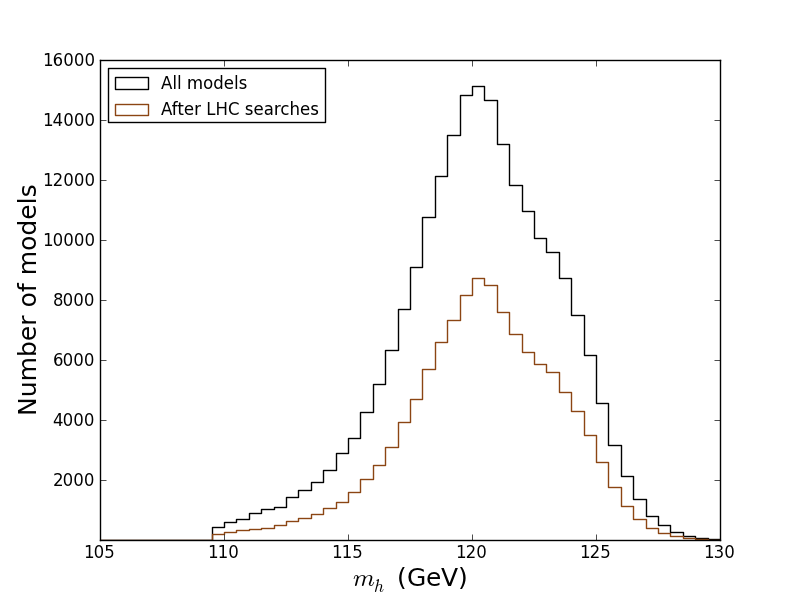}}
\vspace*{-0.10cm}
\caption{The distribution of the predicted Higgs mass before and after the LHC search constraints are applied to the large neutralino LSP model sample as indicated. 
Note that the LHC search efficiency is essentially independent of the Higgs mass.}
\label{figm1}
\end{figure}

\subsection{The Large Neutralino Model Set}
\label{sec:neutlsp}

We will first discuss the results of our analysis for the case of the general neutralino LSP model set. The first important question to address is how well the combined LHC SUSY searches cover the pMSSM parameter space.  One way to approach this is to project the multi-dimensional space of sparticle masses into two-dimensional slices and show the fraction of models excluded by the set of combined LHC searches within specified mass ranges.  These results for a variety of sparticles in the general 
neutralino LSP model sample are presented in Figs.~\ref{fig00}, ~\ref{fig1}, ~\ref{fig2} and ~\ref{fig3}. In addition, Tables~\ref{SearchList7} and~\ref{SearchList8} provide 
further information by listing the fraction of the neutralino pMSSM set that is excluded by each of the 
individual LHC searches (this information is also provided for the low-FT and gravitino LSP model samples). Combining all of the searches we find that $\sim 45.5\%$ of the
large neutralino LSP model sample is currently excluded. Clearly this implies that a 
large fraction of the excluded models are eliminated by more than one search, and, in fact, many models are excluded by several searches.

\begin{figure}[htbp]
\centerline{\includegraphics[width=3.5in]{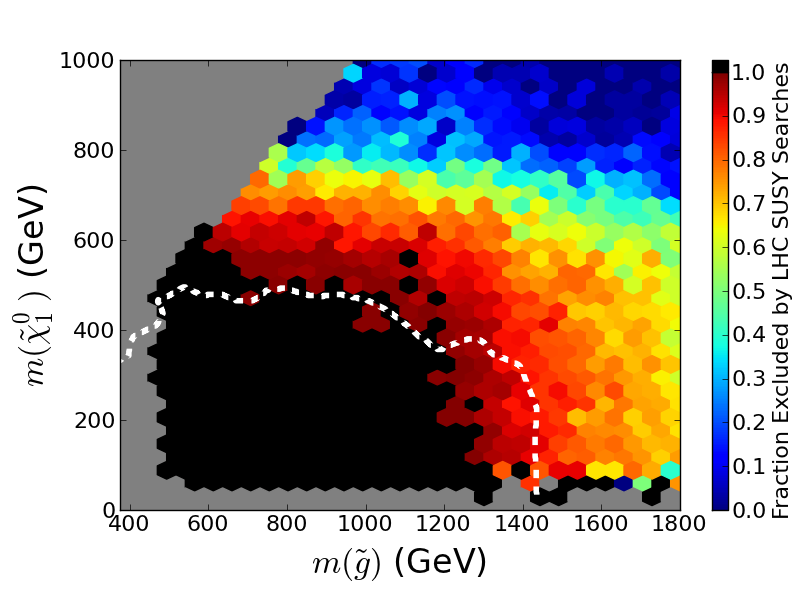}
\hspace{0.20cm}
\includegraphics[width=3.5in]{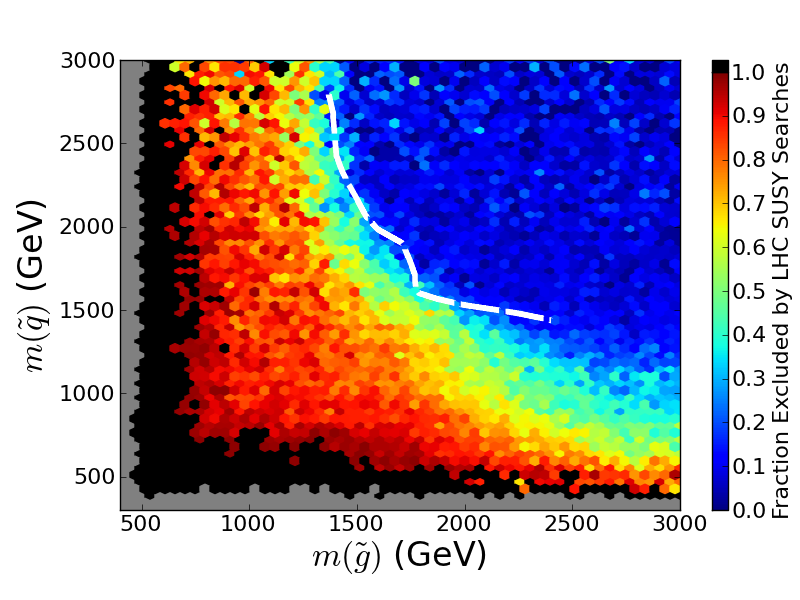}}
\vspace*{-0.10cm}
\caption{Coverage of the pMSSM parameter space for the neutralino LSP model set, showing the fraction of models excluded
in each mass bin by the combined 7 and 8 TeV LHC searches in the gluino-LSP (left) 
and the lightest (1\textsuperscript{st}/2\textsuperscript{nd} generation) squark-gluino (right) mass planes. The color 
code indicates the fraction of models excluded in a specific mass bin. In this and all subsequent figures, the ATLAS results from a simplified model analysis  at 7 (8)
TeV are displayed as the 
solid (dashed) white curves in the various LSP-sparticle mass planes. In particular, the dashed white line in the squark-gluino mass plane is the simplified
model result from the 8 TeV 20 fb$^{-1}$ 2-6 jets + MET 
search~\cite{TheATLAScollaboration:2013fha}, assuming degenerate squarks and a massless LSP.}
\label{fig00}
\end{figure}

Figure~\ref{fig00} shows the efficiency of the combined LHC search channels projected into both the gluino-LSP and the lightest 
(1\textsuperscript{st}/2\textsuperscript{nd} generation) 
squark-gluino mass planes together with the corresponding $95\%$ CL limits from the ATLAS simplified model analysis. Here we see that the 
region excluded by the ATLAS simplified model analysis (below and to the left of the white curve) in the gluino-LSP mass plane roughly encircles 
the all-black region{\footnote{Note that the black region indicates that 100\% of the models in that region are excluded at 95\% C.L.}}
 which is excluded by the combination of analyses. This is interesting, since the ATLAS simplified model limit in the left panel is based solely on the jets + MET channel under the assumption of decoupled squarks, while the pMSSM results arise from combining {\it many} analyses with  
no additional assumptions about the full sparticle spectra. As can be seen here, most of the surviving models with light gluinos have relatively compressed mass spectra, although a few models evade detection by having rather complex decay patterns, often through stops. The lightest squark-gluino 
panel reveals that many models survive far below the ATLAS simplified model bound (where degenerate squarks and a massless LSP have been 
assumed), as might be expected from the more complex spectra in the pMSSM. In particular, many models with heavy gluinos have light squarks far below the simplified model limit, since allowing non-degenerate squarks means that individual squarks may be very light while evading detection if the other squarks are somewhat heavier. 
Note the parabolic bands of colors in this panel, signifying boundaries with increasingly more challenging kinematics.  It is interesting to see that the simplified model
result lies at the pMSSM kinematic boundary, above which essentially no models are excluded.

\begin{figure}[htbp]
\centerline{\includegraphics[width=3.5in]{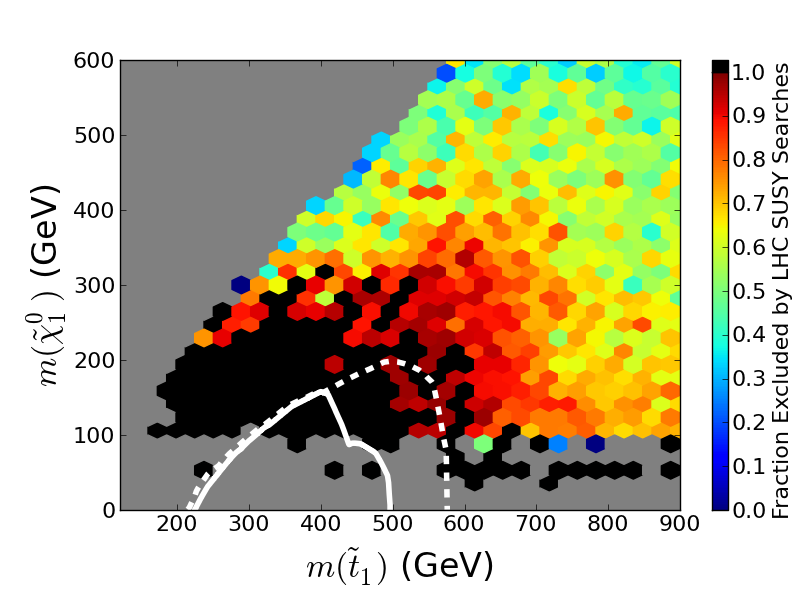}
\hspace{0.20cm}
\includegraphics[width=3.5in]{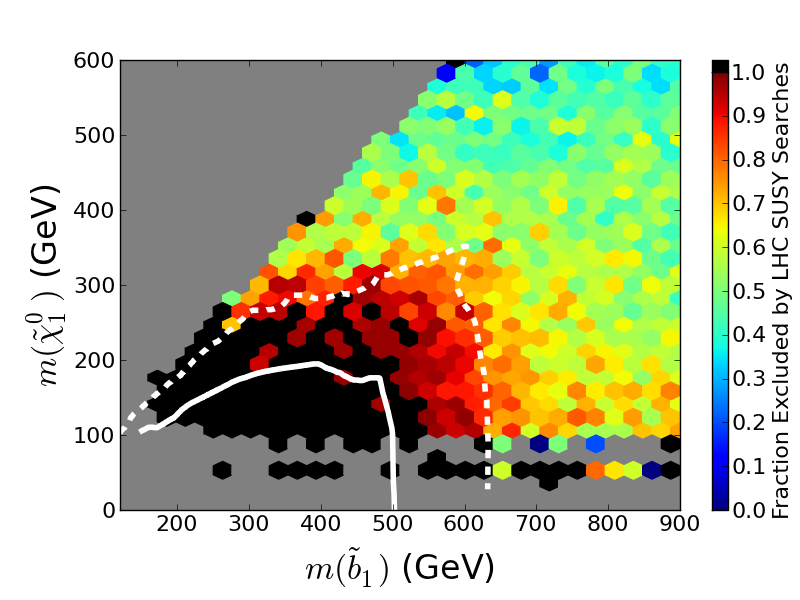}}
\vspace*{0.50cm}
\centerline{\includegraphics[width=3.5in]{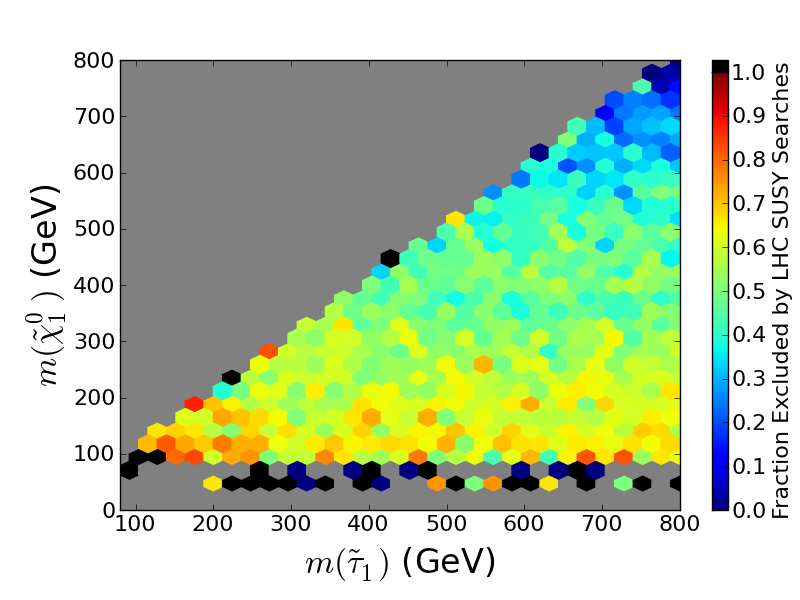}}
\vspace*{-0.10cm}
\caption{Coverage of the pMSSM parameter space for the neutralino LSP set, showing the fraction of models excluded by the combined 7 and 8 TeV LHC searches in the lightest stop-LSP mass plane 
(top left), the lightest sbottom-LSP mass plane (top right) and the lightest stau-LSP mass plane (bottom). The solid and dashed
white lines represent the corresponding 
$95\%$ CL limit results obtained by ATLAS at 7 and 8 TeV, respectively, in the simplified model limit as discussed in the text.}
\label{fig1}
\end{figure}

Searches for 3\textsuperscript{rd} generation sparticles are of particular importance since they have the strongest couplings to the Higgs and are essential 
for solving the `naturalness' and fine-tuning issues associated with the Higgs mass quadratic divergence. At least one of the stop squarks is expected 
to be reasonably light, and if it is mostly left-handed it will also be accompanied by a light sbottom squark with a similar mass. Figure~\ref{fig1} shows the 
impact of the LHC searches in the lightest stop-, lightest sbottom- and lightest stau-LSP mass planes. Note that whereas the simplified model 
treatment by ATLAS arising from searches at 7 (solid) and 8 (dashed) TeV qualitatively describes the coverage in the sbottom-LSP mass 
plane, it is entirely inadequate for placing constraints on the stop squark. 
As we will see below, the difference between the simplified model limit and the pMSSM exclusion in the stop mass plane arises from the fact that the shape of the excluded region is determined by the direct sbottom (two b-jets, zero leptons + MET inclusive) search and
not by the stop searches used to derive the simplified model limit. Additionally,   
non-3\textsuperscript{rd}-generation searches also play an important role in obtaining the excluded regions shown here, especially close to the kinematic boundary. 

When considering the lower panel of figure~\ref{fig1}, it is important to note that we have not incorporated ATLAS searches involving taus in the final state as PGS has a strong tendency to yield a large mis-tag rate while simultaneously having a low tau tagging efficiency. Thus the excluded fraction of models in this panel (which is relatively 
uniform in density as might be expected from these arguments) is the result of the searches without taus. 

\begin{figure}[htbp]
\centerline{\includegraphics[width=3.5in]{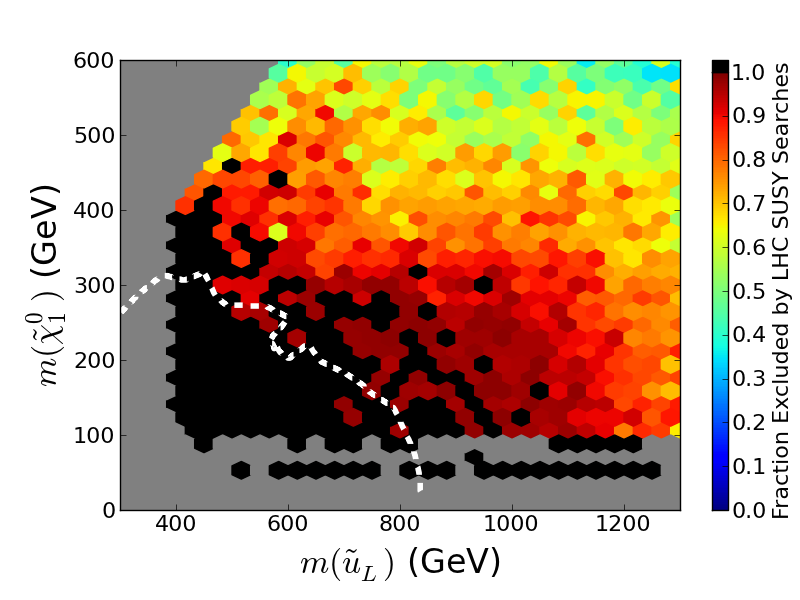}
\hspace{0.20cm}
\includegraphics[width=3.5in]{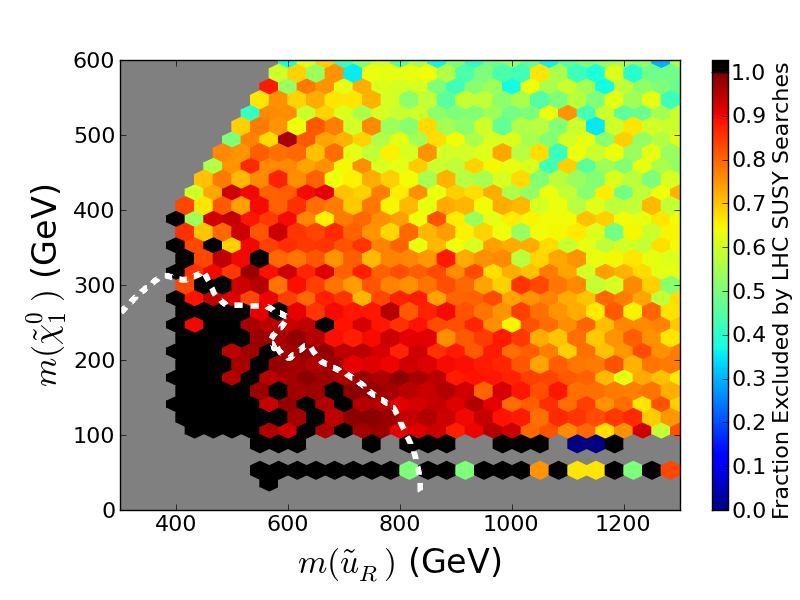}}
\vspace*{0.50cm}
\centerline{\includegraphics[width=3.5in]{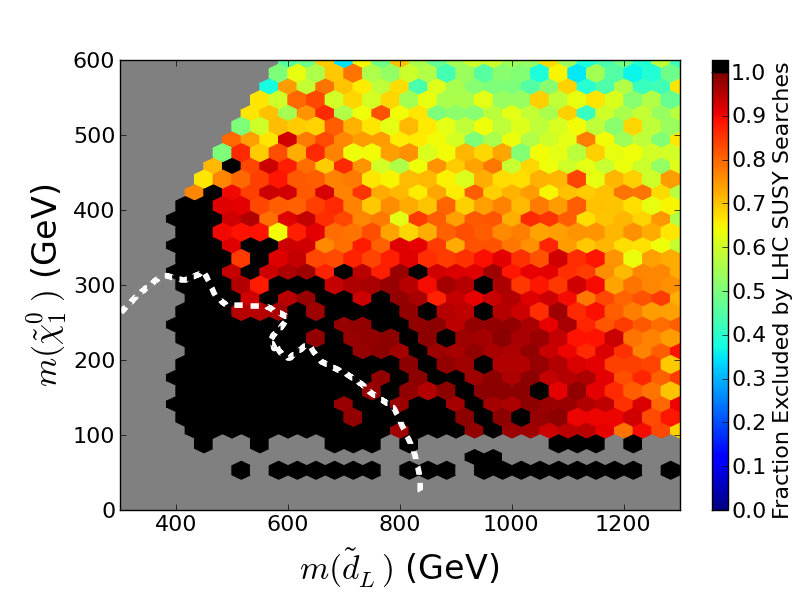}
\hspace{0.20cm}
\includegraphics[width=3.5in]{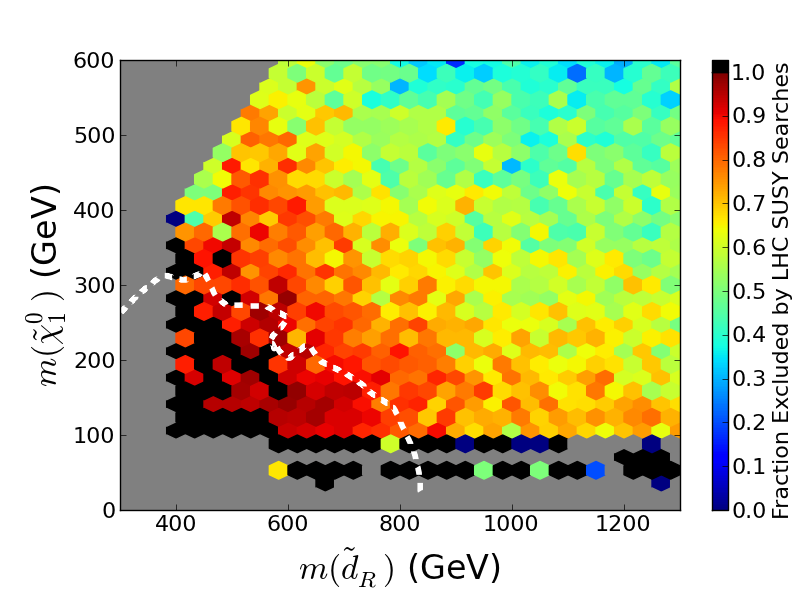}}
\vspace*{-0.10cm}
\caption{Same as the previous figure but now for $\tilde u_L$ (top left), $\tilde u_R$ (top right), $\tilde d_L$ (bottom left) and 
$\tilde d_R$ (bottom right). The ATLAS simplified model limit from~\cite{TheATLAScollaboration:2013fha}, assuming degenerate squarks, is shown for comparison.}
\label{fig2}
\end{figure}

As observed above, a reasonably large number of models survive the LHC searches despite having light squarks, particularly when the gluino is heavy. It is informative to examine these surviving models in 
a bit more detail. The pMSSM coverage in the 1\textsuperscript{st}/2\textsuperscript{nd} generation squark-LSP mass planes are shown individually
for the left- and right-handed up- and down-squarks in Fig.~\ref{fig2}. Conventionally, LHC searches assume that these 
4 squark states are degenerate, but in the pMSSM their masses are uncorrelated (except for the $\tilde u_L$ and $\tilde d_L$ squarks, which are degenerate up to electroweak symmetry breaking effects). From the figure, we see that the LHC search reach is distinct for the different squark types. This is generally easy to understand and is essentially related to the relative sizes of the squark production 
cross sections. Since $\tilde u_L$ and $\tilde d_L$ are relatively degenerate they are produced simultaneously with   
similar rates (although the $\tilde d_L$ production cross section is slightly suppressed due to the smaller $d$-quark parton densities). One might therefore expect similar, though not completely identical, exclusion rates, and indeed we see that is the case, 
with the constraints on $\tilde u_L$ being slightly stronger. In the case of $\tilde u_R$ and $\tilde d_R$, on the other hand, the two masses are uncorrelated, resulting in a lower squark cross-section for the generic case of a light right-handed squark. Once again, for identical masses, $\tilde d_R$ will have a smaller 
production rate due to the PDFs. In figure~\ref{fig2} we see that the sensitivity to either of these right-handed squarks is poor, even though they generally have simple decays (to the LSP or other light neutralinos through their bino component).  In particular, we see that $\tilde d_R$ masses as low as $\sim 450-500$ 
GeV remain possible with LSP masses in the range of $\sim 150$ GeV.  Additional work at the LHC will be needed to close the light squark mass 
loophole.

\begin{figure}[htbp]
\centerline{\includegraphics[width=3.5in]{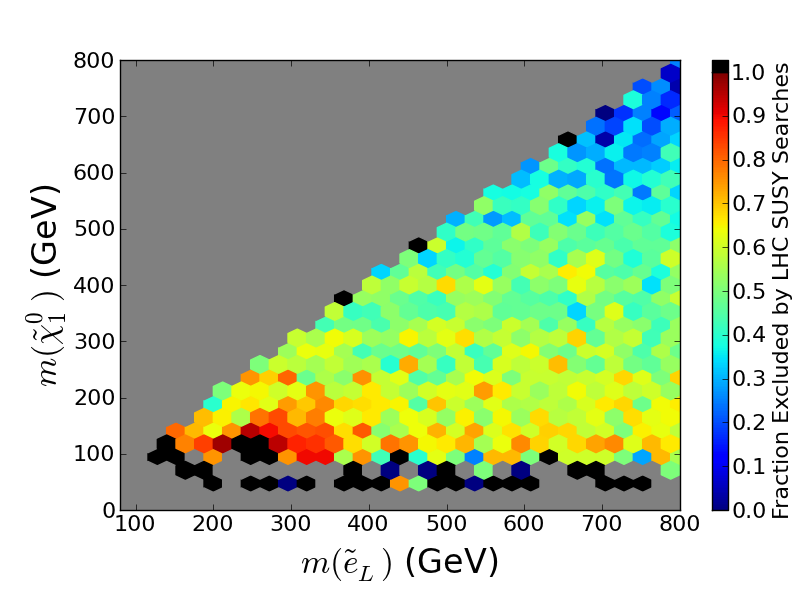}
\hspace{0.20cm}
\includegraphics[width=3.5in]{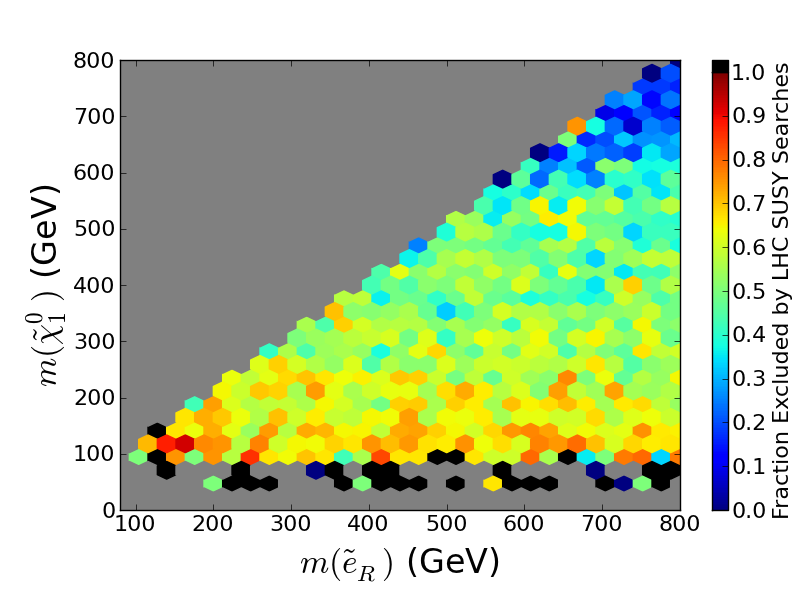}}
\vspace*{0.50cm}
\centerline{\includegraphics[width=3.5in]{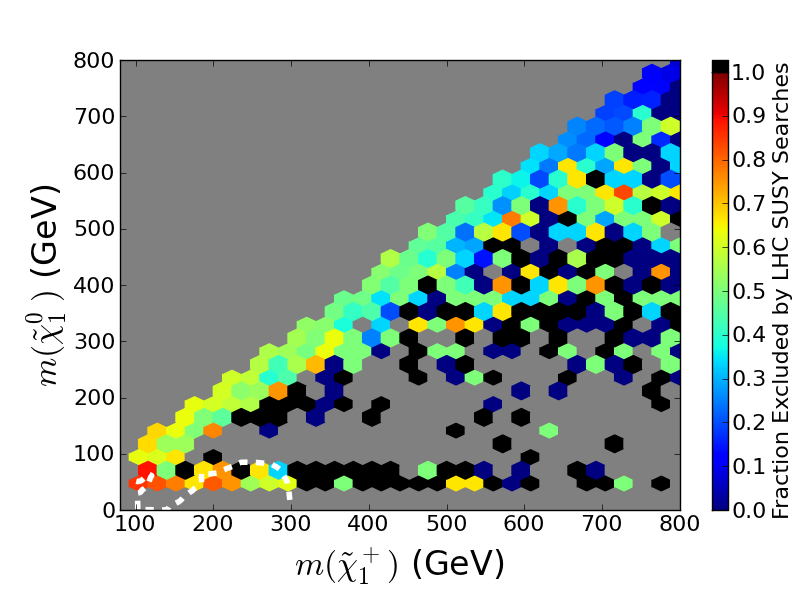}
\hspace{0.20cm}
\includegraphics[width=3.5in]{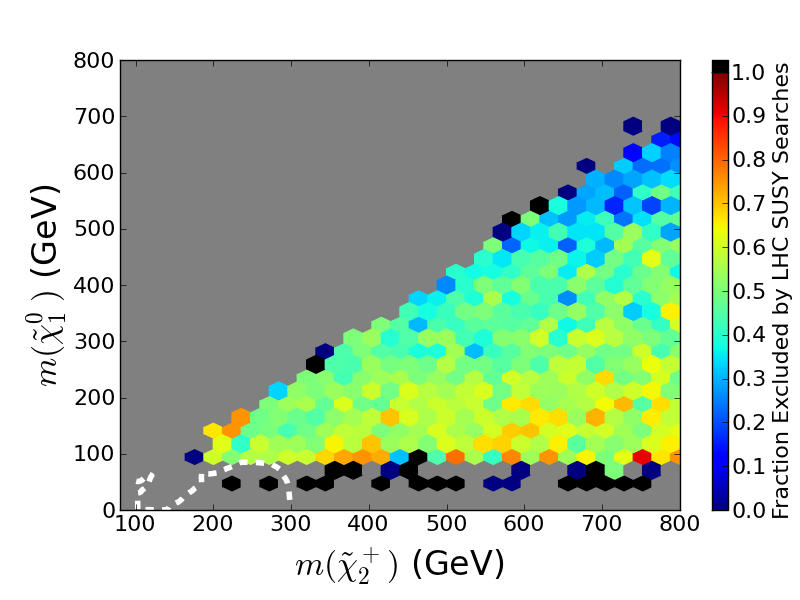}}
\vspace*{-0.10cm}
\caption{Same as the previous figure but now for $\tilde e_L$ (top left),  $\tilde e_R$ (top right), $\tilde \chi_1^\pm$ (bottom left), 
and $\tilde \chi_2^\pm$ (bottom right).}
\label{fig3}
\end{figure}

Figure \ref{fig3} displays the pMSSM coverage in the left- and right-handed selectron and light/heavy chargino - LSP mass planes.  As expected, we see that the coverage is sparse and very light selectrons and charginos are still allowed. It is clear that constraints on pMSSM models are mainly the result of colored sparticle production, and that direct electroweak production remains relatively unexplored, even for sparticle masses near the LEP limit.

Since we simulate a wide range of SUSY searches, it is interesting to compare them and examine which ones dominate in the various regions of parameter space. Figure~\ref{fig4} provides an example of 
this where we compare the impact of the ``vanilla'' jets (+ leptons) + MET analyses (listed in the first 3 entries in Table~\ref{SearchList7} and first 5 entries in Table~\ref{SearchList8}) with the 
3\textsuperscript{rd} generation searches in the lightest stop- and lightest sbottom-LSP mass plane.  The corresponding simplified model results obtained by ATLAS at both 7 (solid) and 8 TeV (dashed) are also displayed. 
Here the red (blue) bins indicate regions in the parameter space where most of the sensitivity arises from the ``vanilla'' (3\textsuperscript{rd} 
generation) searches while green indicates a balance between these two extremes. In fact, we see that most of the regions in both panels are 
green, indicating that both types of searches have comparable impact in probing models. However,  
near the compressed spectrum kinematic boundary we see that the ``vanilla'' searches are more powerful, most likely due to
low b-tagging efficiencies for soft b-jets.  On the other hand, for small LSP masses we see that the 3\textsuperscript{rd} generation searches are 
dominating the exclusion reach. Clearly these results may be modified as additional searches using the full luminosity available at 8 TeV are included. 

As mentioned previously, the 0$\ell$, 1$\ell$ and 2$\ell$ stop searches listed in Tables~\ref{SearchList7} and~\ref{SearchList8} are not very effective in covering the pMSSM
parameter space and are not
responsible for the shape of the excluded region in the LSP-stop mass plane. This is demonstrated in Figure~\ref{fig5}, which compares the fraction of models excluded by jets + MET with the stop searches. From this figure, we see that the fraction of models probed by the stop searches is rather low and has a completely different shape than the region explored by the combined 3\textsuperscript{rd} generation searches in the previous figure.  Using the 7 TeV LHC results, we previously showed~\cite{us2b} that the large majority of models excluded by the 3\textsuperscript{rd} generation searches are caught by the direct sbottom search.  This is because most of our models have stops with mixed decay modes (both $t + \chi^0$ and $b + \chi^+$ final states), and the stop searches quickly become ineffective when the branching fraction to the preferred final state is decreased. Figure~\ref{fig5} demonstrates that this result remains valid when the 8 TeV stop searches are included.

\begin{figure}[htbp]
\centerline{\includegraphics[width=3.5in]{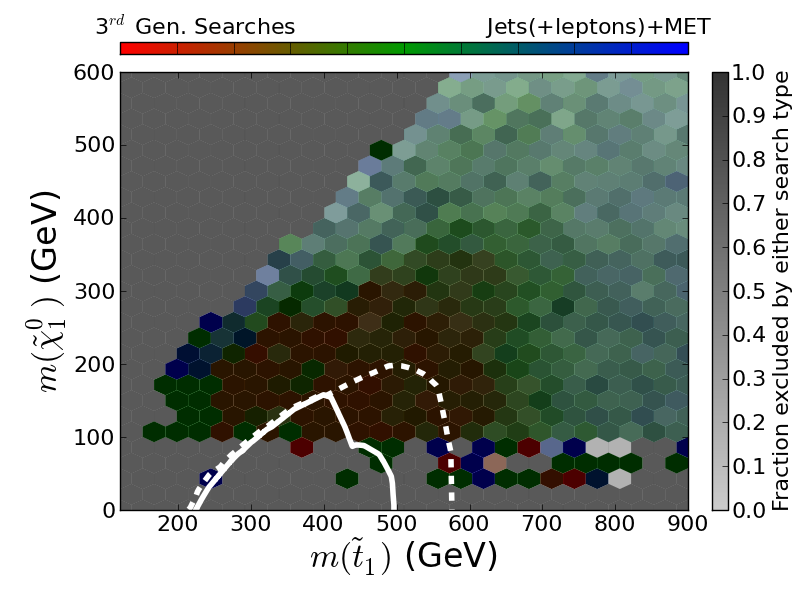}
\hspace{0.20cm}
\includegraphics[width=3.5in]{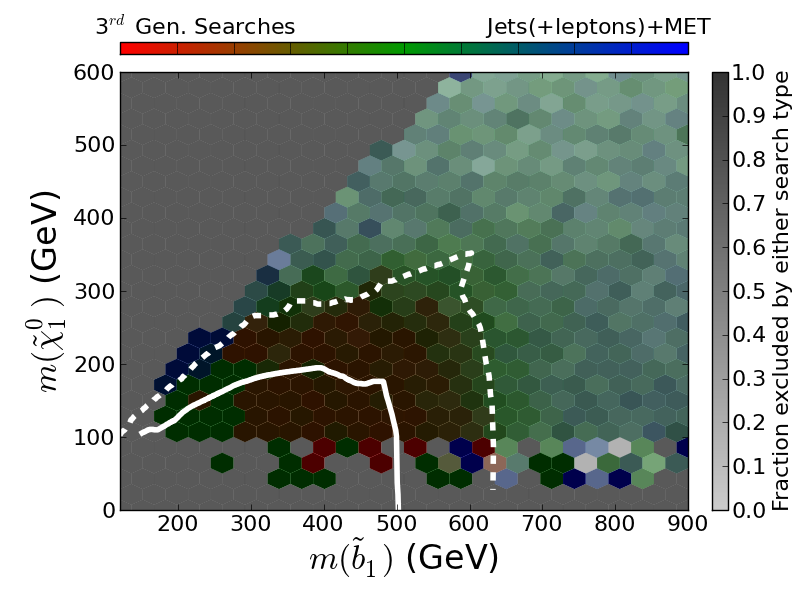}}
\vspace*{-0.10cm}

\caption{Comparison of the contributions to model coverage arising from the inclusive searches, \eg, jets (+ leptons) + MET, with the 3\textsuperscript{rd} generation 
searches as shown in the stop-LSP (left) and sbottom-LSP (right) mass planes for the general neutralino LSP model set. The intensity in each mass bin indicates the fraction of models
that are excluded by the combined searches, while the color indicates which search probes more models as described in the text.}
\label{fig4}
\end{figure}

\begin{figure}[htbp]
\centerline{\includegraphics[width=3.5in]{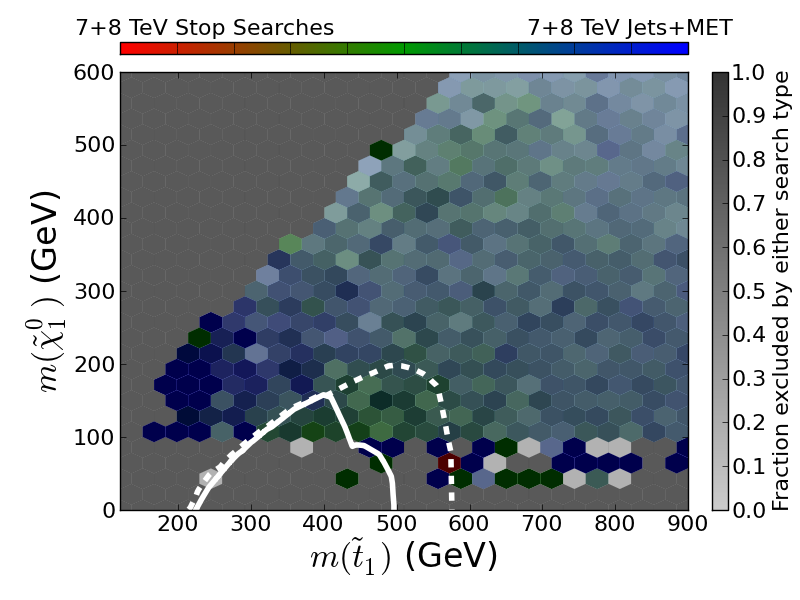}
\hspace{0.20cm}
\includegraphics[width=3.5in]{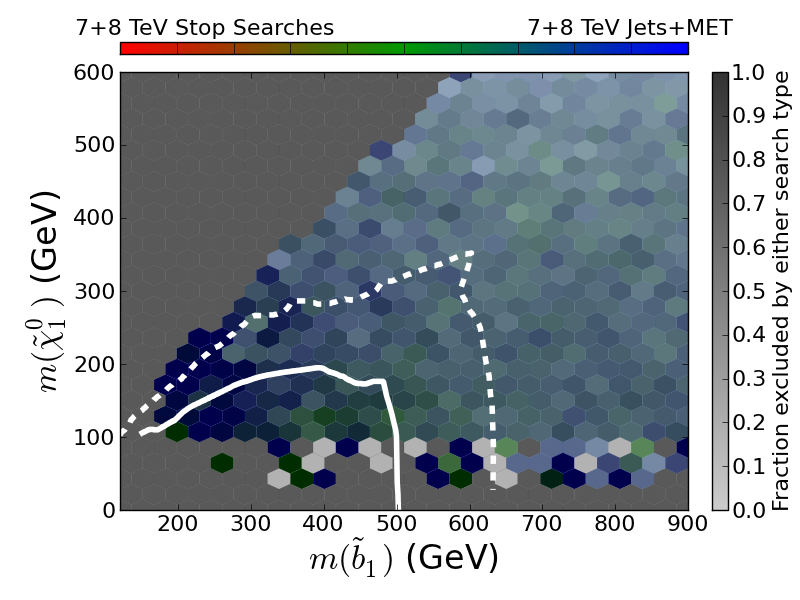}}
\vspace*{-0.10cm}

\caption{Comparison of the contributions to model coverage arising from the 0$\ell$ jets + MET searches at 7 and 8 TeV with the 0$\ell$-2$\ell$ searches for light stops in the stop-LSP (left) and sbottom-LSP (right) mass planes for the general neutralino LSP model set. The intensity in each mass bin indicates the fraction of models
that are excluded by the combined searches, while the color indicates which search probes more models as described in the text.}
\label{fig5}
\end{figure}

\subsection{Low Fine-Tuning Model Set}

As discussed above, we have also generated a small ($\sim 10.2$k) set of pMSSM models with low fine-tuning where the neutralino LSP saturates the thermal relic density and with 
a Higgs mass of $126 \pm 3$ GeV. This low-FT set was selected from an initial sample of $3.3 \times 10^8$ points after the scan ranges were adjusted 
(as shown in Table \ref{ScanRanges}) to target models with 
low-FT spectra. This shows that satisfying the additional constraints of the observed 
relic density and the Higgs mass (in addition to all of the standard collider, precision electroweak, DM search and flavor constraints) is non-trivial to accomplish while
preserving low fine-tuning. 
One reason for this is that while $\sim 20\%$ of the original large neutralino LSP model sample gave the correct Higgs mass of $126\pm 3$ GeV, the range we now allow for the 
relic density ($\Omega h^2 = 0.1153 \pm 0.095$) is quite narrow compared to the range of values present in the large neutralino model set, which extends 
over several orders of magnitude \cite{us1}.  Figure~\ref{fig1xx} displays the 
resulting distributions for the Higgs mass, relic density and amount of fine-tuning, with $\Delta$ being the Ellis-Barbieri-Giudice parameter\cite{Ellis:1986yg,Barbieri:1987fn} 
for this restricted model set. Here we see that the sample is dominated by models 
which have larger values of $\Delta$ and somewhat smaller Higgs masses as we might expect. The smallest value of $\Delta$ we obtain is $\sim 30$ and to go much lower 
would likely require a dedicated Markov chain Monte Carlo study using our lower $\Delta$ points as seeds.  

\begin{figure}[htbp]
\centerline{\includegraphics[width=3.5in]{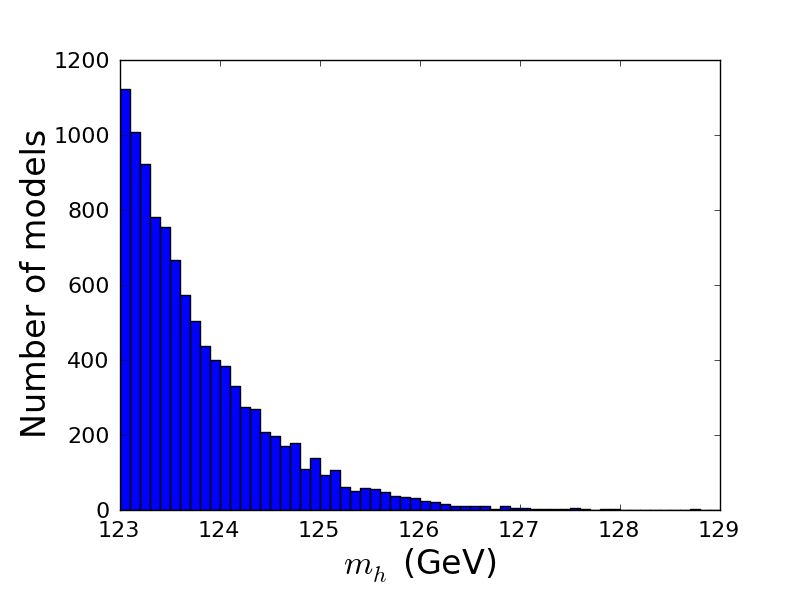}
\hspace{-0.50cm}
\includegraphics[width=3.5in]{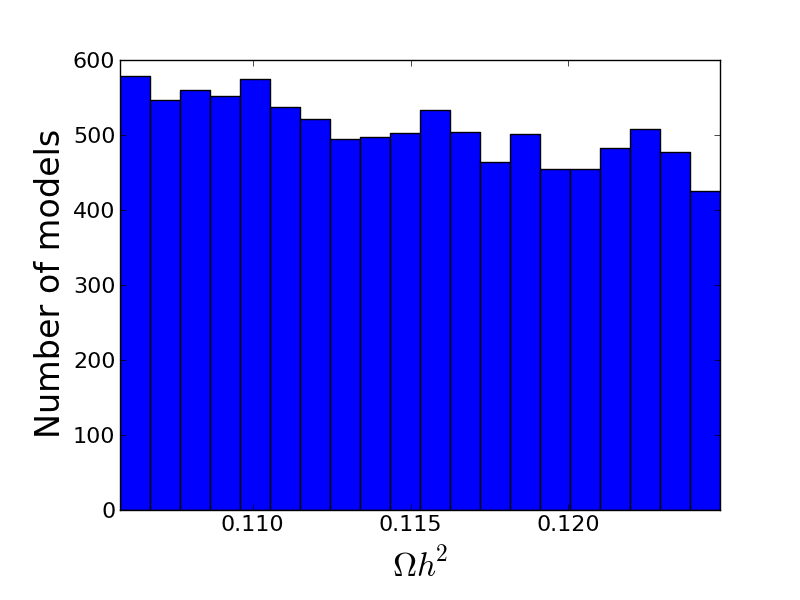}}
\vspace*{0.50cm}
\centerline{\includegraphics[width=3.5in]{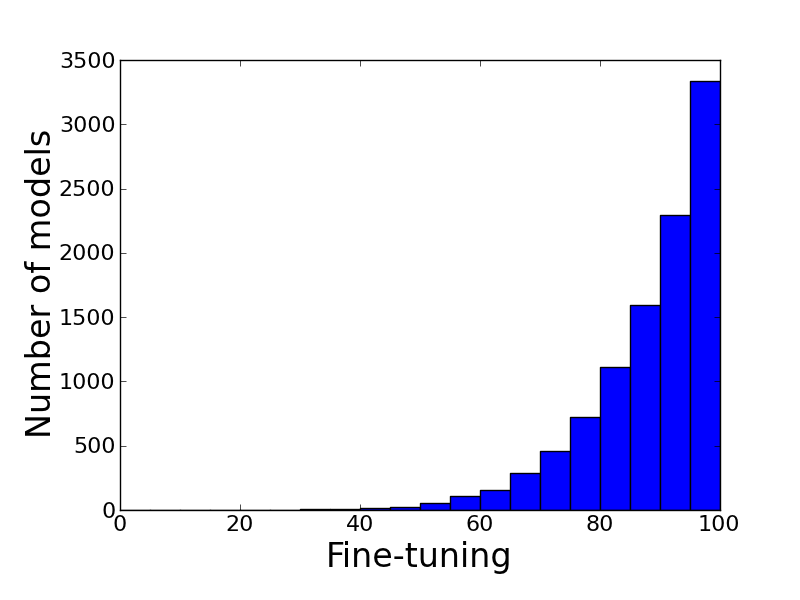}}
\vspace*{-0.10cm}
\caption{Distribution for the Higgs mass (top left), thermal relic density (top right) and the amount of fine-tuning $\Delta$ (bottom) for the 
low-FT model set.}
\label{fig1xx}
\end{figure}
\begin{figure}[htbp]
\centerline{\includegraphics[width=4.50in]{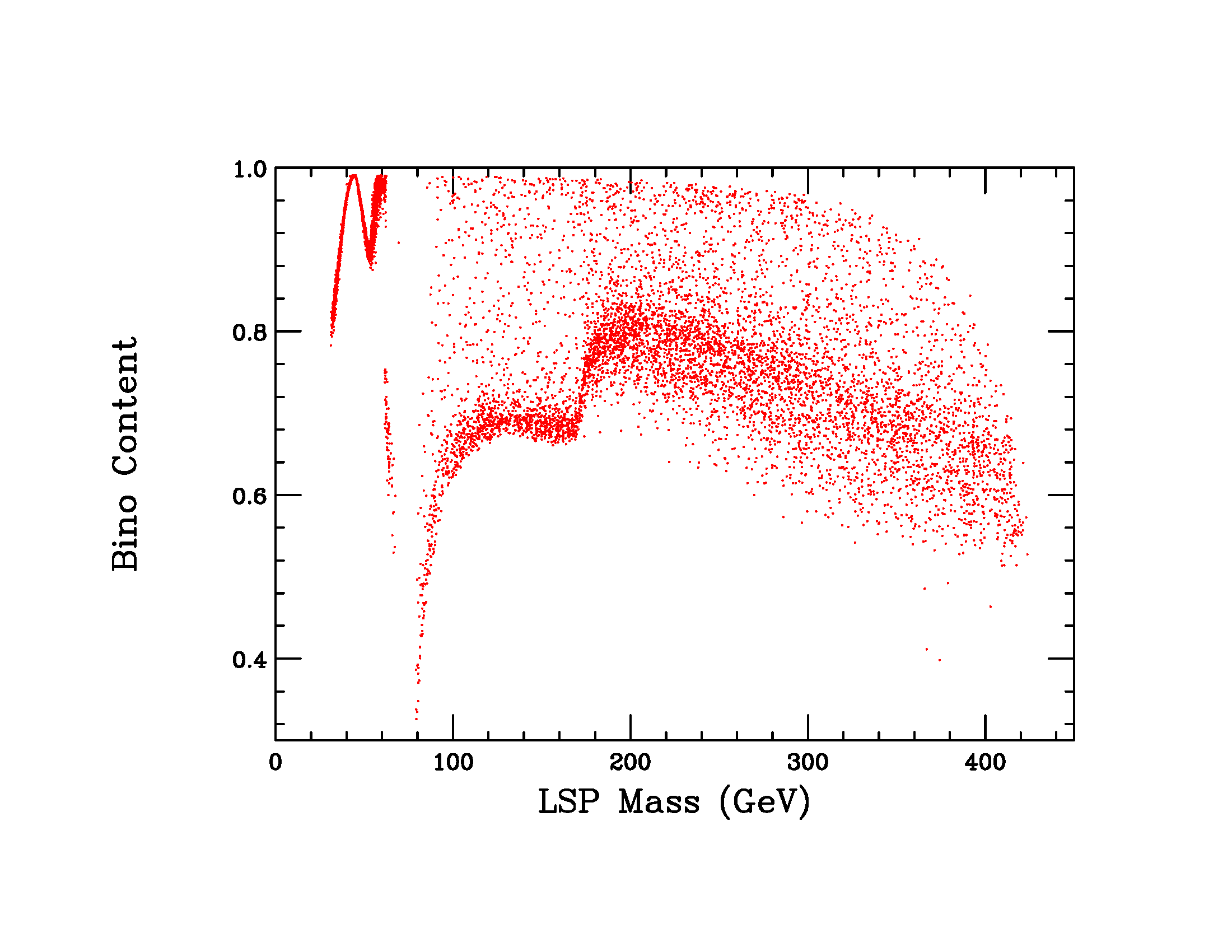}
\hspace{-1.10cm}
\includegraphics[width=4.50in]{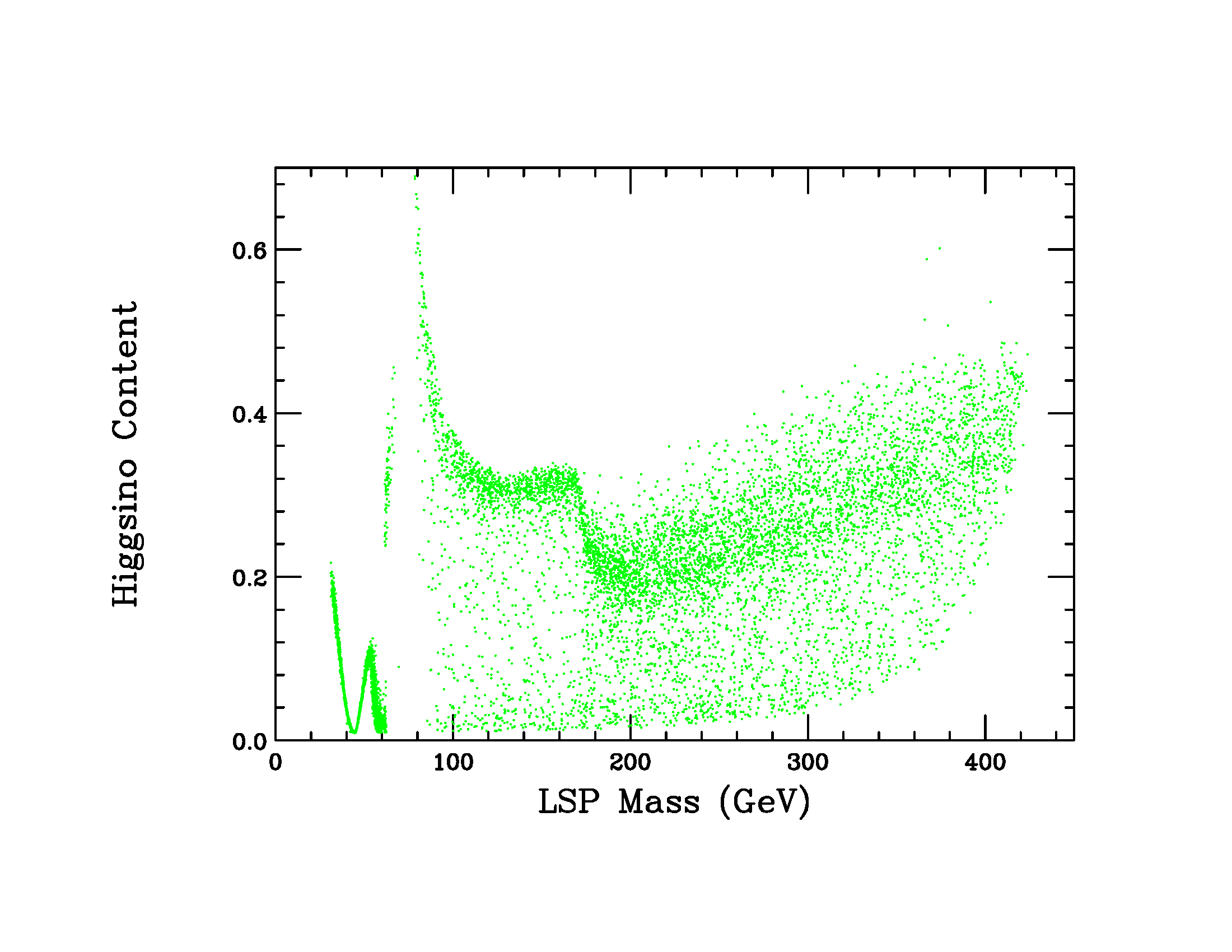}}
\vspace*{0.30cm}
\centerline{\includegraphics[width=4.50in]{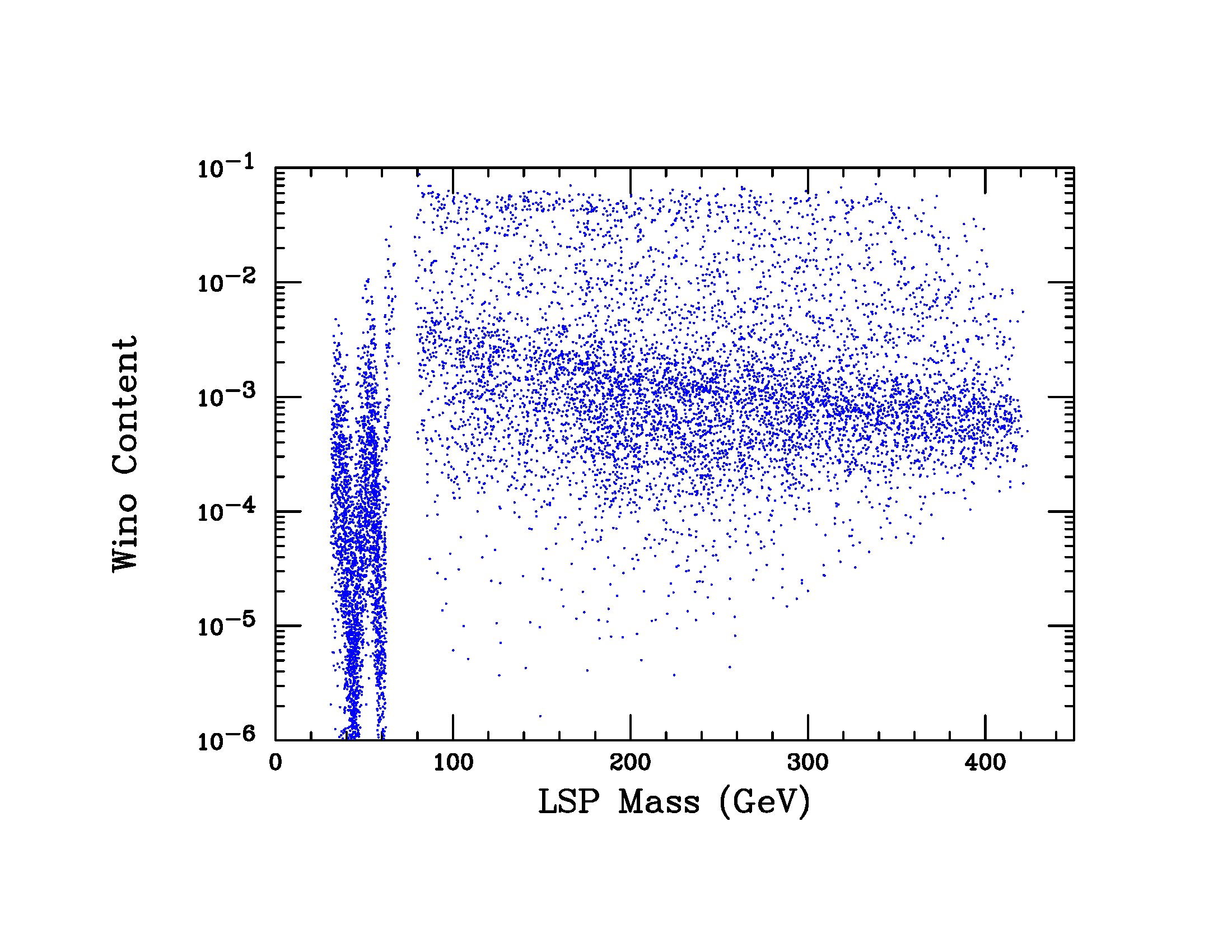}
\hspace{-1.10cm}
\includegraphics[width=4.50in]{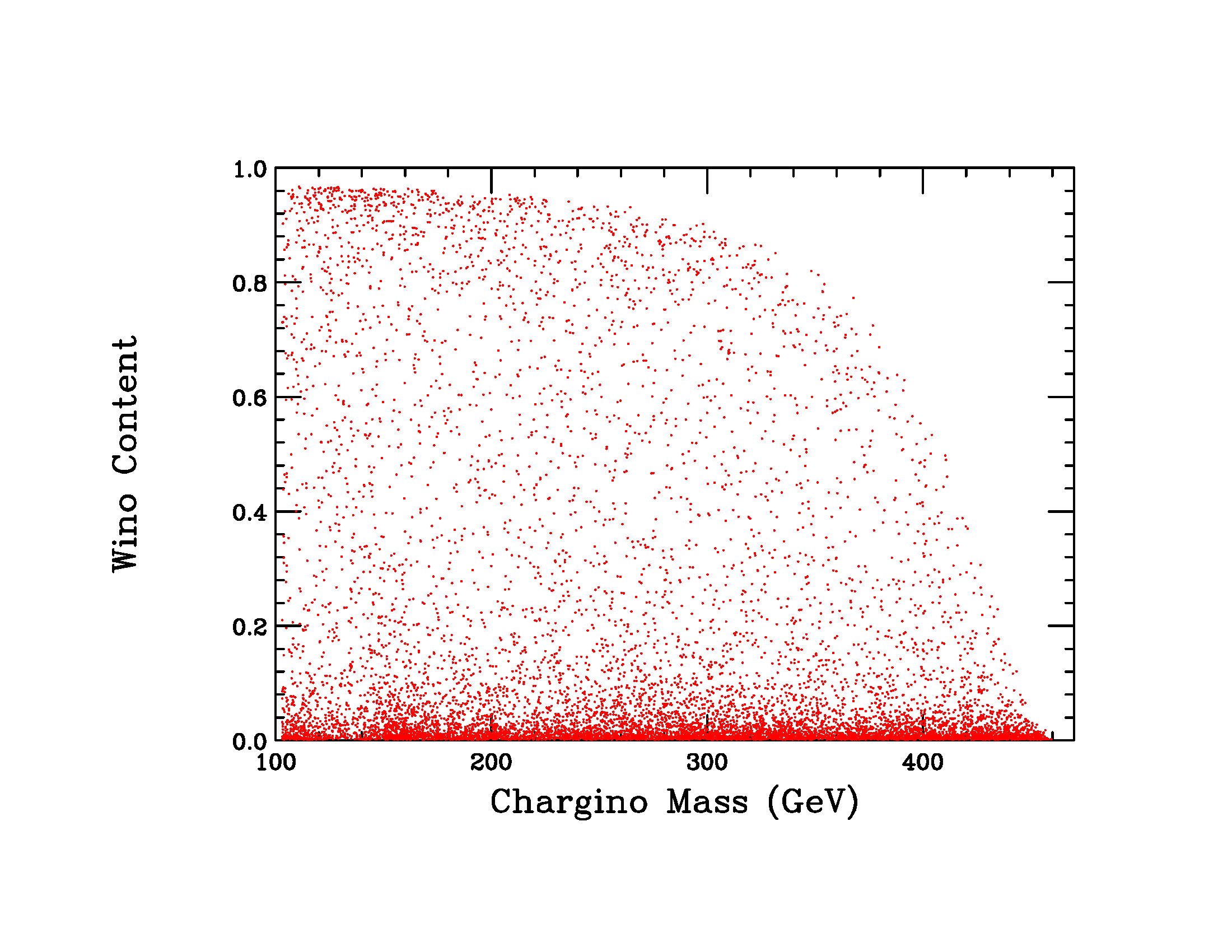}}
\vspace*{-0.3cm}
\caption{The bino (top left), Higgsino (top right) and wino (bottom left) content of the LSP in the low-FT model set as a function of its mass.
The corresponding wino content of the lightest chargino is also shown (bottom right).}
\label{fig1yy}
\end{figure}

These low-FT models have some common features: They necessarily have a relatively light stop and a mostly bino-like LSP (required to achieve the correct relic density), 
as well as Higgsinos with masses below $\sim 450$ GeV. Well-tempered bino-Higgsino 
mixing is the most common mechanism for achieving the observed relic density (accounting for $\sim 53\%$ of the models), with co-annihilation with a 
light slepton, chargino or stop ($\sim$ 14\% of models) or annihilation through either the $Z$ or light Higgs funnel ($\sim 32\%$ of models) and $A$ funnel ($\sim 2\%$ of models) 
also being well-represented annihilation mechanisms.  Figure~\ref{fig1yy} 
shows the electroweak content of the LSP as a function of its mass for the models in the low-FT set; much of the structure associated with this physics 
is directly observed here. Note that for rather light LSPs, co-annihilation is not possible given the constraints from LEP on chargino, squark and slepton masses, so that the 
LSP must be a bino-Higgsino admixture in this case. 
Although our scan ranges allow for somewhat lighter LSPs, we find that they must have masses greater than $\sim 30$ GeV in 
order to satisfy the constraint on the invisible decay width of the $Z$, $\Gamma(Z\to \chi \chi)< 2$ MeV, as shown in Fig.~\ref{fig1xxx}{\footnote {The 
invisible width of the Higgs can in principle also constrain the light neutralino spectrum. However, the model-independent limit on this quantity, $\sim 50-60\%$, is 
not yet sufficiently strong to be meaningful as can be seen in the figure.}}. 
We can, in fact, make an even stronger statement based on our study of both the neutralino and low-FT model sets. Consistency with the following conditions: 
($i$) $m_h = 126 \pm 3$ 
GeV, ($ii$) $\Gamma(Z\to \chi \chi)< 2$ MeV, ($iii$) the LSP as a thermal relic produces a density that either saturates or is {\it below} the WMAP/Planck value and ($iv$) the 
LEP constraints on charged sparticles are trivially satisfied (\ie, their masses can't `tunnel' to values below $\sim 90-100$ GeV for any reason\footnote{see~\cite{Boehm:2013qva} for a discussion of the impact of LEP limits on light LSPs in the pMSSM}),  requires the mass 
of the LSP to exceed $\simeq 30$ GeV. Provided these conditions are shown to be satisfied, if such a lighter LSP were to be discovered it would imply that the pMSSM 
(and more than likely the full MSSM) would be excluded, which would be a very powerful result.

Figure~\ref{fig1yy} also shows the wino content of the lightest chargino. We see that a large fraction of the time the lightest chargino is mostly Higgsino-like  
as we expected due to the requirement of a light Higgsino. However, it is clear that, infrequently, the lightest chargino can be a wino or at least a wino-Higgsino 
hybrid.

\begin{figure}[htbp]
\centerline{\includegraphics[width=3.7in]{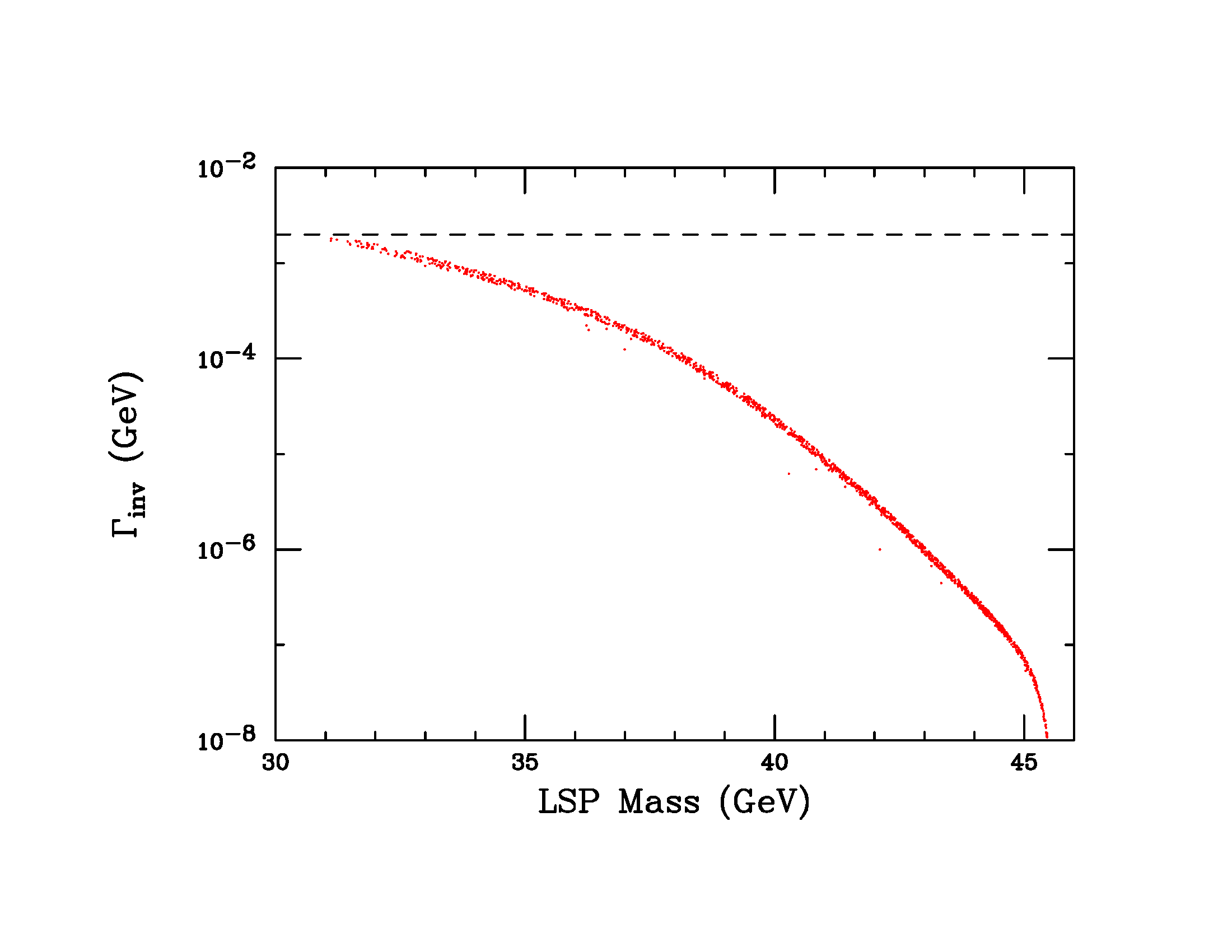}
\hspace{-0.50cm}
\includegraphics[width=3.5in]{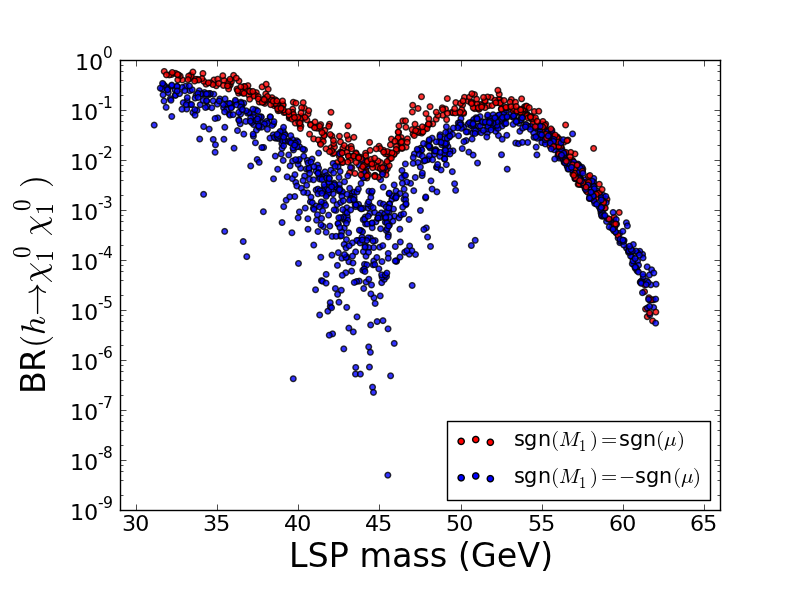}}
\vspace*{-0.10cm}
\caption{Invisible width of the $Z$ (left) and the invisible branching fraction for the light CP-even Higgs (right) for kinematically-accessible LSPs in the low-FT model set. In the left panel the LEP upper 
bound is also shown.}
\label{fig1xxx}
\end{figure}

Continuing our discussion of the characteristics of the low-FT model set, we note that since the lightest stop is most commonly left-handed, light accompanying $\tilde b_1$ squarks are typically present; furthermore, in $\sim 11\%$ of the models the lightest sbottom is lighter than the lightest stop. These features are shown explicitly in Fig.~\ref{fig6789}. Here we see that light stops and sbottoms are frequently reasonably degenerate, as we would expect when they are dominantly left-handed. Interestingly, since $|M_2|<2$ TeV in order to satisfy the 
low-FT requirement, we find that $\sim 60\%$ of the models will also have a wino multiplet below the lightest stop/sbottom. This makes for a rather complex spectrum and 
even more complex decay patterns for these lightest stops and sbottoms. Very infrequently, $\sim 0.5\%$ of the time, only the LSP lies below the lightest stop. 

\begin{figure}[htbp]
\centerline{\includegraphics[width=4.50in]{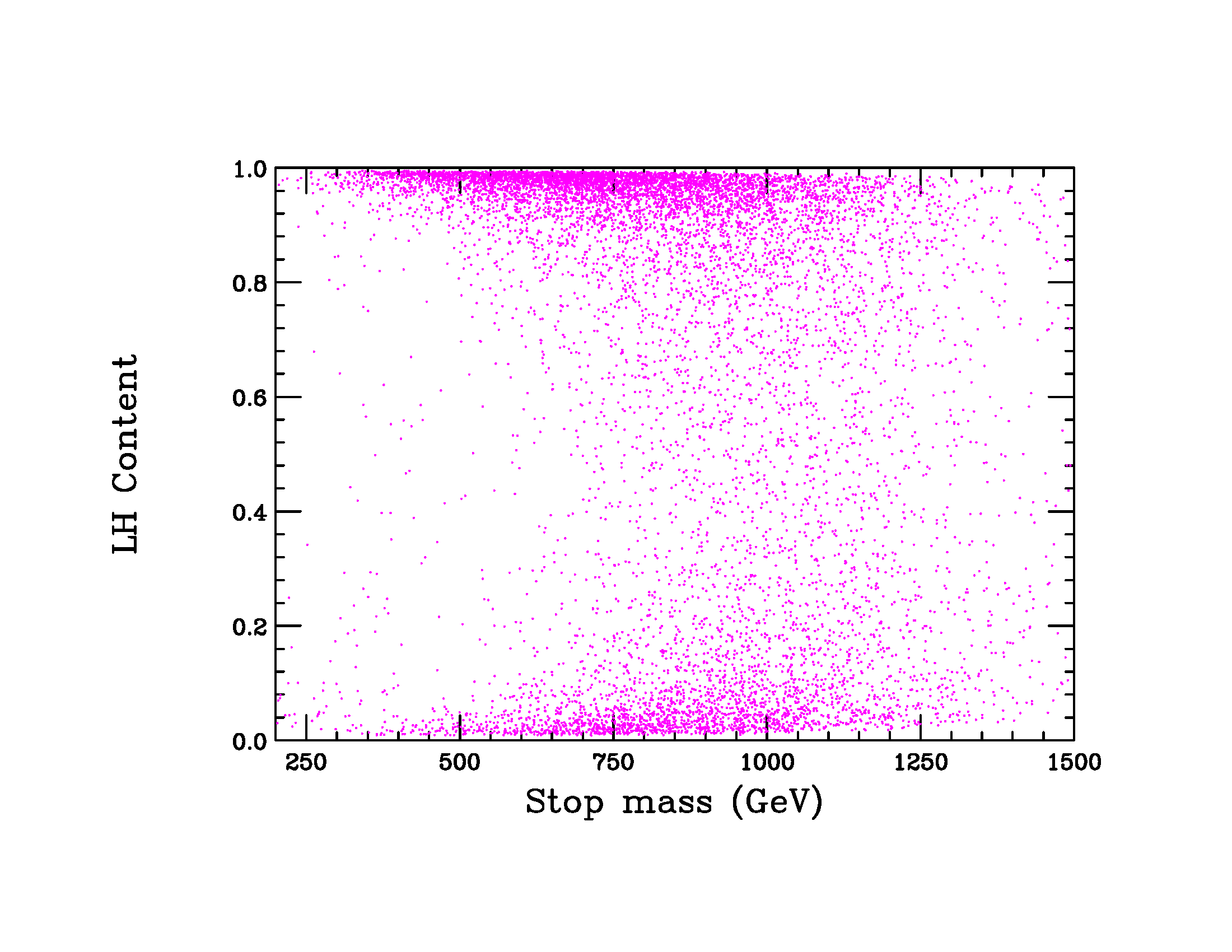}
\hspace{-1.10cm}
\includegraphics[width=4.50in]{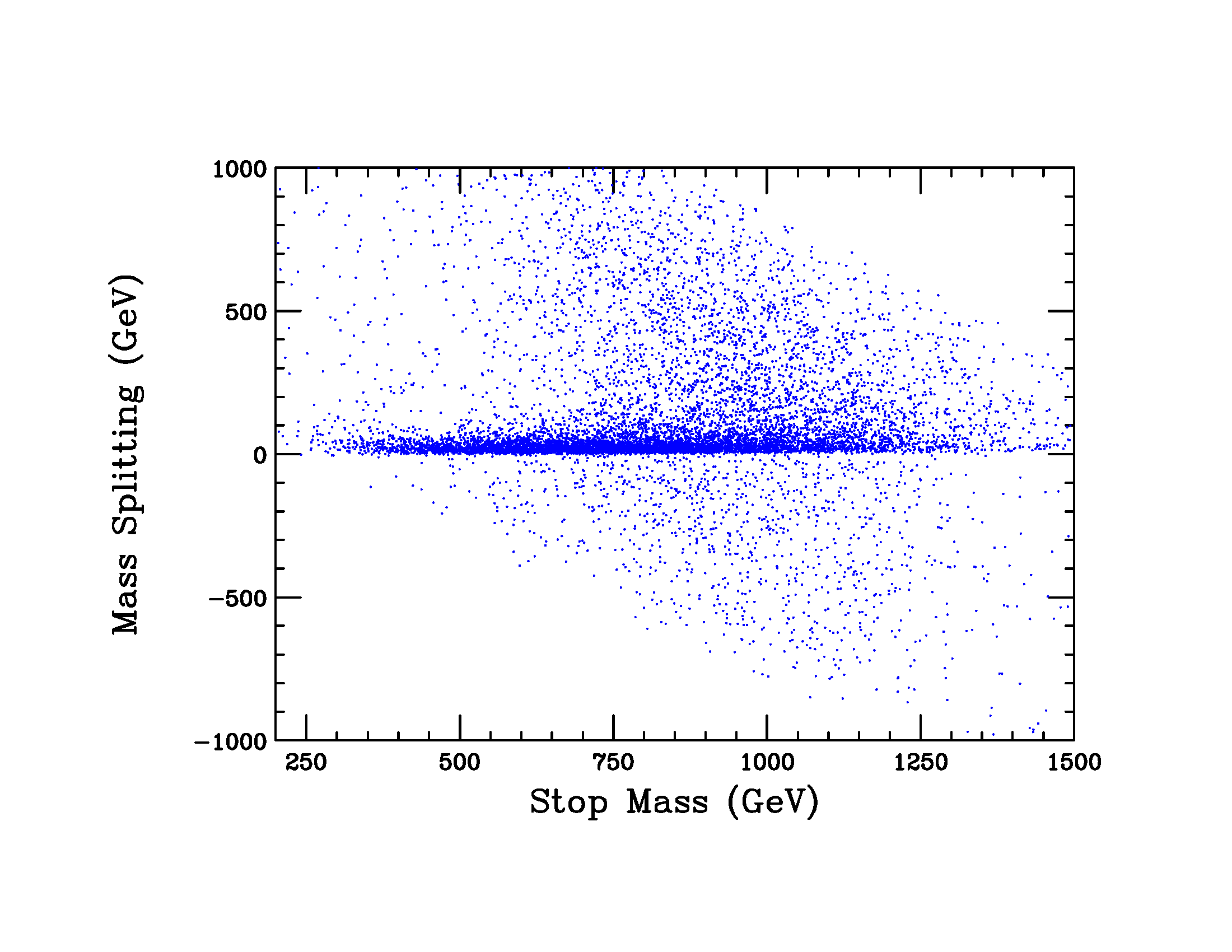}}
\vspace*{-0.3cm}
\caption{(Left) The fraction of $\tilde t_L$ in the lightest stop as a function of its mass. (Right) The mass splitting between the lightest stop and lightest sbottom, 
$m_{t_1}-m_{b_1}$, as a function of the lightest stop mass.}
\label{fig6789}
\end{figure}
\begin{figure}[htbp]

\subfloat{
  \centerline{\includegraphics[width=5.5in]{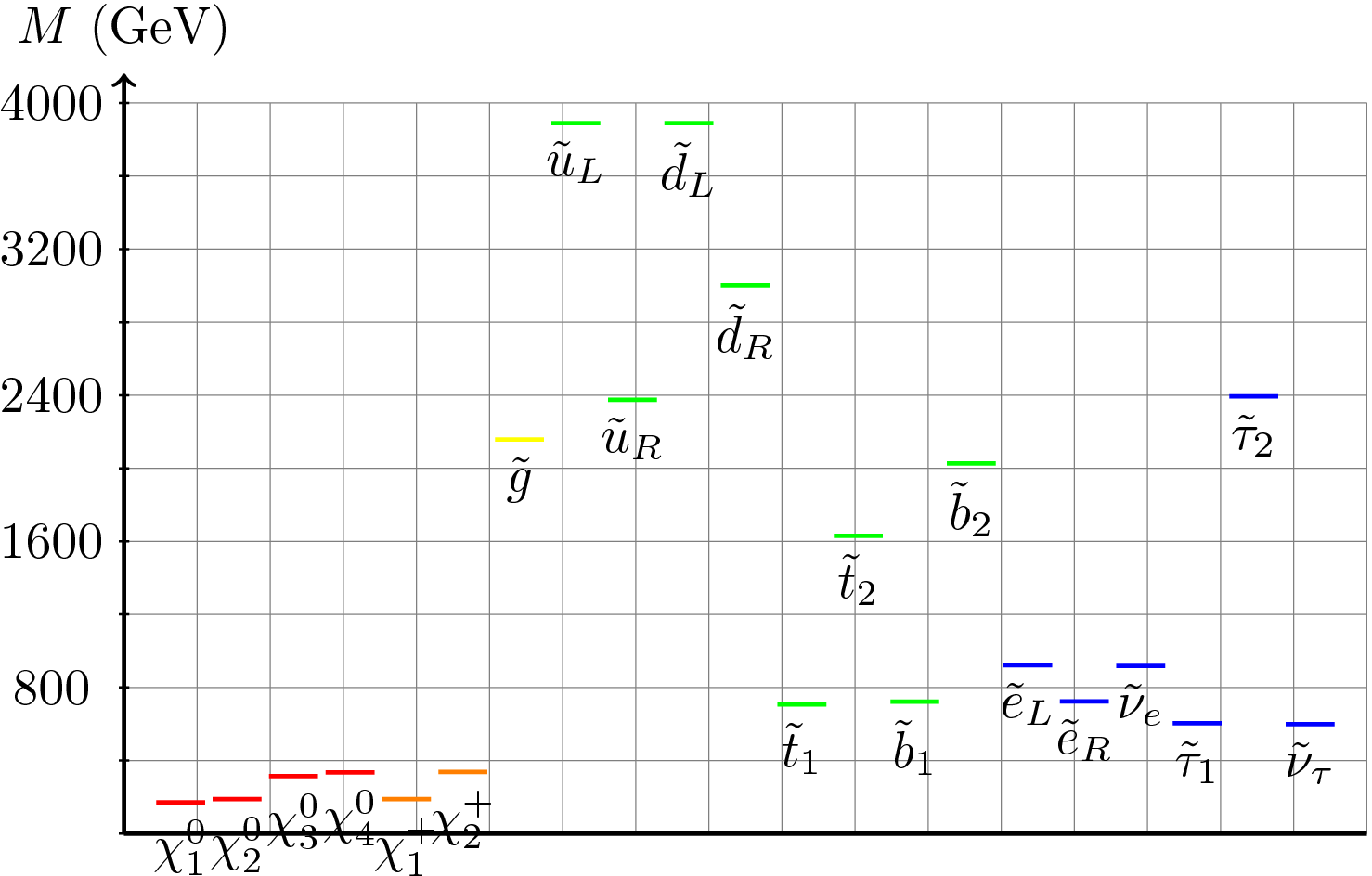}}
  \hspace{-0.50cm}

} 

\subfloat{
  \vspace*{10cm}
}

\subfloat{\centerline{\includegraphics[width=3.5in]{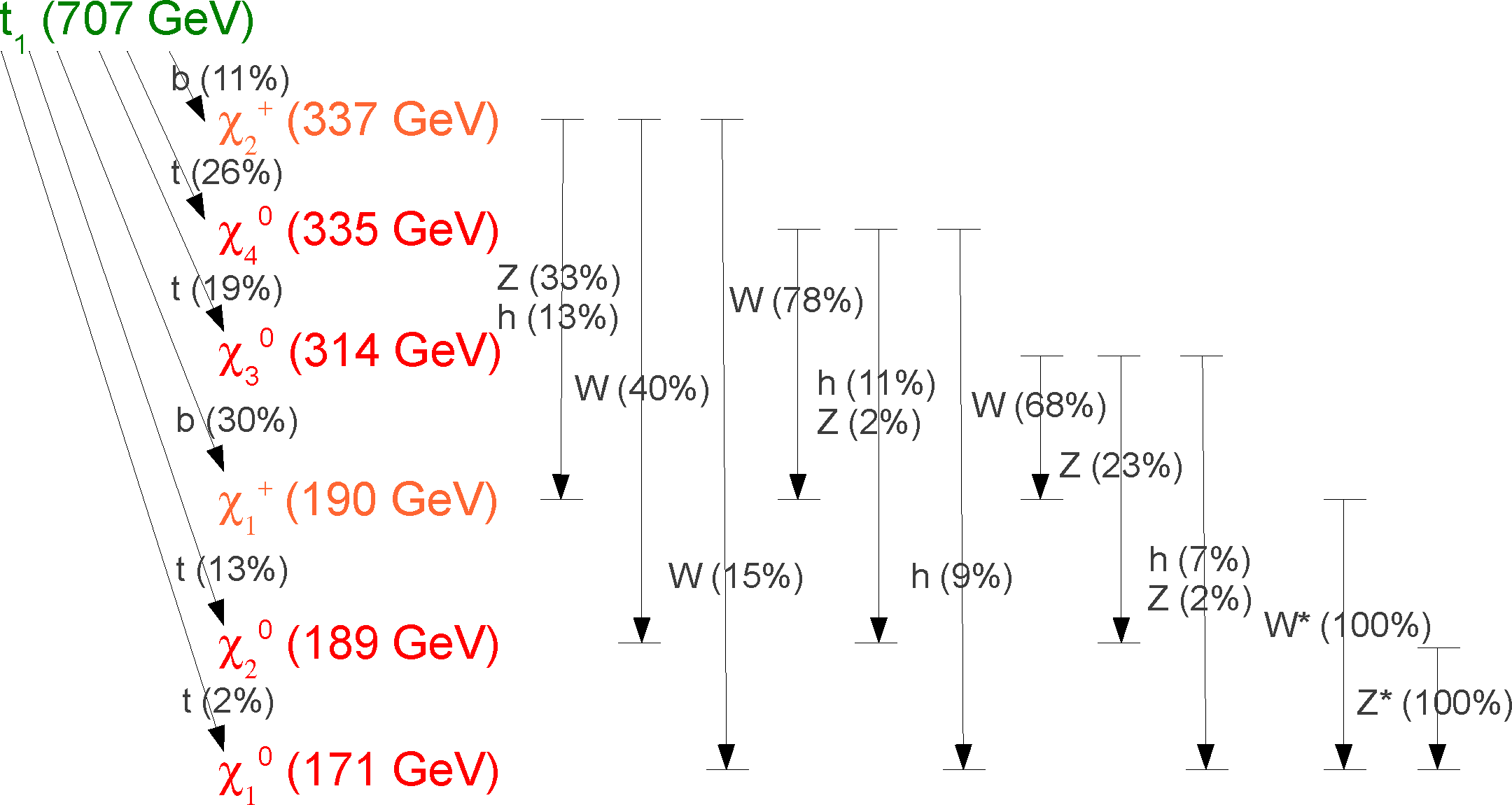}
\vspace*{0.5cm}
\includegraphics[width=3.5in]{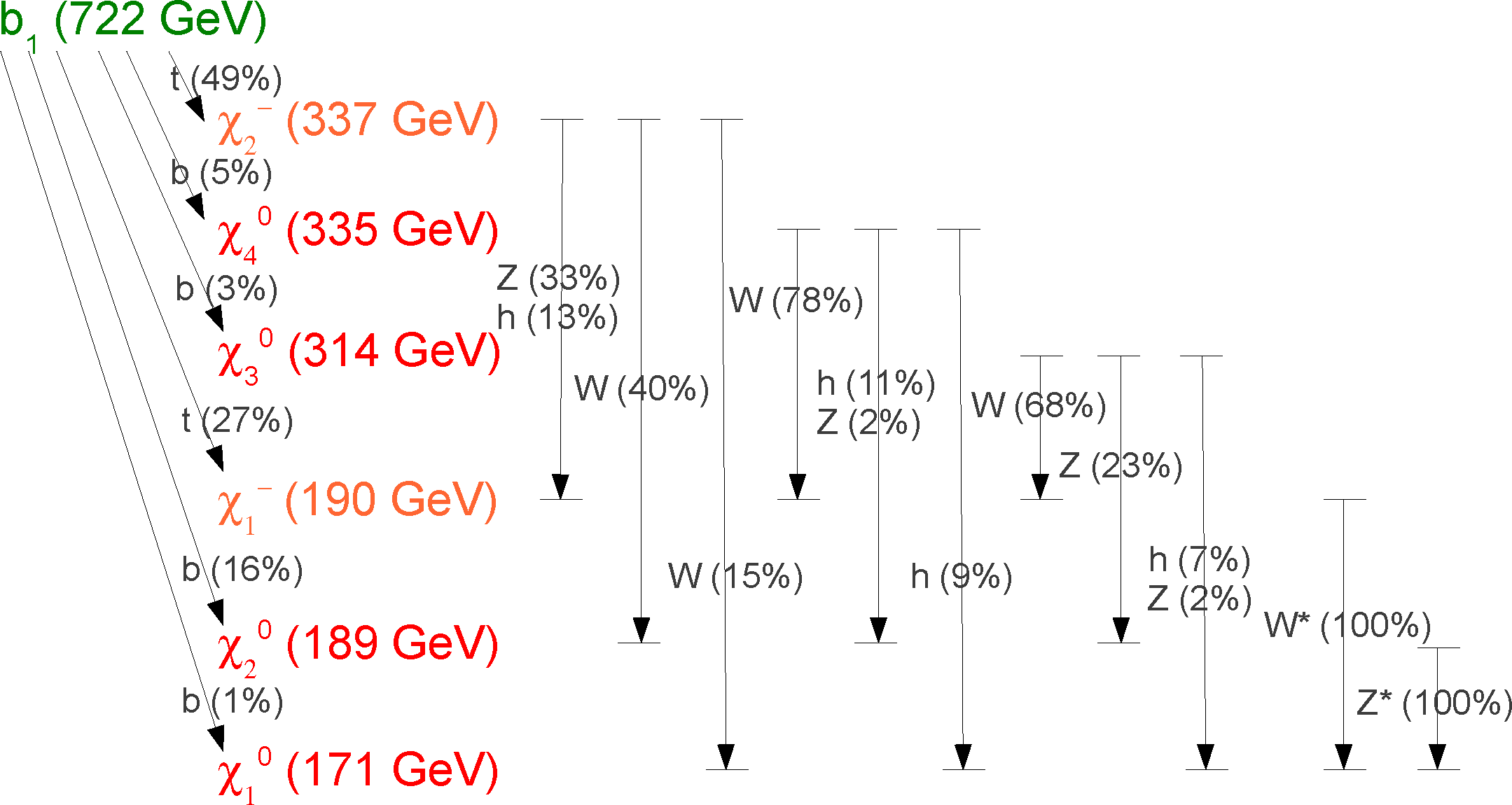}}
}
\vspace*{0.5cm}
\caption{Typical spectrum (top) and decay patterns for stops (bottom left) and sbottoms (bottom right) in a low-FT model.}
\label{fig1x}
\end{figure}

Figure~\ref{fig1x} shows a typical spectrum for one of the low-FT models with very heavy 1\textsuperscript{st}/2\textsuperscript{nd} generation squarks, reasonably heavy 
gluinos and a light stop/sbottom. Here we see that all of the electroweakinos lie below the lighter stop and sbottom. This results in a rather complex decay pattern for both of 
these sparticles, which are depicted in the lower two panels of this figure. Note that the light stop/sbottom can decay to any of the lighter electroweakinos with comparable 
branching fractions; these states then cascade down to lighter ones producing W, Z, and Higgs bosons. We therefore might expect multi-lepton searches to be useful here, although 
the resulting leptons will be rather soft in many cases. Given these 
decay patterns, it is clear that searches for any {\it one} particular final state in stop/sbottom decays will be limited due to the low branching fraction. However, inclusive searches which are sensitive to multiple final states are expected to be more effective.

Tables~\ref{SearchList7} and \ref{SearchList8} above show the ATLAS/CMS SUSY search analyses applied to this model set and the resulting fractions of models excluded by each 
of the individual searches; when combined we find that $\sim 74.0$\% of the low-FT model sample is already excluded by the 7 and 8 TeV results. We note that many of the individual 
searches perform significantly better in the low-FT model set as compared with the general large neutralino set. As a result, the fraction of models probed by the combined set of searches 
is nearly twice as large for the low-FT model set. The increased observability of the low-FT models is unsurprising given that FT constraints generally provide upper bounds on sparticle masses.  We note that 26\% of the low-FT models remain feasible!  

The most important upper bound is on the LSP itself, which can't be heavier than $\sim$ 400 GeV. This cap on the LSP mass reduces the potential for spectrum compression, as can be clearly seen by inspecting Fig.~\ref{fig00ft}, which shows the low-FT model coverage in both the gluino-LSP and lightest 
squark-gluino mass planes. In particular, we see that the lower limit on the gluino mass has increased substantially, now that the compressed region is removed.  
We also note that the existence of light stops implies that the gluinos typically decay to final states with a 
profusion of top and bottom quarks, for which the ATLAS 3b search was designed.
This targeted search may be somewhat less sensitive to the onset of spectrum compression than the general jets + MET searches. Figure~\ref{fig00ft} shows, given our level of statistics, 
that this results in the exclusion of all of our low-FT models with gluinos below $\sim 1.2$ TeV! Note that this is in better agreement with the simplified model result.
As in the general neutralino set, we see that 
1\textsuperscript{st}/2\textsuperscript{nd} generation squarks can be relatively light, provided the gluino is heavy. The fraction of models excluded is also enhanced by the low-FT requirement biasing the stop, sbottom, and (to a lesser extent) gluino masses towards lighter values, improving their visibility.

\begin{figure}[htbp]
\centerline{\includegraphics[width=3.5in]{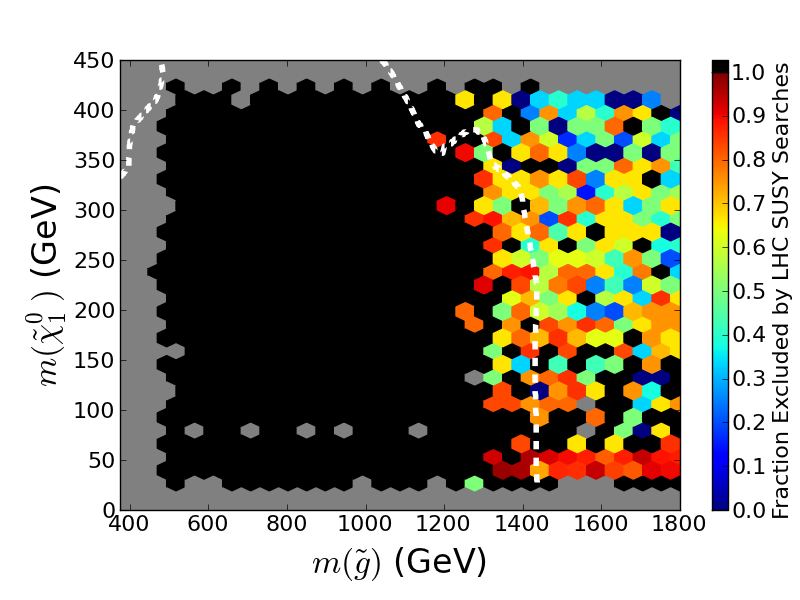}
\hspace{0.20cm}
\includegraphics[width=3.5in]{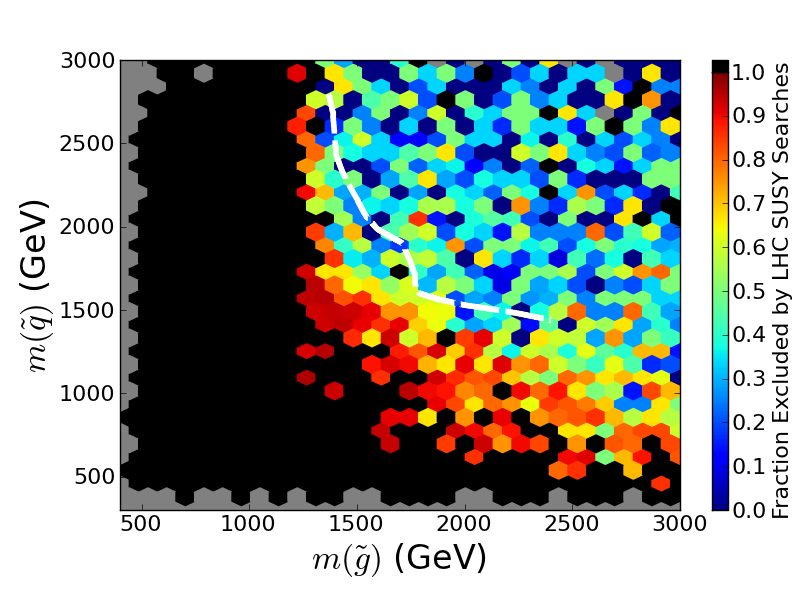}}
\vspace*{-0.10cm}
\caption{Coverage of the pMSSM low-FT model parameter space from the 7 and 8 TeV LHC searches in the gluino-LSP (left) 
and the lightest (1\textsuperscript{st}/2\textsuperscript{nd} generation) squark-gluino (right) mass planes. The simplified model analysis results from ATLAS are also shown for comparison as white lines. 
The grey holes in these 
panels arise from the rather low statistics of the low-FT model sample.}
\label{fig00ft}
\end{figure}

Since a much smaller variety of spectra are allowed in the low-FT model set, we might expect the experimental reach in the stop-LSP plane to be improved compared with the general neutralino LSP model sample, as some of the most challenging scenarios may not be consistent with the low-FT requirements. Fig.~\ref{fig1ft} shows that the region excluded by LHC searches in this plane is indeed somewhat larger for the low-FT model set. Of course, the overall fraction of models excluded across the plot is significantly enhanced as well, in agreement with the results in Tables~\ref{SearchList7} and~\ref{SearchList8}. This is in contrast to the case of light staus, also presented in this figure, where we see that the overall exclusion fraction is again enhanced, but here, the improvement is essentially uniform.  This indicates that the improved sensitivity is coming from searches for sparticles other than the stau, as we might expect.  As we saw for the general neutralino sample, the simplified model limit does not accurately depict the pMSSM coverage for stops, indicating that other searches are making important contributions to the excluded region.

\begin{figure}[htbp]
\centerline{\includegraphics[width=3.5in]{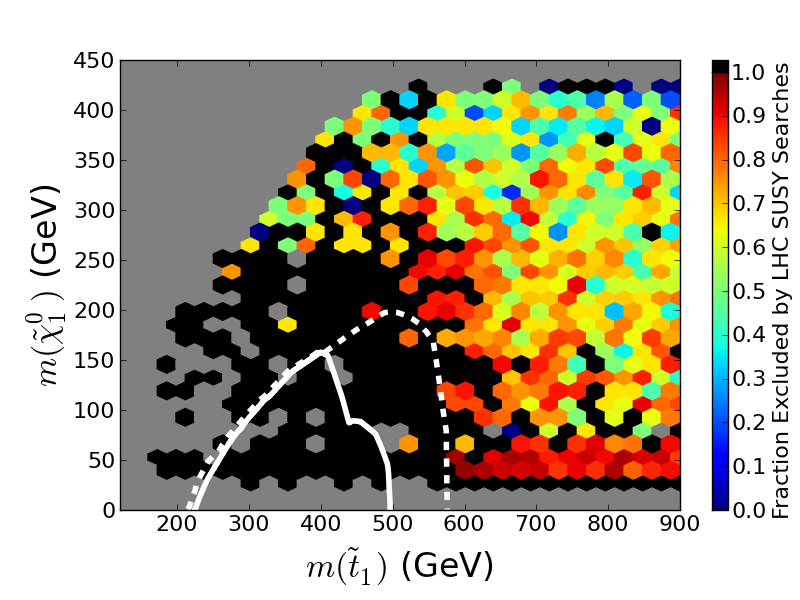}
\hspace{0.20cm}
\includegraphics[width=3.5in]{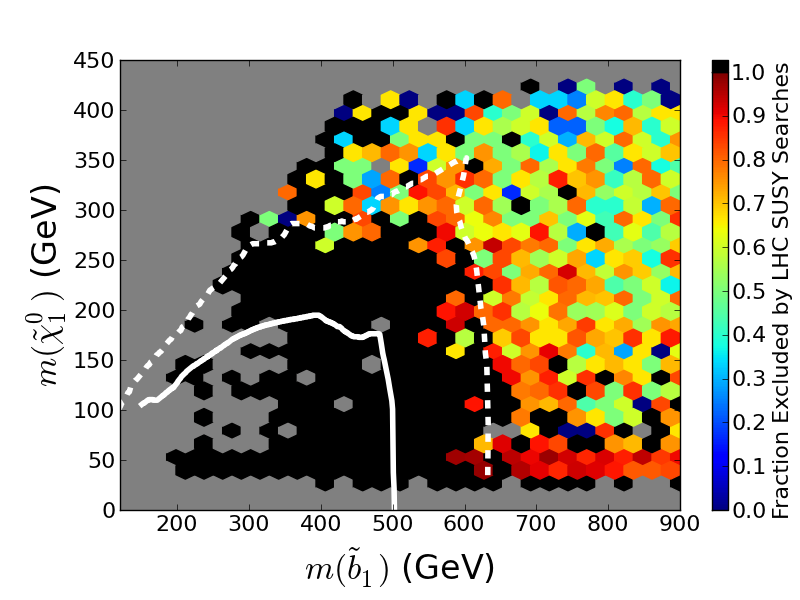}}
\vspace*{0.50cm}
\centerline{\includegraphics[width=3.5in]{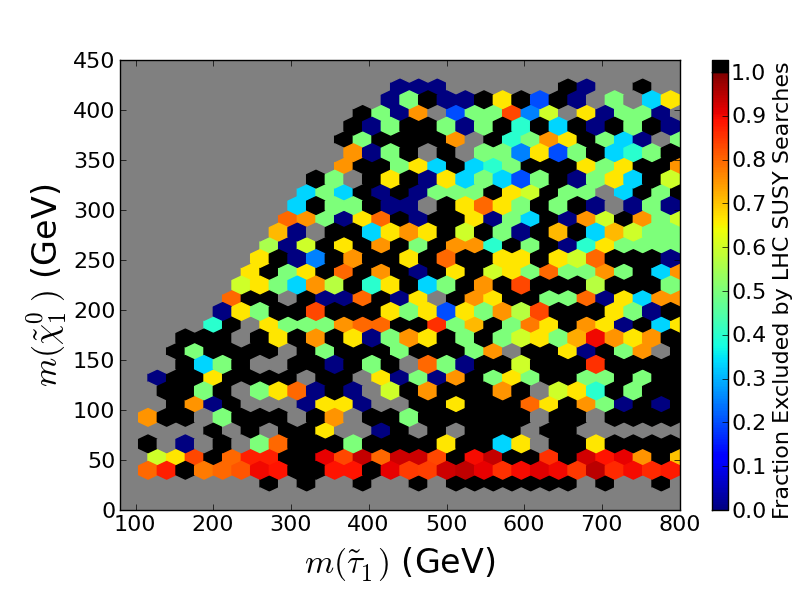}}
\vspace*{-0.10cm}
\caption{Coverage of the pMSSM low-FT model parameter space from the 7 and 8 TeV LHC searches in the lightest stop-LSP mass plane 
(top left), the lightest sbottom-LSP mass plane (top right) and for the lightest stau-LSP mass plane (bottom). 
The white lines represent the corresponding $95\%$ CL limit results obtained by ATLAS in the simplified model limit as discussed in the text.}
\label{fig1ft}
\end{figure}

Figure~\ref{fig2ft} shows the coverage of the 1\textsuperscript{st}/2\textsuperscript{nd} generation squark-LSP mass plane for the low-FT set which should be compared with 
the analogous results for the general neutralino model sample in Fig.~\ref{fig2}, shown above. As before, we see that the coverage is greatest for $\tilde u_L$ and 
$\tilde d_L$, followed by $\tilde u_R$ with the least coverage in searches for $\tilde d_R$ squarks. 
However, in all cases, we see that the coverage is far more complete for the low-FT 
set and is  generally more uniform across the mass plane as compared to the general neutralino model sample.

\begin{figure}[htbp]
\centerline{\includegraphics[width=3.5in]{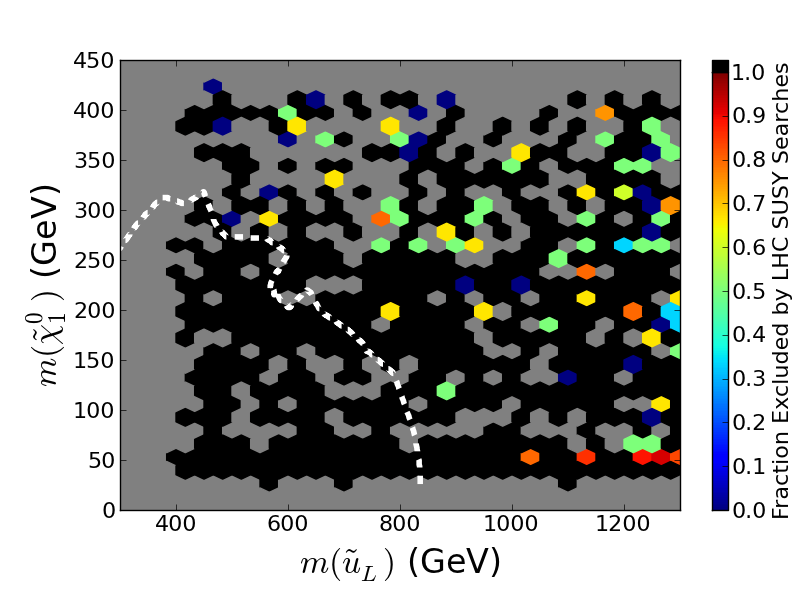}
\hspace{0.20cm}
\includegraphics[width=3.5in]{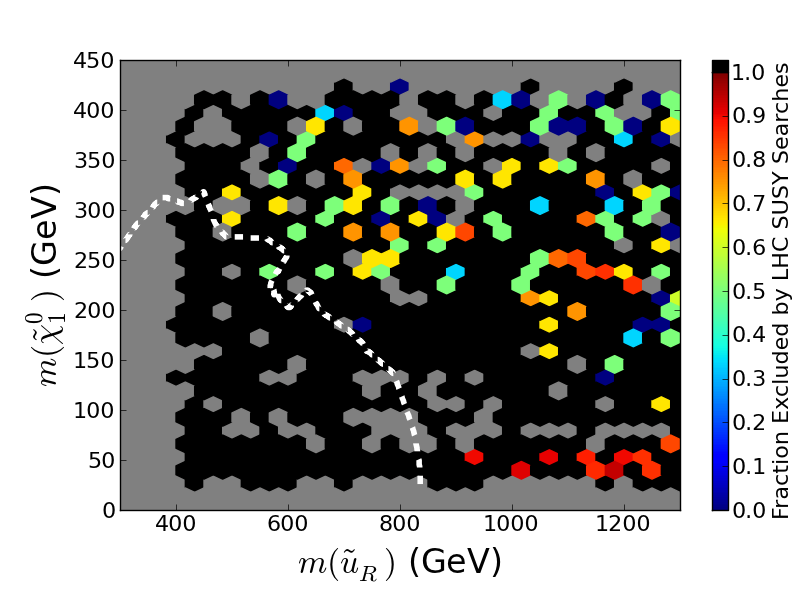}}
\vspace*{0.50cm}
\centerline{\includegraphics[width=3.5in]{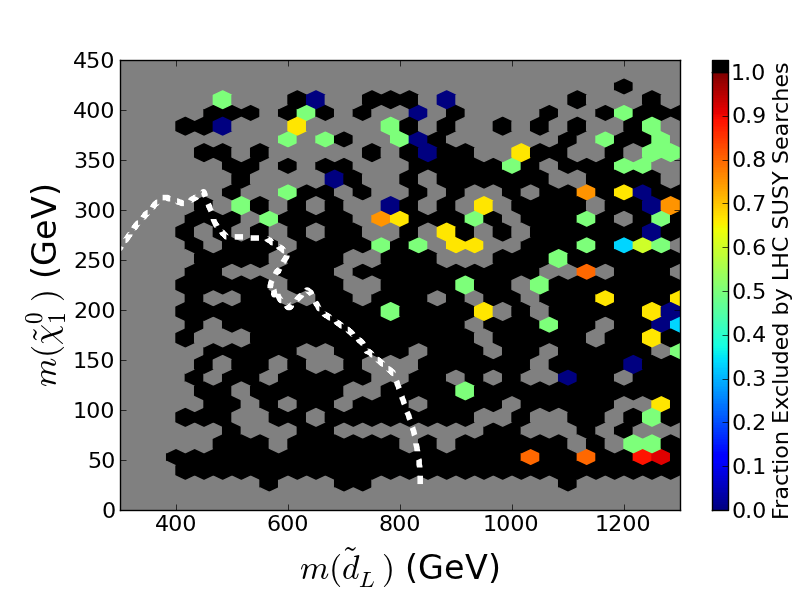}
\hspace{0.20cm}
\includegraphics[width=3.5in]{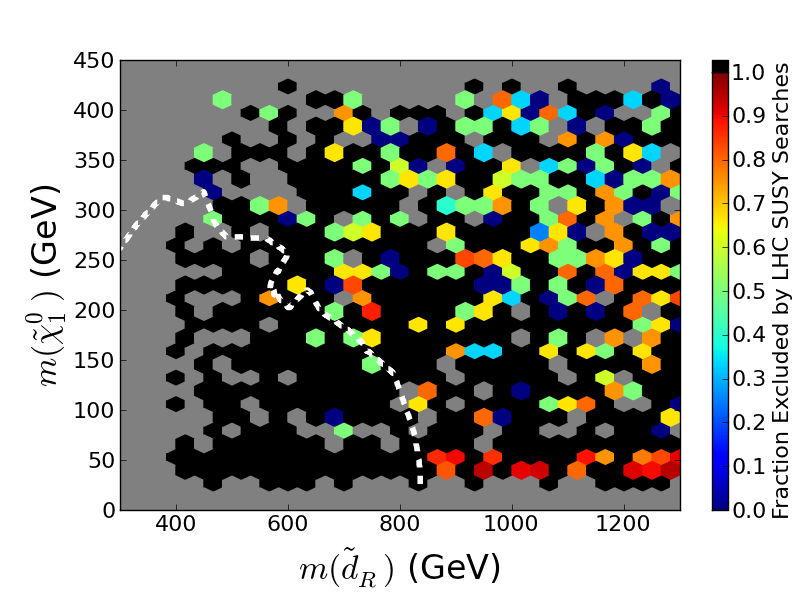}}
\vspace*{-0.10cm}
\caption{Same as the previous figure but now for $\tilde u_L$ (top left),  $\tilde u_R$ (top right), $\tilde d_L$ (bottom left), 
and $\tilde d_R$ (bottom right).}
\label{fig2ft}
\end{figure}

Figure~\ref{fig3ft} displays the analogue of Fig.~\ref{fig3} for the low-FT model set. In all sparticle searches the 7 and 8 TeV LHC coverage is, of course, more complete for this
model set. In the upper 
panels, we see that the sensitivity to light sleptons is reasonably good in the low-FT set (although the mass regions that are completely excluded remain small). The enhanced 
sensitivity to light sleptons most likely arises from the ubiquitous presence of a light chargino (with a mass below $\sim 460$ GeV) in this model sample. 
Having a light chargino means that 
light sleptons can be excluded not only via slepton pair production, but also by enhancing the detectability of charginos. The latter possibility occurs when the slepton 
is an intermediate in the chargino decay cascade, producing a much more distinctive signature (hard leptons) than the soft gauge bosons typically produced in electroweakino 
cascades. This is demonstrated in the bottom panels of Figure~\ref{fig3ft}, where we see that the exclusion efficiency for models with light charginos is reasonably good; this 
may result from an increased frequency of light sleptons (which are more common for the low-FT model set because of their role as co-annihilators) enhancing 
the chargino visibility through the cascade decays with final state leptons. The LHC searches are particularly sensitive to models containing light second charginos, in which case all 
6 electroweak gauginos are light (in contrast with the large neutralino model set, where the bino is frequently heavier than both charginos){\footnote {The second 
chargino is always found to be at least $\sim 100$ GeV heavier than the lighter one, however the distribution peaks near this value due to the nature of the parameter 
scan.}}. In this scenario, the four lepton search is highly effective, since a large number of leptons are frequently produced in cascades between the gaugino multiplets; although 
some of these leptons may be rather soft, they can still pass the low $p_T$ thresholds allowed by the high multiplicity lepton searches. Recall that in many cases, the 
charginos may be produced dominantly through decays of light stops and sbottoms, boosting their production cross sections and making them even more accessible to searches 
for multilepton final states.

\begin{figure}[htbp]
\centerline{\includegraphics[width=3.5in]{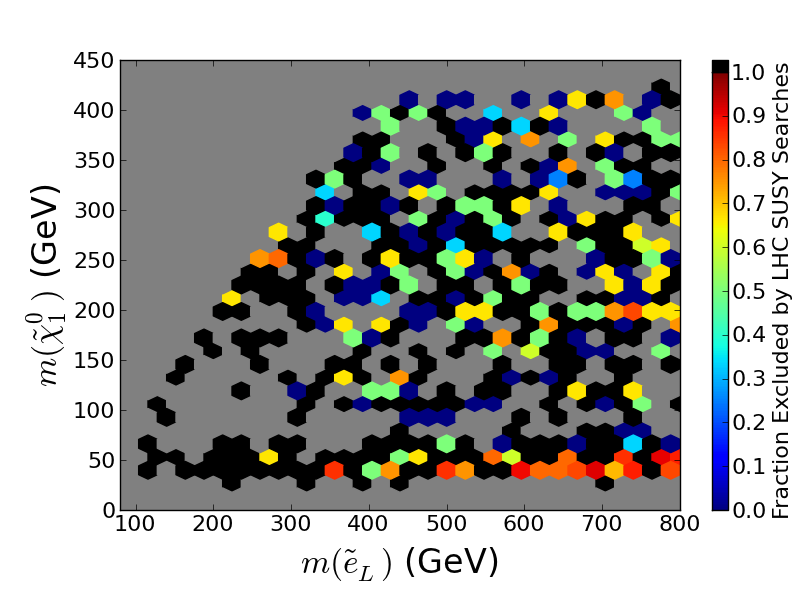}
\hspace{0.20cm}
\includegraphics[width=3.5in]{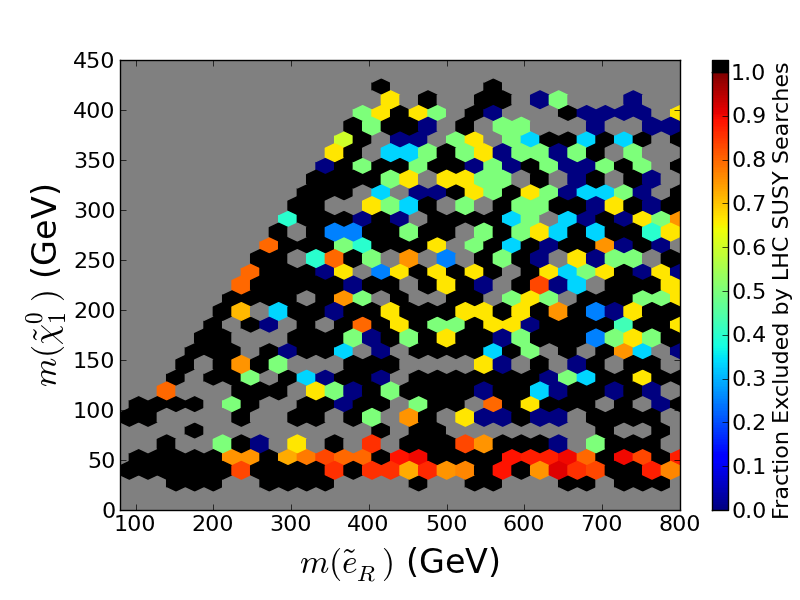}}
\vspace*{0.50cm}
\centerline{\includegraphics[width=3.5in]{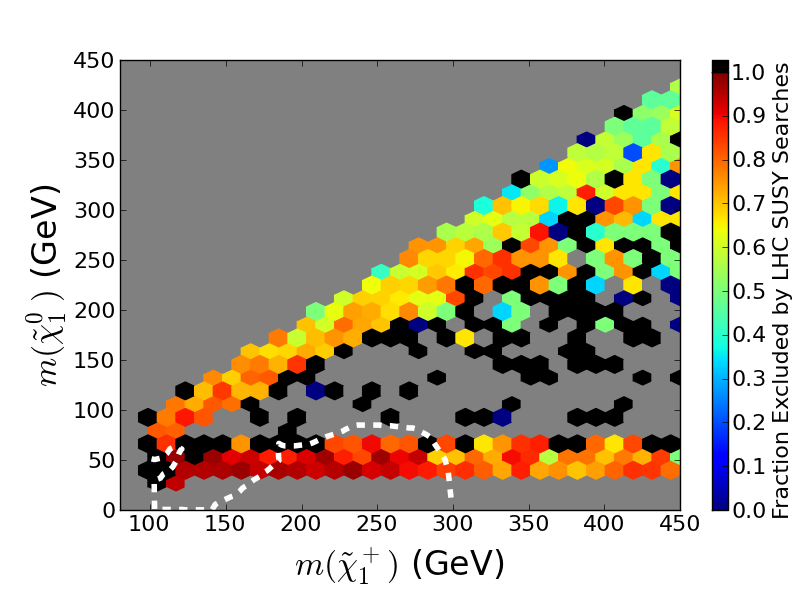}
\hspace{0.20cm}
\includegraphics[width=3.5in]{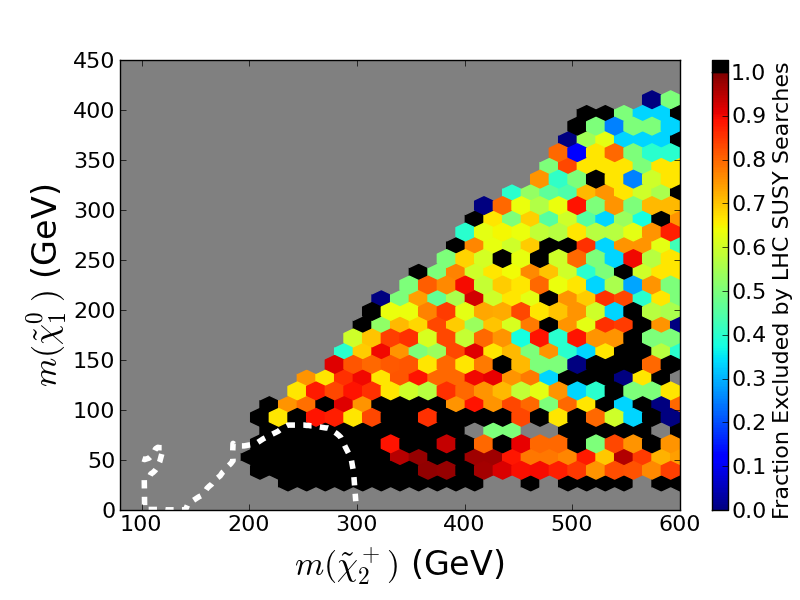}}
\vspace*{-0.10cm}
\caption{Same as the previous figure but now for $\tilde e_L$ (top left),  $\tilde e_R$ (top right), $\tilde \chi_1^\pm$ (bottom left), 
and $\chi_2^\pm$ (bottom right).}
\label{fig3ft}
\end{figure}

As before, we now examine the model coverage according to search category. Figure~\ref{fig4ft} shows the analogue of Fig.~\ref{fig4}, which compared the reach of targeted 3\textsuperscript{rd} generation searches with that of the more standard ``vanilla'' jets+MET(+leptons) searches. 
As we saw above in Tables~\ref{SearchList7} and~\ref{SearchList8}, both sets of searches are significantly more effective in the low-FT model set. We see that while the fraction of models excluded by the searches is much higher, the same pattern (with the ``vanilla'' searches most effective in the compressed region and the 3$^{rd}$ generation searches most effective in the uncompressed region) holds for the low-FT model sample.

Finally, we can again examine the effectiveness of the stop searches, which target stops decaying to $t + \chi^0$; we expect these searches may play a more significant role
in the low-FT model set. The effectiveness of these searches are compared to that of the zero-lepton+ jets + MET channel in Figure~\ref{fig5ft}. We see that the stop searches are much more effective in the low-FT model set, as expected, compared with the general neutralino LSP model sample, particularly in the light LSP region. Although stops decaying directly to $t + \chi^0$ with nearly 100\% branching fractions remain rare in the low-FT model set, there is frequently a sufficiently large splitting between charginos and the LSP to produce an on-shell gauge boson. Stop decays to $b + \chi^+$ can therefore mimic the $t + \chi^0$ final state if the chargino produces an on-shell W boson. The prevalence of this decay pattern in the low-FT model set accounts for the relative effectiveness of the stop searches when applied to this scenario. Interestingly, the common presence of multiple electroweakino multiplets below the stop mass greatly enhances the importance of multilepton searches (which detect leptons coming from cascade decays between neutralinos and charginos), and the direct sbottom search retains an important role, especially in the slightly compressed region (where decays to tops are kinematically disfavored). Overall, stops in the low-FT model set have a large variety of decay modes and are therefore susceptible to a much larger variety of searches than in the general neutralino LSP model sample.

\begin{figure}[htbp]
\centerline{\includegraphics[width=3.5in]{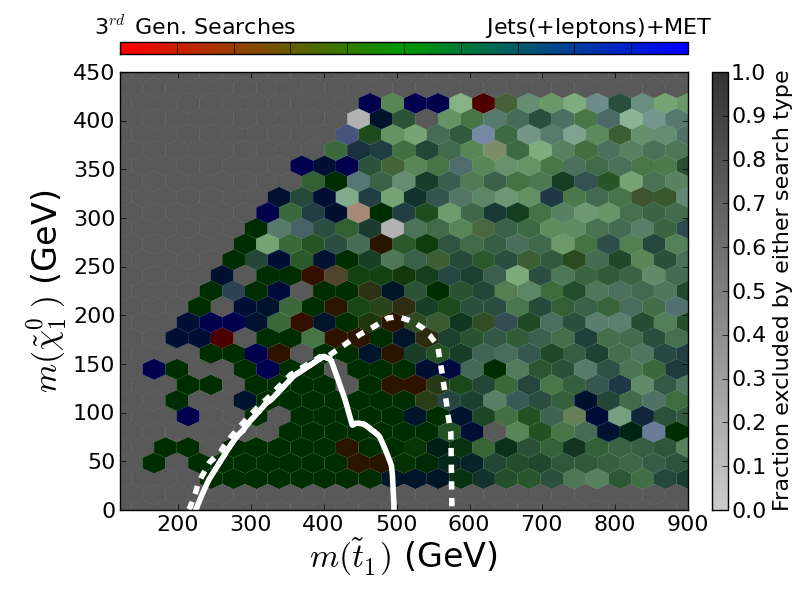}
\hspace{0.20cm}
\includegraphics[width=3.5in]{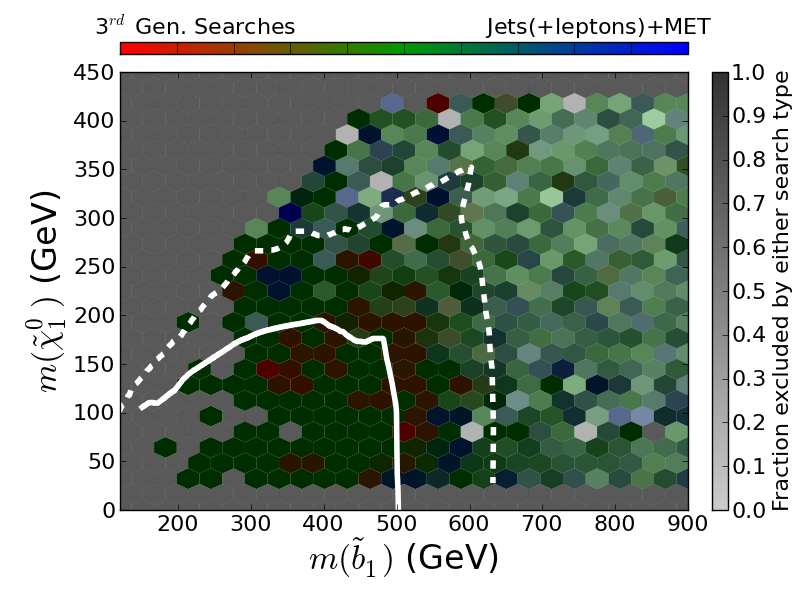}}
\vspace*{-0.10cm}
\caption{Comparison of the contributions to model coverage arising from jets (+ leptons) + MET and 3\textsuperscript{rd} generation searches in the in the stop-LSP (left) and sbottom-LSP (right) mass planes for the low-FT model set. The color-coding seen here is as described above.}
\label{fig4ft}
\end{figure}
\begin{figure}[htbp]
\centerline{\includegraphics[width=3.5in]{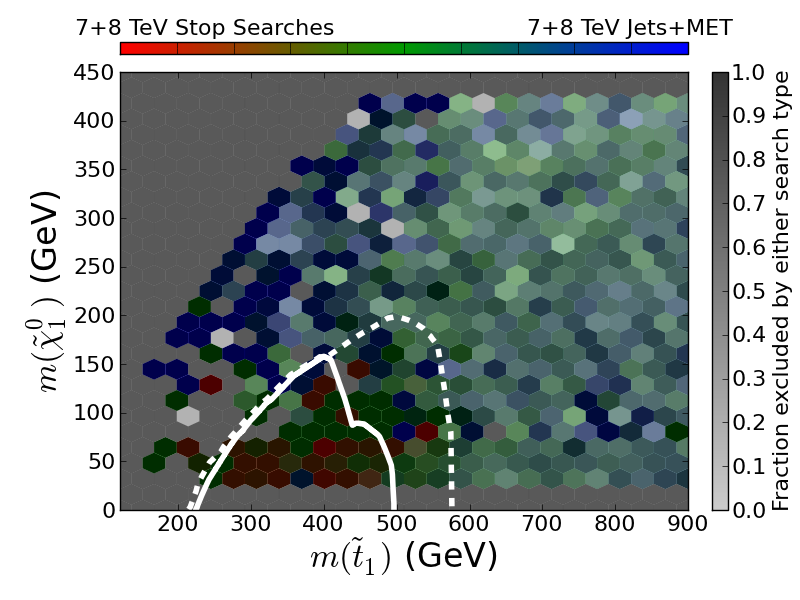}
\hspace{0.20cm}
\includegraphics[width=3.5in]{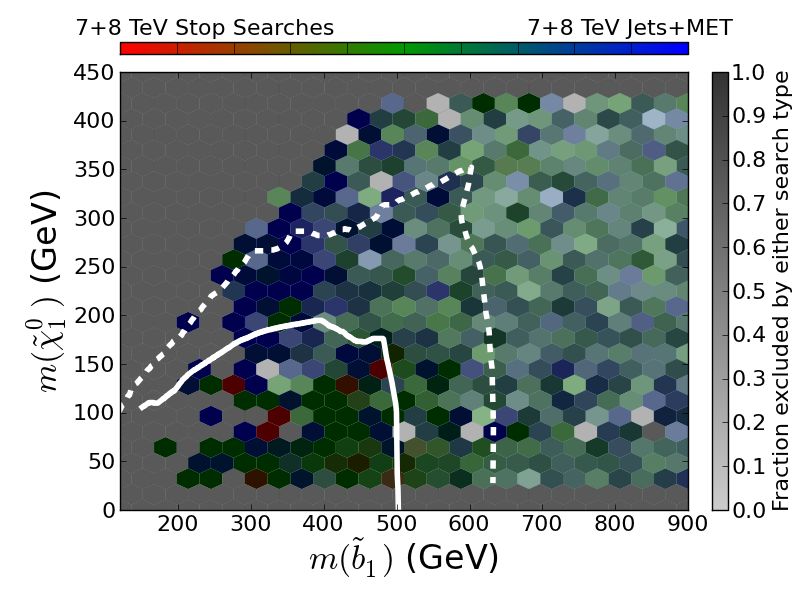}}
\vspace*{-0.10cm}
\caption{Comparison of the contributions to model coverage arising from the zero-lepton + jets + MET searches at 7 and 8 TeV and from the 0$\ell$-2$\ell$ searches for light stops in the stop-LSP (left) and sbottom-LSP (right) mass planes for the low-FT model set. The color coding seen here is as described above.}
\label{fig5ft}
\end{figure}

\section{Expectations for 14 TeV LHC SUSY Searches}

In addition to the 7 and 8 TeV LHC searches, future operations at $\sim$ 14 TeV, together with improved analysis techniques, will greatly extend the expected coverage of the pMSSM parameter space. 
In this section, we consider the impact of two of the more powerful of these searches to be performed by ATLAS, namely the zero-lepton jets + 
MET~\cite{ATLAS-EP} and the zero- and one-lepton stop analyses~\cite{ATLAS-EP2}{ \footnote{These analyses feature sliding missing energy and transverse mass cuts for optimal sensitivity to different stop masses, creating a very large effective 
number of signal regions. We chose to examine the signal regions that were optimized for stop masses of 800 GeV and 1 TeV with 3 ab$^{-1}$ of integrated luminosity, deriving $95\%$ CL$_s$ limits from the expected ATLAS background numbers and scaling these limits to estimate the sensitivity at 300 fb$^{-1}$ as well.}}. 
ATLAS investigated their expectations for these channels as part of the European Strategy for Particle Physics and the U.S. Snowmass Summer Study. 
We have performed our own simulation of these channels 
in a manner identical to that employed above for the 7 and 8 TeV LHC by following ATLAS as closely as possible. 
We note that in extrapolating from 300 fb$^{-1}$ to 3 ab$^{-1}$ of integrated luminosity, scaling of the required signal rate has been employed 
to obtain the results shown below.

\begin{table}
\centering
\begin{tabular}{|l|l|l|c|c|c|} \hline\hline
Search & Lumi & Reference & Neutralino & Gravitino &  Low-FT    \\
\hline

2-6 jets & 300 fb$^{-1}$  & ATLAS-PHYS-PUB-2013-002 & 90.74\% & 79.58\% & 97.35\% \\
Stop (0l) & 300 fb$^{-1}$  & ATLAS-PHYS-PUB-2013-011 & 3.88\% & 5.03\% & 1.90\% \\
Stop (1l) & 300 fb$^{-1}$  & ATLAS-PHYS-PUB-2013-011 & 16.98\% & 33.43\% & 52.09\% \\
2-6 jets & 3000 fb$^{-1}$  & ATLAS-PHYS-PUB-2013-002 & 97.08\% & 90.57\% & 99.96\% \\
Stop (0l) & 3000 fb$^{-1}$  & ATLAS-PHYS-PUB-2013-011 & 18.81\% & 14.9\% & 39.27\% \\
Stop (1l) & 3000 fb$^{-1}$  & ATLAS-PHYS-PUB-2013-011 & 43.45\% & 61.77\% & 93.43\% \\

\hline\hline
\end{tabular}
\caption{Percentages of the neutralino and low-FT model sets that \textit{survive} the 7 \& 8 TeV LHC searches but are expected to be excluded by the 14 TeV ATLAS jets + MET 
and stop searches. With 300 (3000) fb$^{-1}$ of data, the combination of these searches covers 90.83\% (97.15\%) of the neutralino model set and 97.69\% (100\%) of the 
low-FT model set.}
\label{SearchList14}
\end{table}

In order to simplify our analysis, and to obtain the results presented here within a reasonable amount of CPU time, we consider only the $\sim 29.8$k neutralino LSP models that survive the 7 and 8 TeV LHC analyses 
above and also predict a Higgs mass of $126\pm 3$ GeV, as well as the analogous subset of 2645 surviving low-FT models. Given the high luminosities, these subsets of 
models required $\sim 2 \cdot 10^6$ core-hrs of CPU to generate 14 TeV signal events and perform the necessary analysis{\footnote {Note that this represents less 
than $\sim 10\%$ of our total set of models, implying that a study of these sets in their entirety would have required $\sim 20-25 \cdot 10^6$ core-hrs of CPU which is far beyond 
current capabilities.}}. Since the dominant direct impact of these search channels is on the production of squarks and gluinos, including stops and sbottoms, we restrict 
the discussion below to the gluino-squark-LSP sector and the stop- and sbottom-LSP sectors.  We note that 
since the results of the 7 and 8 TeV analyses are essentially independent of the Higgs mass, it is expected that the results for this narrow Higgs mass range 
would in fact be applicable, at least to a very good approximation, to the entire general neutralino model set. (Recall the Higgs mass cut is already applied to the low-FT model set.) Our conclusions are summarized in 
Table~\ref{SearchList14} and will be discussed in detail below.

\begin{figure}[htbp]
\centerline{\includegraphics[width=3.5in]{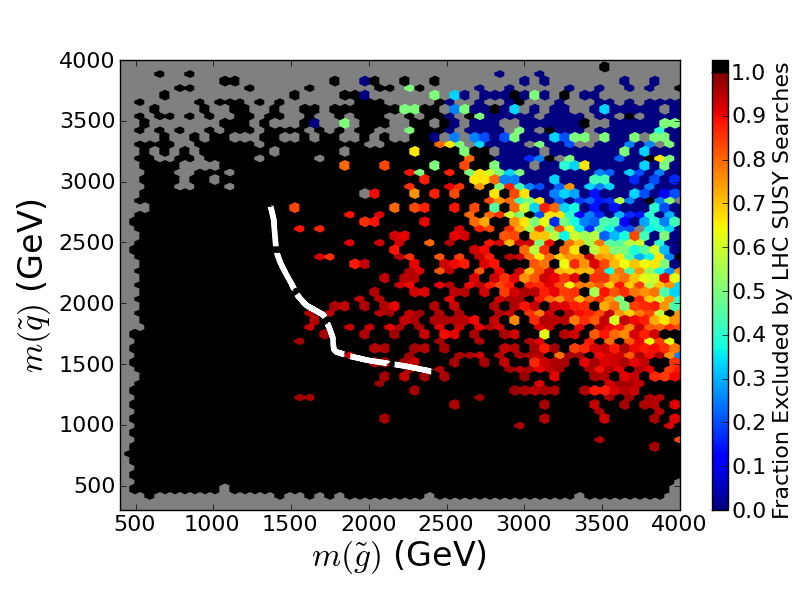}
\hspace{0.20cm}
\includegraphics[width=3.5in]{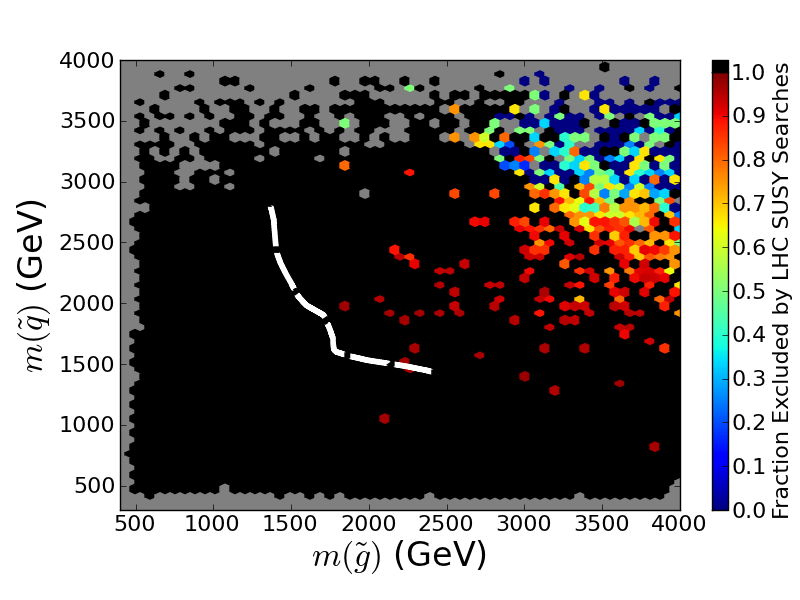}}
\vspace*{0.50cm}
\centerline{\includegraphics[width=3.5in]{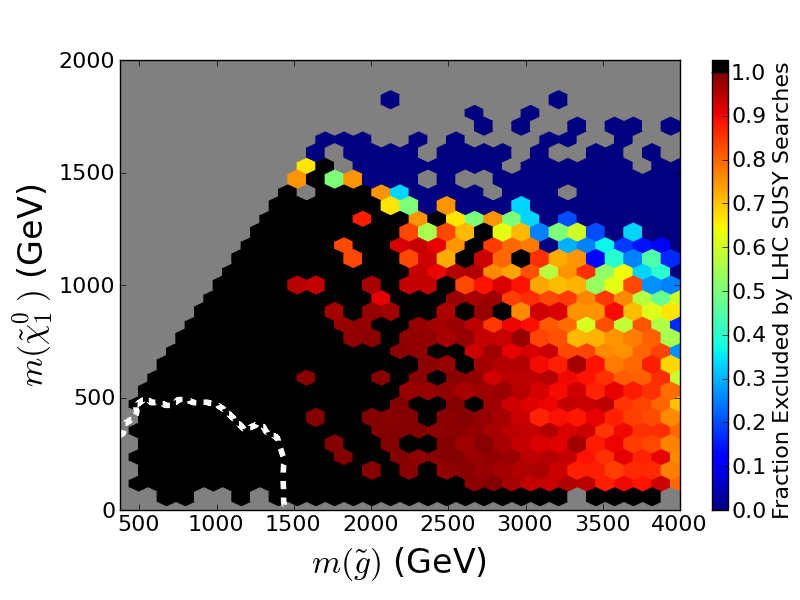}
\hspace{0.20cm}
\includegraphics[width=3.5in]{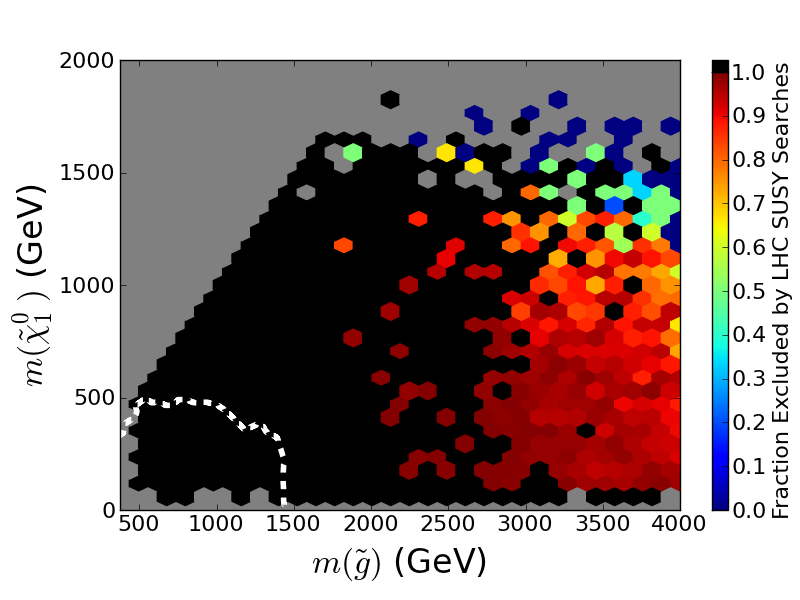}}
\vspace*{-0.10cm}
\caption{Expected results from the combined jets + MET plus 0$\ell$ and 1$\ell$ stop searches at the 14 TeV LHC assuming an integrated luminosity of 300 fb$^{-1}$ (left) and 
3000 fb$^{-1}$ (right), in the lightest squark-gluino and LSP-gluino mass planes for the general neutralino model set.  The white curves represent the corresponding
7/8 TeV simplified model results.}
\label{figxx1}
\end{figure}

Let us first consider the general neutralino LSP model set. In Fig.~\ref{figxx1} we present the pMSSM coverage in the lightest squark-gluino and the gluino-LSP mass planes for the general neutralino model set at 14 TeV arising from the combination of the jets + MET and 0-lepton and 1-lepton stop analyses with either 300 or 3000 fb$^{-1}$ of integrated luminosity. Here we see that even with the lower value of the integrated luminosity, the reach of the 14 TeV searches is extensive. Table~\ref{SearchList14} shows the model coverage provided by the 
individual searches. Specifically, we find that 90.8 (97.4)$\%$ of the models in this subset can be probed by these analyses assuming 300 (3000) fb$^{-1}$ of data.
We expect these fractions to be roughly valid for the entire neutralino model 
set. In particular, in these figures we see that increasing the integrated luminosity makes a reasonable impact on the overall pMSSM model coverage.  Although this coverage is indeed 
very significant, we observe that models with 1\textsuperscript{st}/2\textsuperscript{nd} generation squarks as light as $\sim 700-800$ GeV and/or gluinos as light as $\sim 1.5$ TeV can 
still survive these searches, even at high integrated luminosities, and that these models fall below the 7,8 TeV simplified model constraints. 
Interestingly, surviving models with light squarks and gluinos remain undetected not only because of spectrum compression, 
but also because of specific decay patterns for the squark and/or gluino which nearly always produce high-$p_T$ leptons. In such cases, the models will immediately fail the lepton veto of 
the powerful jets + MET search and so remain undetected. Conversely, these models may fail to produce $b$-jets and therefore do not produce a signal in the stop searches. Clearly adding additional analyses, specifically those targeting final state leptons, will only increase the model coverage and will compensate for the underestimated systematics.

\begin{figure}[htbp]
\centerline{\includegraphics[width=3.5in]{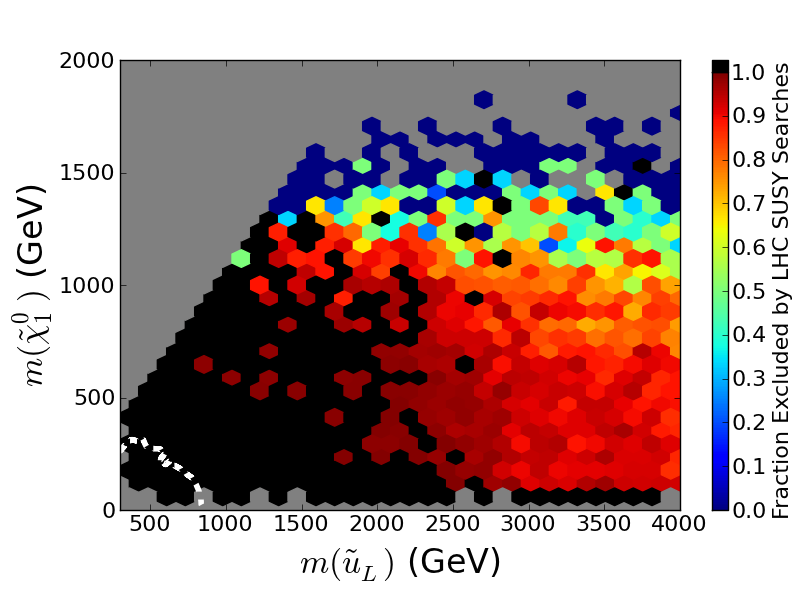}
\hspace{0.20cm}
\includegraphics[width=3.5in]{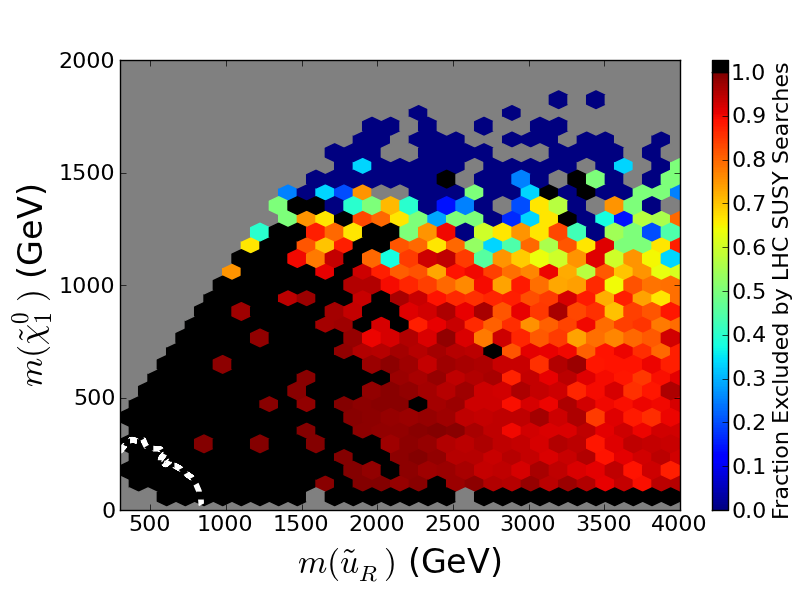}}
\vspace*{0.50cm}
\centerline{\includegraphics[width=3.5in]{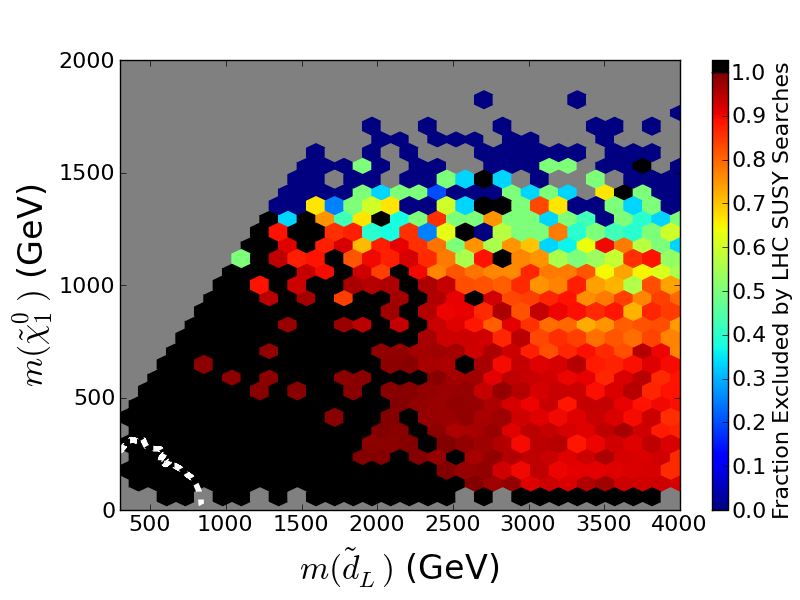}
\hspace{0.20cm}
\includegraphics[width=3.5in]{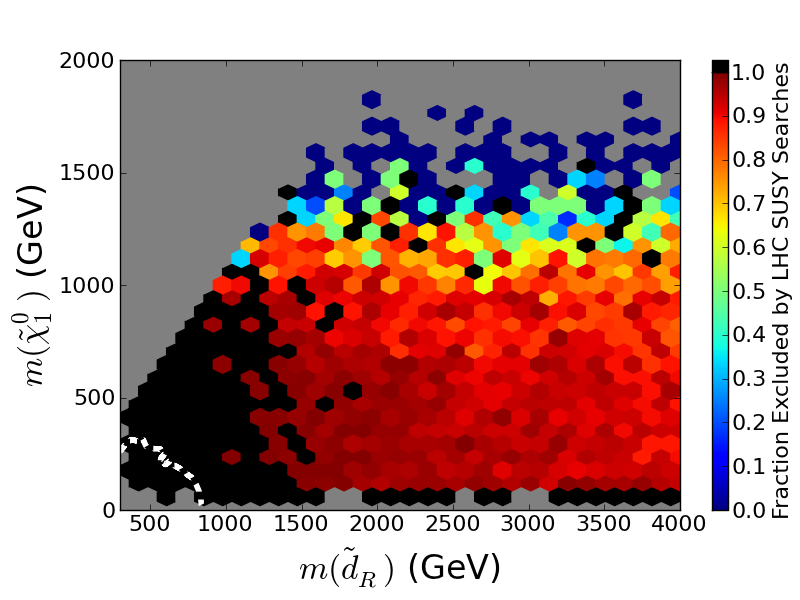}}
\vspace*{-0.10cm}
\caption{Combined jets + MET plus 0$\ell$ and 1$\ell$ stop search results at 14 TeV, assuming an integrated luminosity of 300 fb$^{-1}$, shown in the various LSP-squark mass planes for the general neutralino model sample.}
\label{figxx2}
\end{figure}

In order to further elucidate the potential of the LHC to probe light squarks, Figs.~\ref{figxx2} and ~\ref{figxx3} show the search efficiencies in the squark-LSP mass plane separately 
for the $\tilde u_L,~ \tilde u_R,~ \tilde d_L$ and $\tilde d_R$ squarks at 14 TeV for an integrated luminosity of 300 and 3000 fb$^{-1}$, respectively. Here we see a number of things: 
($i$) since $\tilde u_L$ and $\tilde d_L$ are similar in mass they are produced together and increase the corresponding signal rate as observed at 7,8 TeV. 
Thus it is quite rare (but not 
impossible) for light $\tilde u_L, \tilde d_L$ to remain undetected by the 14 TeV jets + MET and the 0l and 1l stop searches. ($ii$) Since $\tilde u_R$ and $\tilde d_R$ have 
uncorrelated masses and are iso-singlets, the LHC exclusion for right-handed squarks is again reduced compared with that for left-handed squarks. In particular, the $\tilde d_R$ production is also further suppressed by the PDFs and we see that quite light $\tilde d_R$ squarks would remain viable after these searches. It would be interesting to see 
how models with light squarks would fare if additional channels incorporating hard leptons were included in a more complete analysis. 
In Figures~\ref{figxx1},~\ref{figxx2}, and~\ref{figxx3}, we see that while the limits on colored sparticles are quite impressive, models with light LSPs, below $\sim$ 150 GeV, remain viable provided that the colored sparticles are heavy. This is not surprising as placing direct constraints on the electroweak sector will require dedicated analyses beyond the jets + MET and stop searches considered here.

\begin{figure}[htbp]
\centerline{\includegraphics[width=3.5in]{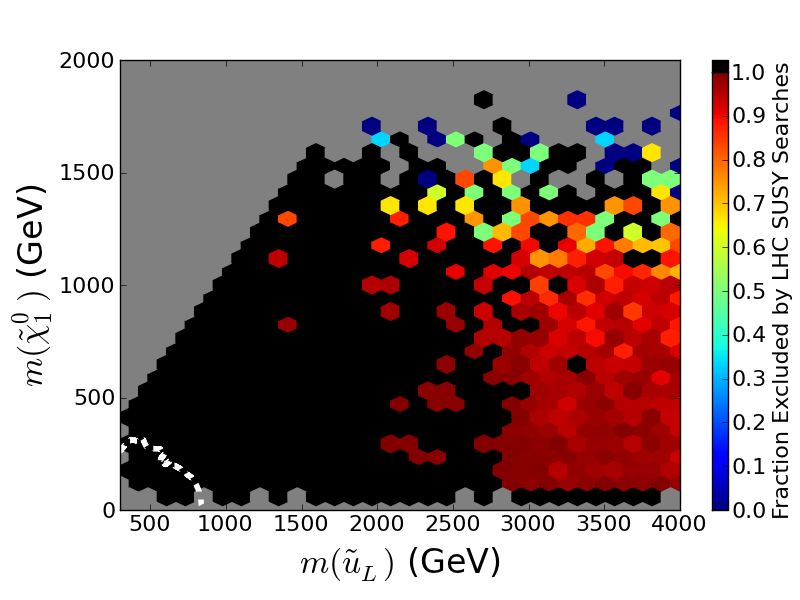}
\hspace{0.20cm}
\includegraphics[width=3.5in]{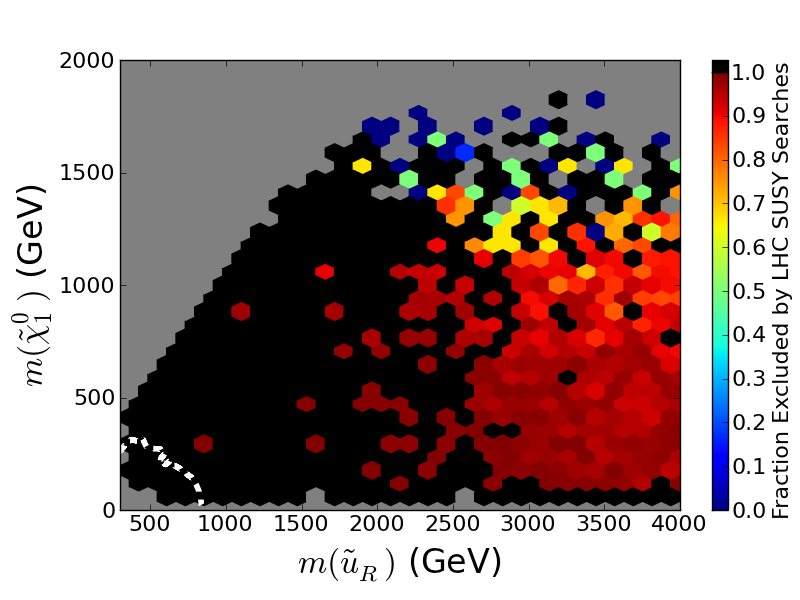}}
\vspace*{0.50cm}
\centerline{\includegraphics[width=3.5in]{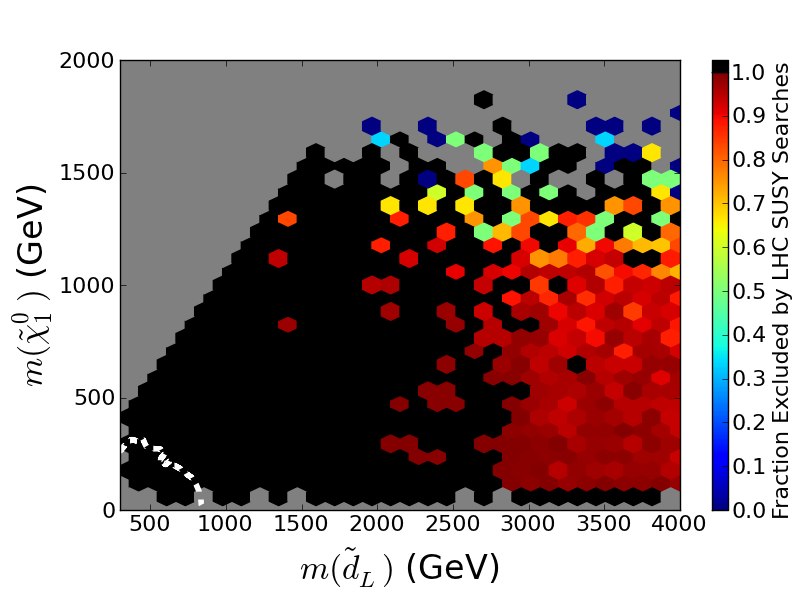}
\hspace{0.20cm}
\includegraphics[width=3.5in]{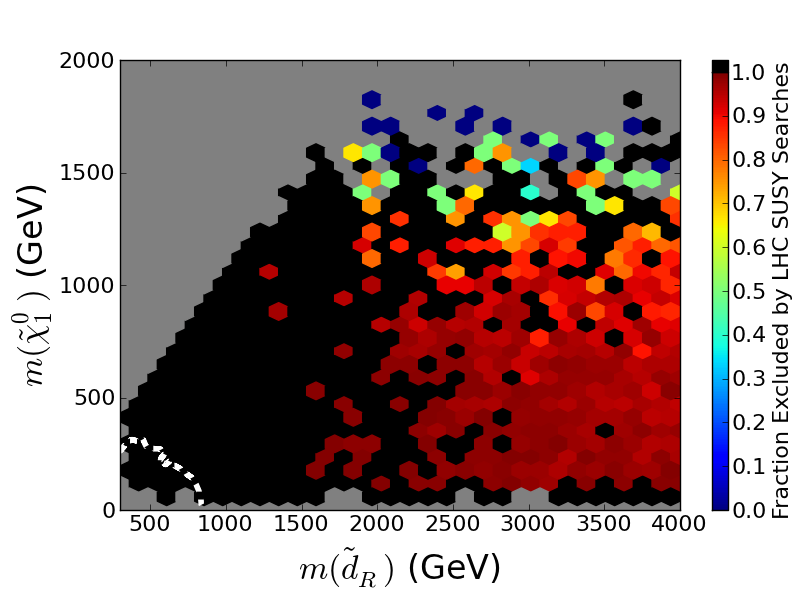}}
\vspace*{-0.10cm}
\caption{Same as the previous figure but now for an integrated luminosity of 3000 fb$^{-1}$.}
\label{figxx3}
\end{figure}

Since the 14 TeV 0$\ell$ and 1$\ell$ stop searches specifically probe third generation sparticles, it is particularly interesting to examine the 14 TeV LHC exclusion reach in both the lightest stop-LSP and lightest sbottom-LSP mass planes, as shown in Fig.~\ref{figrr1}. We note that as in the previous figures, this illustration shows the fraction of models excluded by the {\it combination} of all the searches. Interestingly, the regions covered by the 14 TeV searches are similar for stops and sbottoms, which suggests that the coverage is determined mainly by the total production cross-section and spectrum compression, rather than by details of the decay process. This in turn indicates that a generic search (jets + MET) is driving the reach.  This hypothesis is confirmed in Fig.~\ref{figrr2}, which compares the fraction of models covered by jets + MET to that covered by the targeted stop searches in the stop-LSP mass plane. We see that at 300 fb$^{-1}$ the jets + MET search is indeed essentially doing all of the work. Although the stop searches become more effective at higher integrated luminosities, the jets + MET exclusion is still dominant (or both searches are equally effective) in most of the parameter space. The lackluster performance of the 14 TeV stop searches is not altogether surprising; we saw previously that the 8 TeV stop searches also had a very limited effectiveness because their sensitivity plummets when the stop branching fraction to t + $\chi^0$ is significantly less than 100\%, which is typically the case in our model sample. By analogy with our 8 TeV results, we expect that a 14 TeV direct sbottom search would substantially improve the reach in the stop-LSP mass plane. Figure~\ref{figrr2} also shows that most of the models which are detected by the stop searches have stops and sbottoms with masses below $\sim 1.1-1.2$ TeV. This is not surprising as the direct stop/sbottom pair production cross section is falling 
rapidly for masses significantly larger than this, resulting in too low of a signal rate when the relevant branching fractions are accounted for. This is in essential agreement with 
the ATLAS analysis presented in Ref.~\cite{ATLAS-EP2}. Finally, we note the existence of a handful of models with stop/sbottom masses below $\sim 700$ GeV which remain viable even after 3000 fb$^{-1}$ of luminosity. These models generally have a small splitting between the LSP and the stop or sbottom, meaning that the branching fraction to tops is suppressed with the overall spectrum being compressed. These models are prime targets for the direct sbottom search, and are unlikely to remain viable once it is included.

\begin{figure}[htbp]
\centerline{\includegraphics[width=3.5in]{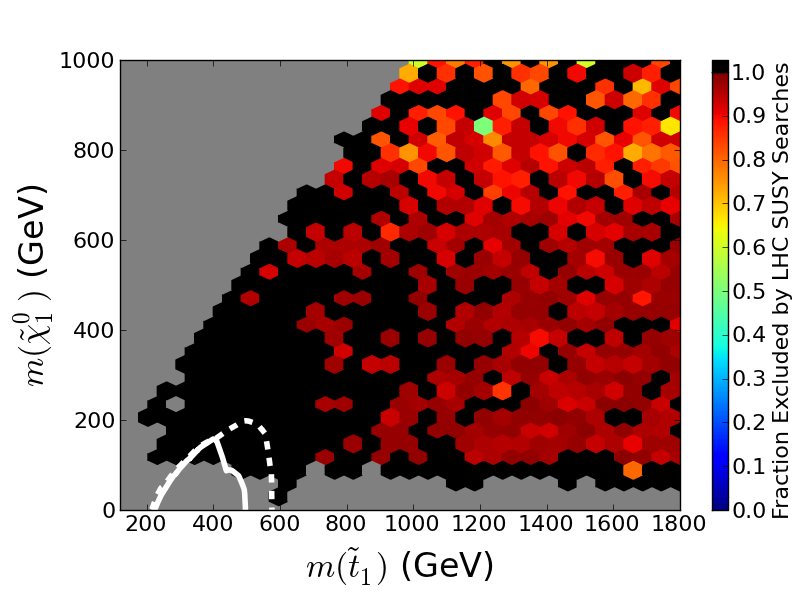}
\hspace{0.20cm}
\includegraphics[width=3.5in]{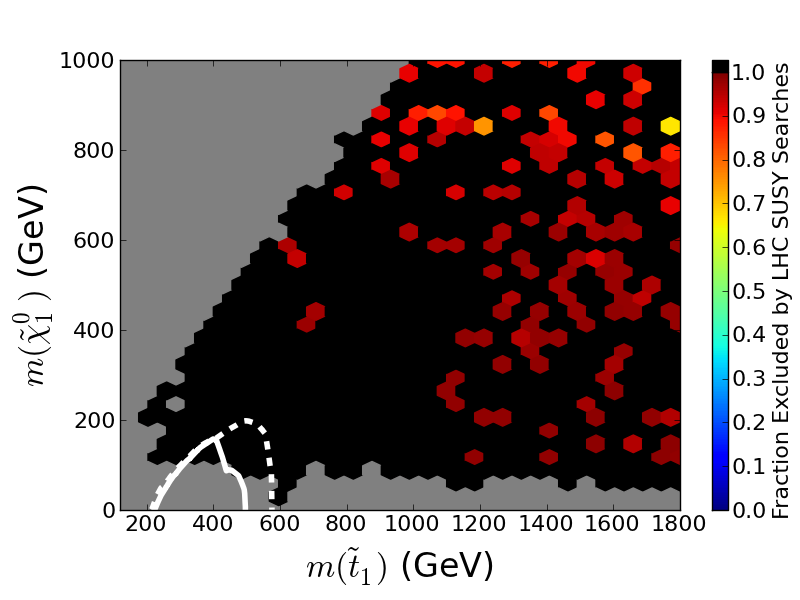}}
\vspace*{0.50cm}
\centerline{\includegraphics[width=3.5in]{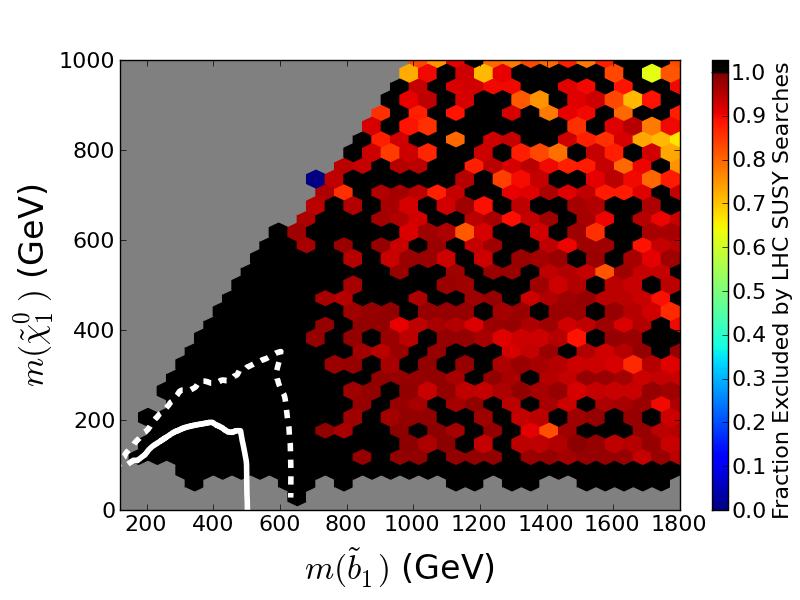}
\hspace{0.20cm}
\includegraphics[width=3.5in]{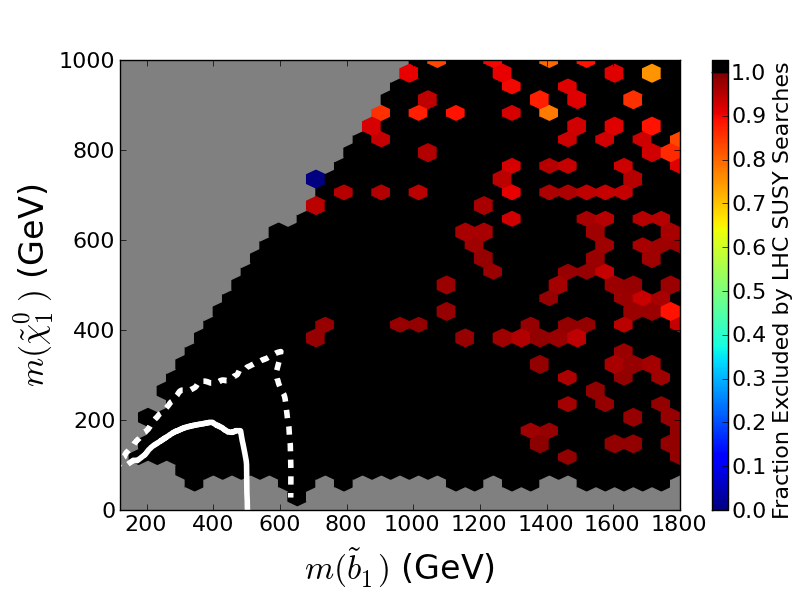}}
\vspace*{-0.10cm}
\caption{Expected results from the combined jets + MET and 0,1-$\ell$ stop searches at the 14 TeV LHC assuming an integrated luminosity of 300 fb$^{-1}$ (left) and 
3000 fb$^{-1}$ (right), in the LSP-stop (top) and LSP-sbottom (bottom) mass planes for the general neutralino model sample. 
The same simplified model exclusions as shown above are also displayed.}
\label{figrr1}
\end{figure}
\begin{figure}[htbp]
\centerline{\includegraphics[width=3.5in]{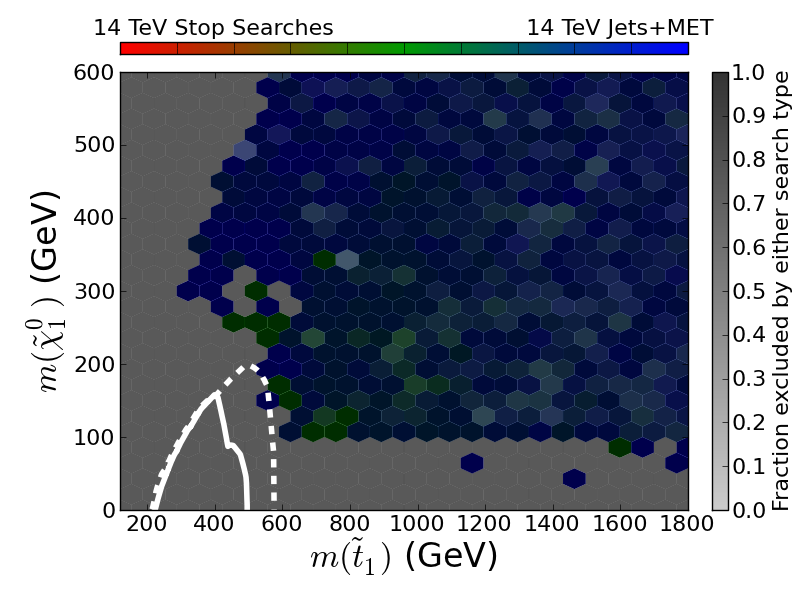}
\hspace{0.20cm}
\includegraphics[width=3.5in]{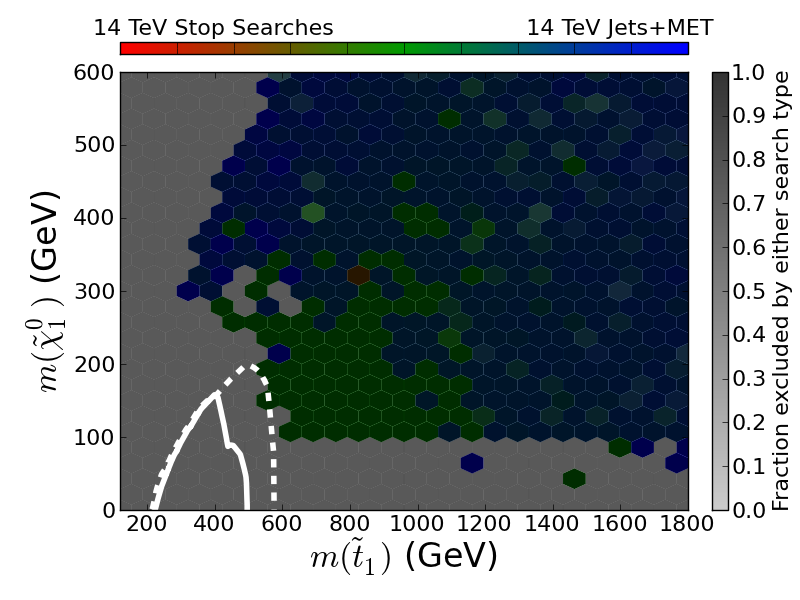}}
\vspace*{0.50cm}
\centerline{\includegraphics[width=3.5in]{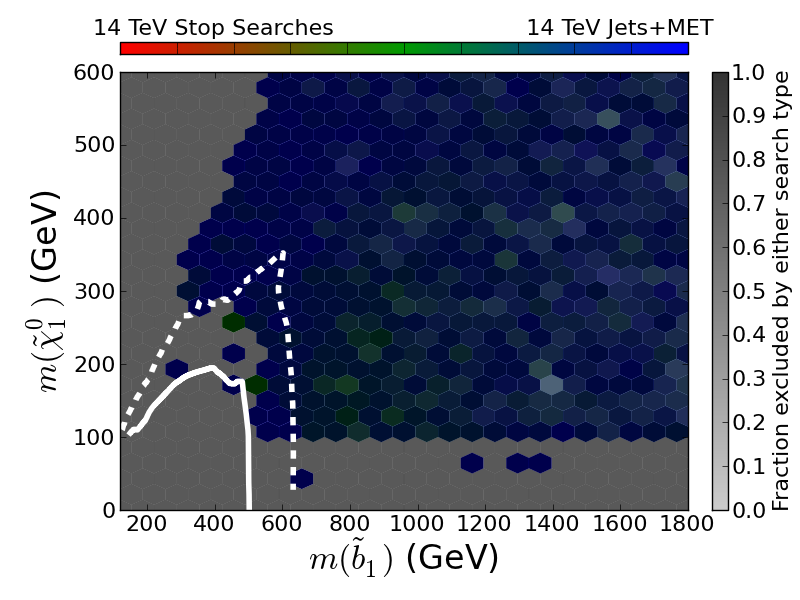}
\hspace{0.20cm}
\includegraphics[width=3.5in]{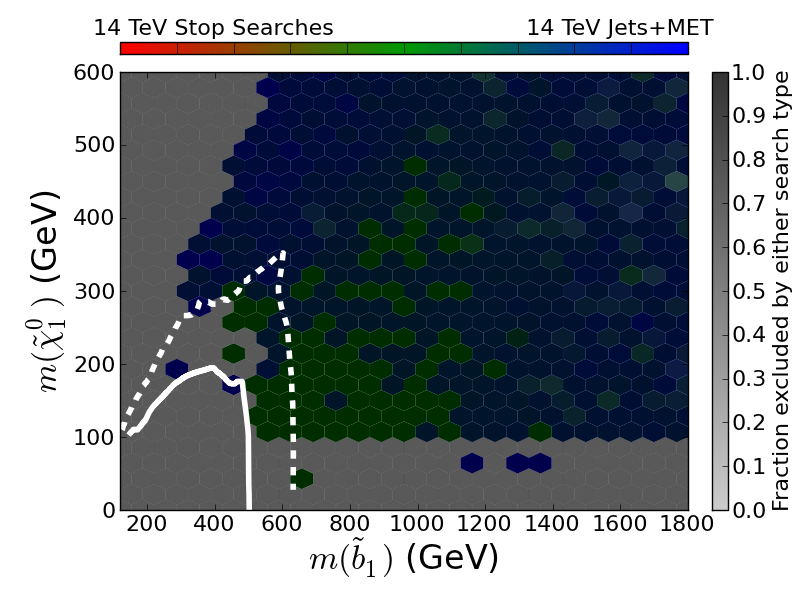}}
\vspace*{-0.10cm}
\caption{Comparison of the search efficiencies for the jets + MET and 0,1-$\ell$ stop searches at the 14 TeV LHC assuming an integrated luminosity of 300 fb$^{-1}$ (left) and 
3000 fb$^{-1}$ (right), for stops (top) and sbottoms (bottom) in the general neutralino model sample.} 
\label{figrr2}
\end{figure}

Although the model coverage is seen to be quite significant from just the combination of the jets + MET and 0$\ell$ and 1$\ell$ stop channels alone, further detailed study of the 
neutralino pMSSM model set at the 14 TeV LHC, incorporating more channels, is certainly warranted.

Finally, we turn to the $\sim 2.6$k model subset of low-FT models that survive the 7 and 8 TeV searches and  
see how they fare at 14 TeV.  The impact of the searches on the lightest 
squark-gluino and the LSP-gluino mass planes are shown in Fig.~\ref{figxxlowFT} for an integrated luminosity of 300 fb$^{-1}$. Here we see that, except for a handful of pixels, the 
panels are found to be almost entirely black, indicating that essentially {\it all} of the remaining low-FT model set would be covered by these two sets of analyses. As 
Table~\ref{SearchList14} shows, both the jets + MET as well as the 0$\ell$ and 1$\ell$ stop searches perform extremely well when applied to the low-FT set, as we might have expected based on the 7 
and 8 TeV results above. In particular we see from the Table that {\it none} of the low-FT models survive all of these searches at the higher integrated luminosity (and hence the corresponding panels 
are absent from Fig.~\ref{figxxlowFT}). Indeed, even with the lower integrated luminosity (300 fb$^{-1}$), only 56 of the low-FT models are found to survive the combined 14 TeV analyses! (This 
corresponds to a fractional coverage for the remaining low-FT models of $97.4\%$ at this integrated luminosity.)  This result demonstrates that the 14 TeV LHC can provide
a more definitive statement on the existence of natural SUSY.

\begin{figure}[htbp]
\centerline{\includegraphics[width=3.5in]{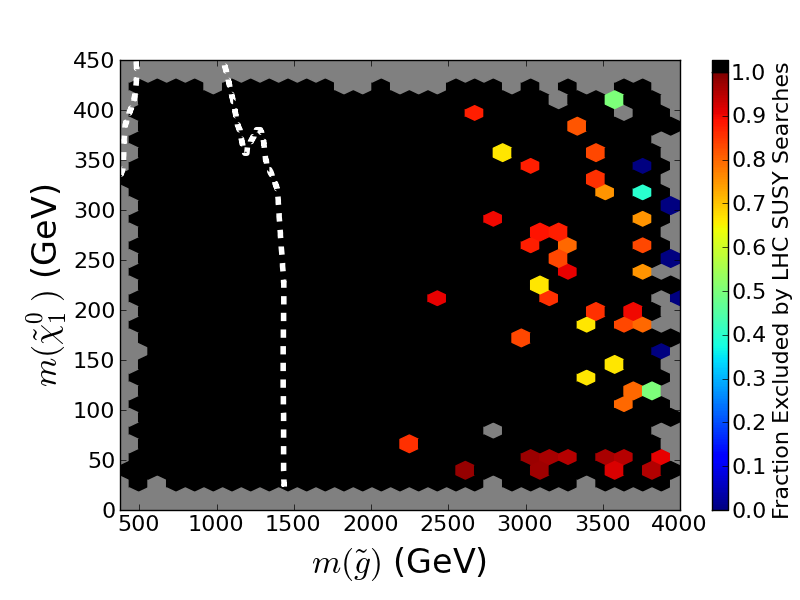}
\hspace{0.20cm}
\includegraphics[width=3.5in]{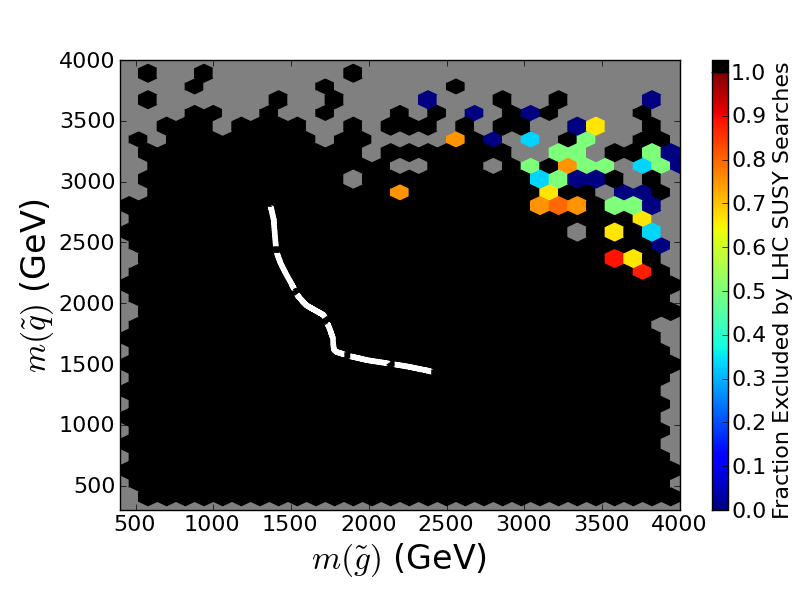}}
\vspace*{-0.10cm}
\caption{Results similar to those as shown in Fig.~\ref{figxx1} above but now for the low-FT model set for a luminosity of 300 fb$^{-1}$ as described in the text.}
\label{figxxlowFT}
\end{figure}

As can be seen from Fig.~\ref{figxxlowFT}, the few surviving models generally have very heavy squarks and/or gluinos. The allowed region is far smaller than in the general neutralino 
model set for two reasons. First, the decay patterns that produce high-$p_T$ leptons in the general neutralino model sample 
generally involve a bino-like intermediate state, which is incompatible 
with the necessity of having a bino-like LSP in the low-FT model set. Second, the low-FT spectra are required to be relatively uncompressed since fine-tuning places an upper limit 
of $\sim$ 400 GeV on the LSP mass, in contrast to cases in the general neutralino model set where the LSP can be heavier than 1 TeV. These effects combine to allow the nearly complete 
coverage of the low-FT model set at the 14 TeV LHC with only 300 fb$^{-1}$ as can be seen from Table~\ref{SearchList14}. The addition of the 0$\ell$ and 1$\ell$ stop searches captures the single model remaining after the jets + MET search at 3000 fb$^{-1}$, as can be seen from Table~\ref{SearchList14}.

It is interesting to examine what the coverage in the individual squark-LSP mass planes looks like at 300 fb$^{-1}$; Fig.~\ref{figxxlowFT2} shows 
these results. As one might expect, here we see that the lightest surviving squarks tend to be $\tilde d_R$ since they have the smallest production cross section due to both their isospin and the PDFs. But even in this case all surviving models have right-handed down squarks above $\sim 2.3$ TeV.

\begin{figure}[htbp]
\centerline{\includegraphics[width=3.5in]{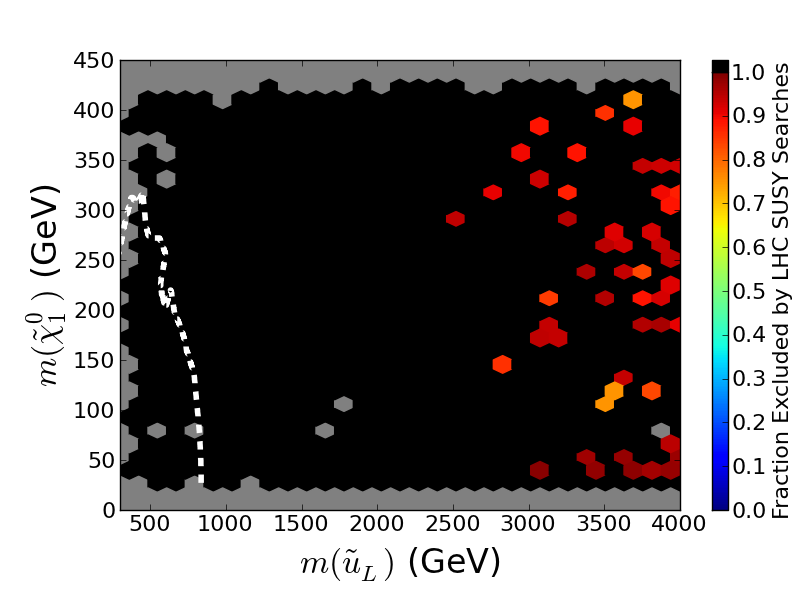}
\hspace{0.20cm}
\includegraphics[width=3.5in]{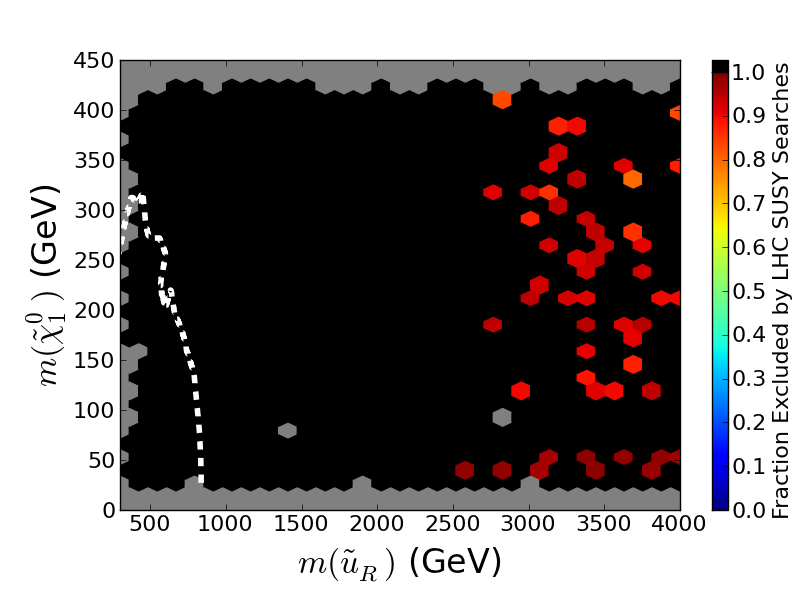}}
\vspace*{0.50cm}
\centerline{\includegraphics[width=3.5in]{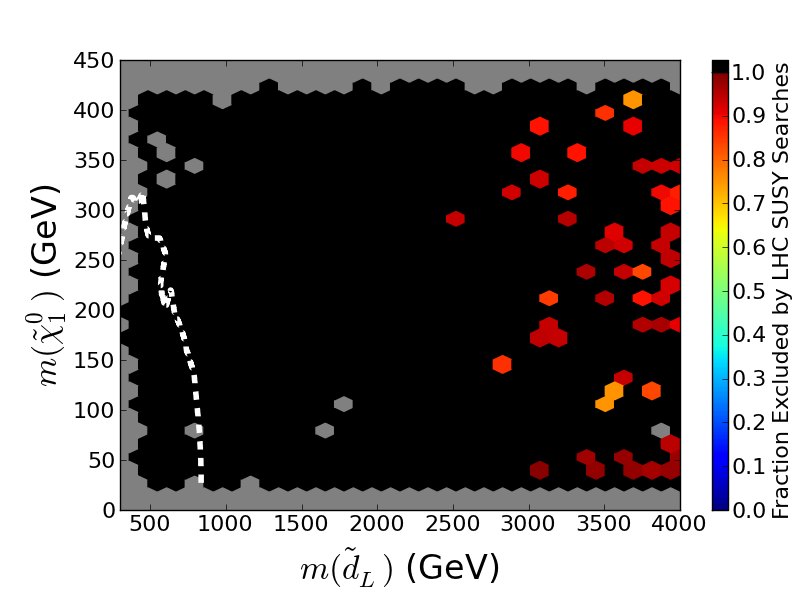}
\hspace{0.20cm}
\includegraphics[width=3.5in]{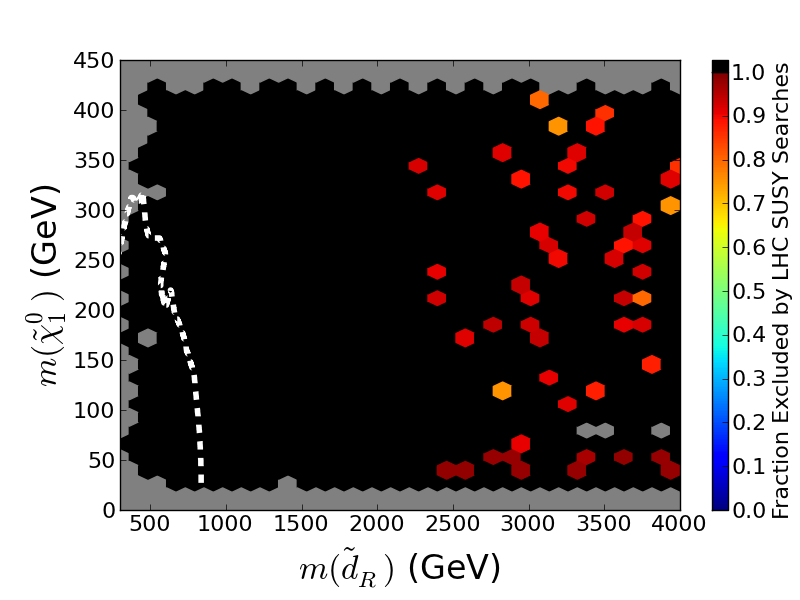}}
\vspace*{-0.10cm}
\caption{Results similar to those shown in Fig.~\ref{figxx2} above, but now for the low-FT model set for a luminosity of 300 fb$^{-1}$ as described in the text.} 
\label{figxxlowFT2}
\end{figure}

As we saw in the 7 and 8 TeV study above, third generation searches are especially powerful when acting upon the low-FT model sample since stops must now be relatively light. 
Table~\ref{SearchList14} shows that this is indeed the case and this is reflected in the stop-LSP and sbottom-LSP mass planes displayed in Fig.~\ref{figxxlowFT3}. However, we still see that 
models with stops and sbottoms as light as $\sim 750$ GeV remain viable after 300 fb$^{-1}$ for a wide range of LSP masses. Clearly additional third generation searches, particularly the direct sbottom search, will be required to probe such models at these luminosities. Additionally, by analogy with our 8 TeV results, we would expect multilepton searches to significantly increase the search reach in this plane.

\begin{figure}[htbp]
\centerline{\includegraphics[width=3.5in]{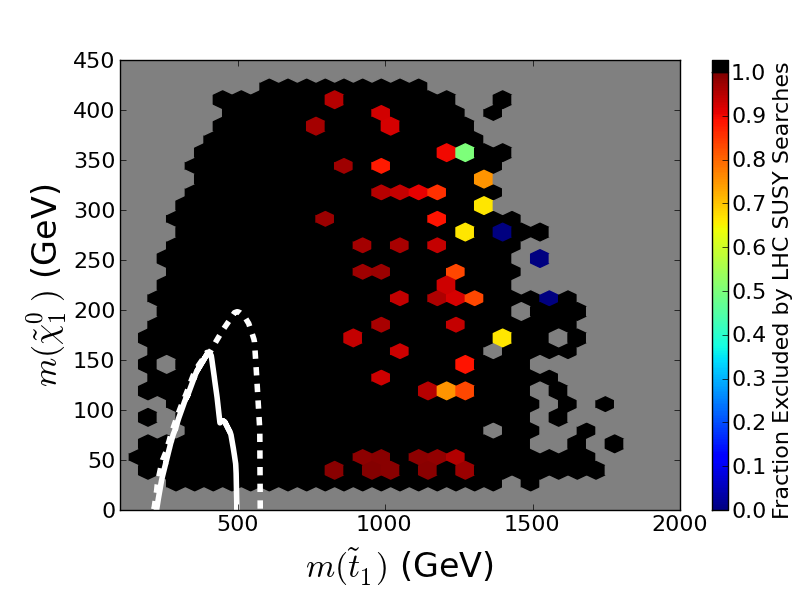}
\hspace{0.20cm}
\includegraphics[width=3.5in]{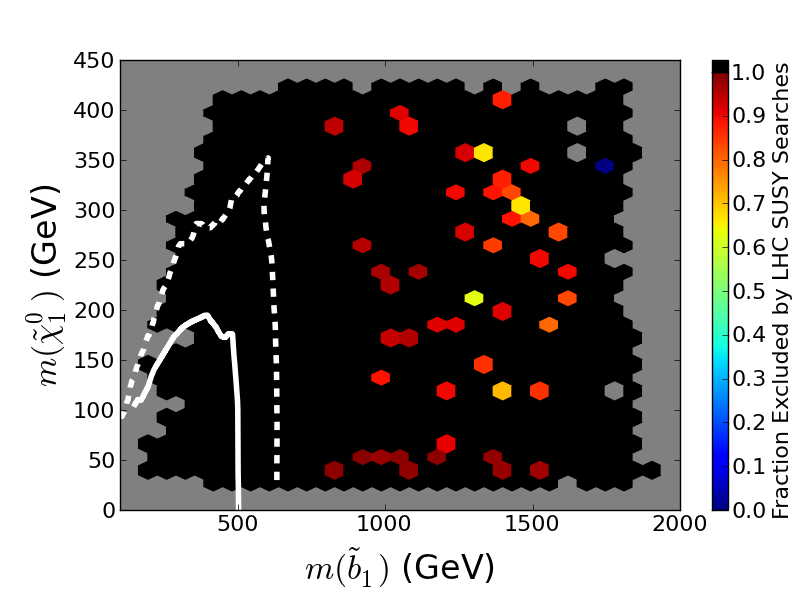}}
\vspace*{-0.10cm}
\caption{Results similar to those shown in Fig.~\ref{figrr1} above, but now for the low-FT model set for a luminosity of 300 fb$^{-1}$ as described in the text.}
\label{figxxlowFT3}
\end{figure}

Again, since the jets + MET search by itself is quite powerful, it is useful to know how much of the model coverage for the stop and sbottom squarks arose from this vanilla search in 
comparison to that arising from the 0$\ell$ and 1$\ell$ stop searches; Fig.~\ref{figxxlowFT4} addresses this question. Table~\ref{SearchList14} shows that the stop searches are substantially more effective in the low-FT model set than for the general neutralino LSP model set, and that once again they gain tremendously in sensitivity when the integrated luminosity is increased. This behavior is illustrated in Fig.~\ref{figxxlowFT4}. At 300 fb$^{-1}$ we see that the exclusion fractions are completely dominated by jets + MET for all stop and sbottom masses, while at 
3000 fb$^{-1}$ we see that the coverage is rather balanced between the two types of searches, particularly for stop masses below $\sim 1.2$ TeV. The same pattern is observed for sbottoms. The addition of other third generation searches would obviously improve the coverage of the stop mass plane at the lower luminosity.

\begin{figure}[htbp]
\centerline{\includegraphics[width=3.5in]{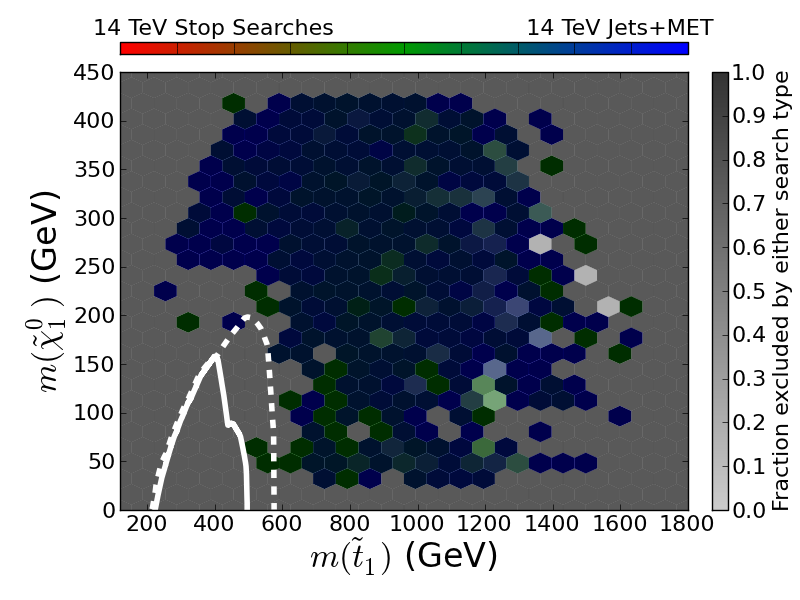}
\hspace{0.20cm}
\includegraphics[width=3.5in]{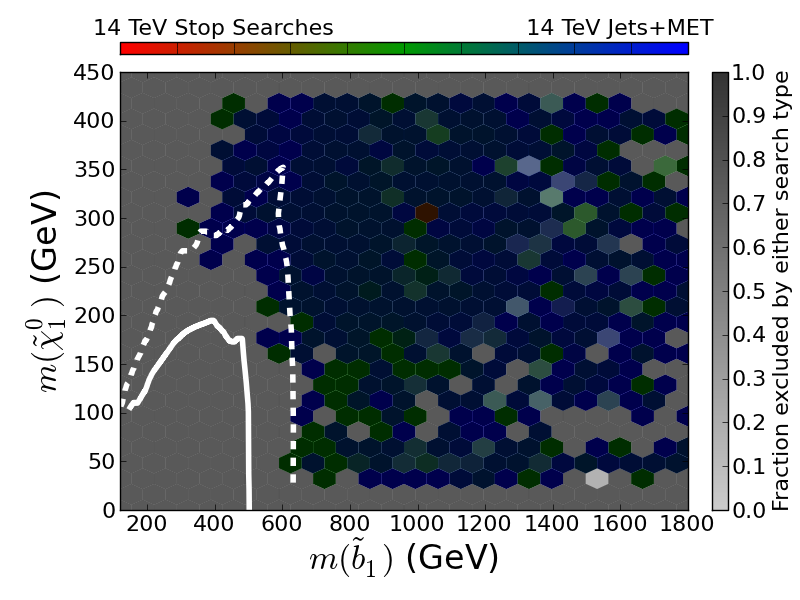}}
\vspace*{0.50cm}
\centerline{\includegraphics[width=3.5in]{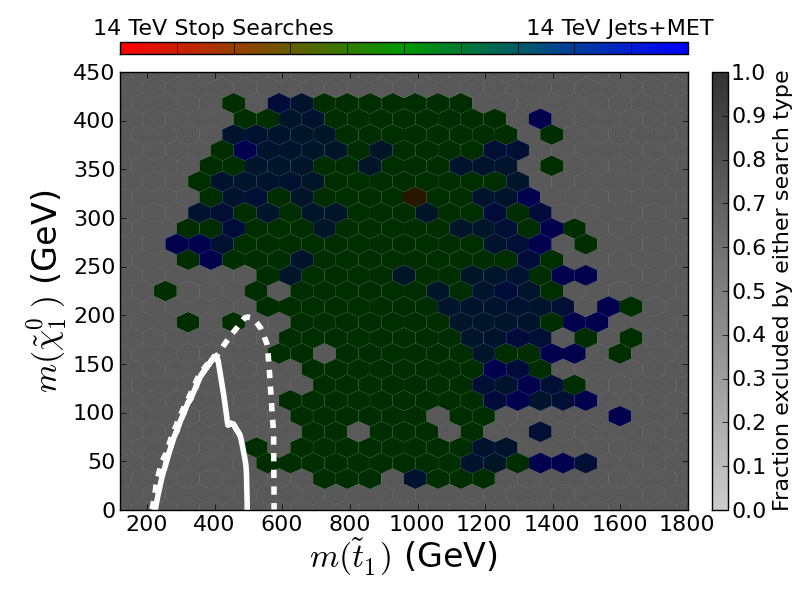}
\hspace{0.20cm}
\includegraphics[width=3.5in]{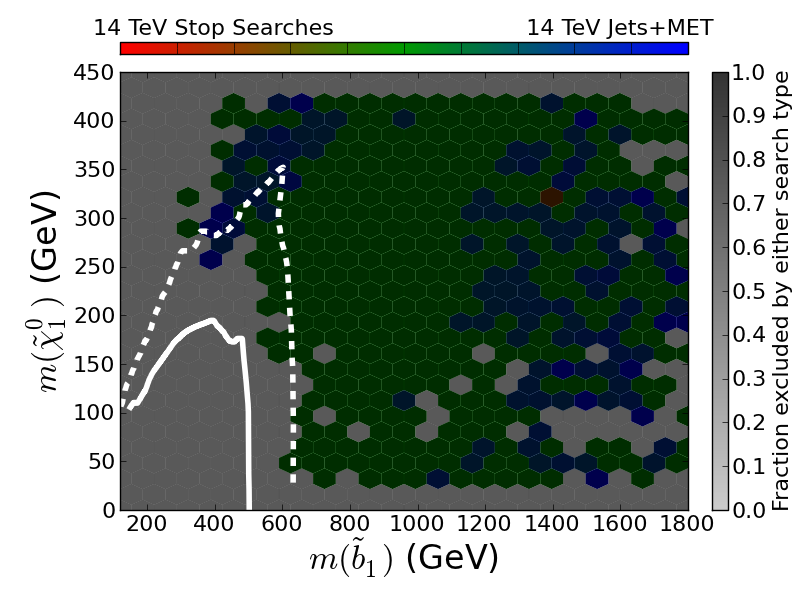}}
\vspace*{-0.10cm}
\caption{Results similar to those shown in Fig.~\ref{figrr2} above, but now for the low-FT model set as described in the text.}
\label{figxxlowFT4}
\end{figure}

\section{Summary}

The flexibility of the 19/20-parameter pMSSM provides a very powerful tool to combine, compare and contrast the various searches for SUSY at the LHC (and elsewhere), 
even those which employ different collision energies.  Here we have examined how the pMSSM parameter space is probed by the suite of ATLAS SUSY searches by replicating 
these analyses using fast Monte Carlo, and then determining how these searches impact two distinct pMSSM model samples, both with the lightest neutralino being
identified as the LSP.   The first is a large generic model set, and the second corresponds to a smaller specialized model set with low-FT and a thermal LSP saturating the relic density. We have shown that the models in these sets 
generally respond quite differently to the various SUSY searches. However, in both cases, we see that the combination of results obtained from the many LHC searches can 
significantly augment the total coverage of the model space.  Furthermore, not knowing the exact form that the SUSY spectrum might take {\it a priori}, all of the 
searches can play important roles in constraining the pMSSM model parameters. For models in either the neutralino or low-FT sets, we also found that the zero lepton, 
jets + MET search combined with the 0$\ell$ and 1$\ell$ stop searches at the 14 TeV LHC are very likely to be able to exclude (or discover!) the bulk of models that have survived the 7 and 
8 TeV searches. Indeed, complete coverage was found for the low-FT set with 3 ab$^{-1}$ of luminosity. Augmenting these searches with others at 14 TeV would be of significant 
interest and can only increase the already excellent reach that will be obtained by the searches considered here. 

In summary, we find that much phase-space is left to accommodate the existence of natural Supersymmetry (as defined above) after the conclusion of the 7,8 TeV LHC
operations.  Specifically, we find that the simplified model results do not describe the LHC search results in more complex forms of SUSY, and that numerous models
are currently viable that allow for light (500-1000 GeV) squarks and gluinos.  However, we show that the power of the 14 TeV LHC can provide a more definitive statement
on the existence of natural supersymmetry, even in  a complex form such as the pMSSM.  We conclude that the discovery space of the upcoming run is significant.

\section{Acknowledgments}

The authors would like to thank Alan Barr, David C\^{o}t\'{e}, Joe Lykken, and Brian Petersen for invaluable discussions.  We also thank Richard Dubois and Homer Neal for
computational support with the SLAC PPA batch farm system. This work was supported by the Department of Energy, Contracts DE-AC02-06CH11357, DE-AC02-76SF00515 and DE-FG02-12ER41811.

\end{document}